\documentclass[12pt]{article}
\pdfoutput=1

\usepackage[geometry=yurie,physics=yurie,caption=yurie,subfigure=off,tikz=yurie,biblatex=yurie,hyperref=yurie]{yurie}
\usepackage{yurie-math}
\usepackage{yurie-qft}

\newcommand{\phat}{\hat{p}}

\newcommand{\qhat}{\hat{q}}

\newcommand{\mathred}[1]{\mathcolor{red}{#1}}
\newcommand{\mathblue}[1]{\mathcolor{blue}{#1}}
\newcommand{\mathgreen}[1]{\mathcolor{orange}{#1}}
\newcommand{\complexDelta}[1]{2\pi \delta_{c}\left(i(#1)\right)}
\newcommand{\wilsonPolynomial}{W}

\usepackage{multirow}
\usepackage{tabularx}
\usepackage{float}

\newcommand{\constOfDiscreteSeries}{\alpha_{d}}

\begin{document}

\title{Massive celestial amplitudes and celestial amplitudes beyond four points}

\author{Reiko Liu$^{a}$, Wen-Jie Ma$^{b,c}$}

\maketitle

\address{\small${}^a$Yau Mathematical Sciences Center (YMSC), Tsinghua University, Beijing, 100084, China}

\address{\small${}^b$Fudan Center for Mathematics and Interdisciplinary Study, Fudan University, Shanghai, 200433, China}

\address{\small${}^c$Shanghai Institute for Mathematics and Interdisciplinary Sciences (SIMIS), Shanghai, 200433, China}

\email{{{\tt reiko@tsinghua.edu.cn, \tt wenjie.ma@simis.cn}}}

\vfill

\begin{abstract}

We compute scalar three-point celestial amplitudes involving two and three massive scalars. The three-point coefficient of celestial amplitudes with two massive scalars contains a hypergeometric function, and the one with three massive scalars can be represented as a triple Mellin-Barnes integral. Using these three-point celestial amplitudes, we investigate the conformal block expansions of five- and six-point scalar celestial amplitudes in the comb channel. We observe the presence of two-particle operators in the conformal block expansion of five-point celestial amplitudes, which confirms the previous analysis by taking multi-collinear limit. Moreover, we find that there are new three-particle operators in the conformal block expansion of six-point celestial amplitudes.  
Based on these findings, we conjecture that exchanges of $n$-particle operators can be observed by considering the comb channel conformal block expansion of $(n+3)$-point massless celestial amplitudes.
Finally, we show that a new series of operators appears when turning on the mass of the first incoming particle. The leading operator in this series can be interpreted as a two-particle exchange in the OPE of one massive and one massless scalars.

\end{abstract}

\date{}

\newpage
\tableofcontents
\newpage

\section{Introduction}

The celestial holography program suggests that a $(d+2)$-dimensional quantum gravity in asymptotic Minkowski spacetime is dual to a hypothetical $d$-dimensional celestial conformal field theory (celestial CFT, CCFT) on the celestial sphere \cite{Pasterski:2016qvg,Raclariu:2021zjz,Pasterski:2021raf}. 
In this duality, the celestial amplitudes - the bulk scattering amplitudes expanded in the conformal basis - should serve as correlation functions in the boundary CCFT.

As a CFT, it is expected that all the information of CCFT is encoded in the operator product expansion (OPE), which can be extracted from the three-point celestial amplitudes and the conformal block expansions of four- and higher-point celestial amplitudes. The three-point celestial amplitudes of three massless scalars were computed in \cite{Chang:2022seh} and the ones involving one massive scalar were computed in \cite{Lam:2017ofc}. 
In CCFT$_2$, the three-point celestial amplitudes of three massive scalars with mass $m$, $m$, and $2m(1+\epsilon)$ were obtained in \cite{Pasterski:2016qvg} to the leading order of $\epsilon$.\footnotemark{} 
However, the complete results of three-point celestial amplitudes involving more than one massive scalar remain unknown.

\footnotetext{By performing the integral in position space, the authors of \cite{Iacobacci:2022yjo} computed three-point celestial amplitudes of three massive scalars with generic mass $m_1$, $m_2$ and $m_3$ in CCFT$_d$. However, when setting $m_1=m_2=m$, $m_3=2m(1+\epsilon)$ and $d=2$, their results differ from the results of \cite{Pasterski:2016qvg} by a term that is regular in $\epsilon$.
Besides, in \cite{Sleight:2023ojm,Iacobacci:2024nhw} the authors proposed a modified conformal basis and computed three-point celestial amplitudes of three massive scalars in this new basis.}

For the conformal block expansions, the authors of \cite{Lam:2017ofc,Atanasov:2021cje,Chang:2021wvv,Fan:2021pbp,Fan:2021isc,Fan:2022kpp,Garcia-Sepulveda:2022lga,Chang:2022jut,Chang:2023ttm,Fan:2023lky,Himwich:2023njb} have investigated the conformal partial wave expansions and/or conformal block expansions of four-point massless scalar amplitudes. It is worth mentioning that four-point massless celestial amplitudes, which are expanded in the conformal primary basis given by a Mellin transform on the plane
waves, do not have proper conformal block expansions \cite{Chang:2021wvv}. This issue can be solved by choosing the conformal basis and shadow conformal basis for the incoming and outgoing particles respectively \cite{Chang:2022seh}. Furthermore, in order to obtain the complete OPE, it is necessary to further explore the conformal block expansions of celestial amplitudes involving massive particles. Unfortunately, such conformal block expansions are still absent in the literature since the computations are more involved.

Moreover, recent studies have revealed the presence of two-particle operators in the leading order OPE of two massless operators, in addition to the single-particle operators \cite{Ball:2023sdz,Guevara:2024ixn}. 
These two-particle exchanges can only be observed when higher-point celestial amplitudes are considered. 
Following the work of \cite{Ball:2023sdz,Guevara:2024ixn}, two questions arise.
Firstly, are there other types of exchanged operators appearing in the OPE of two massless operators? Secondly, do the two-particle exchanges also appear in the OPE of massive operators?
The answers to these questions can be obtained by studying the conformal block expansions of higher-point celestial amplitudes involving massless and/or massive particles.
Unfortunately, obtaining these expansions is a challenging task, and no results are currently available in the literature.

The split representation introduced in \cite{Chang:2023ttm} is a powerful tool to systematically compute the conformal block expansions of the celestial amplitudes. The split representation can be derived by dividing the Fourier integral in momentum space into integrals over the regions inside and outside the lightcone at the origin, where the regions inside the past or future lightcones are foliated by $(d+1)$-dimensional Euclidean anti-de Sitter (EAdS) slices, and the region outside the lightcones is foliated by $(d+1)$-dimensional de Sitter (dS) slices.\footnote{Similar foliation has been recently discussed in \cite{Melton:2023bjw,Melton:2024gyu} to obtain the leaf amplitudes.}
Using the split representation, the Feynman propagator can be factorized into a product of two conformal primary wavefunctions with a common boundary point integrated over the celestial sphere. 
This decomposition enables the direct derivation of the conformal partial wave expansions for celestial amplitudes. Subsequently, 
it is straightforward to obtain the conformal block expansions by the standard techniques in CFT.

In this paper, we focus on the scattering process of scalar particles. 
Firstly, we calculate the three-point celestial amplitudes involving more than one massive scalar. Using these three-point celestial amplitudes and the split representation, we then investigate the conformal block expansion of the four-point celestial amplitude involving two massive scalars. Additionally, we obtain the conformal block expansions of five- and six-point massless celestial amplitudes by choosing the conformal basis and shadow conformal basis for the incoming and outgoing particles, respectively. Finally, we determine the conformal block expansions of five-point celestial amplitudes involving one massive scalar.

The rest of this paper is organized as follows. 
Section \ref{sec:preliminaries} sets up conventions, and reviews the shadow formalism, the dictionary of celestial holography and the split representation.
Section \ref{sec:CPW} derives higher-point conformal partial wave expansions in the comb channel.
Section \ref{sec:3pt} evaluates the three-point coefficients of massive celestial amplitudes.
Section \ref{sec:4pt} and Section \ref{sec:5&6pt} discuss conformal block expansions of four- and higher-point celestial amplitudes.
Section \ref{sec:conclusion} ends with concluding remarks.
Appendix \ref{sec:formulas} collects useful formulas.
Appendix \ref{sec:tables} contains tables of operator spectra.

\section{Preliminaries}
\label{sec:preliminaries}

\subsection{Notations and conventions}

Throughout this paper we adopt the following notations and conventions. 
We choose the most positive signature for the $(d+2)$-dimensional Minkowski spacetime $\RR^{1,d+1}$, \ie,
\begin{align}
    g_{\mu\nu}=\operatorname{diag}(-,+,\dots,+)
    \; .
\end{align}
The bulk point in the Minkowski spacetime is denoted as $X\in\RR^{1,d+1}$, while the boundary point on the celestial sphere is denoted as $x\in\RR^{d}$. The difference between two boundary points $x_{i}$ and $x_{j}$ is denoted as
\begin{align}
    x_{ij}\equiv x_i-x_j
    \; .
\end{align}

For a $(d+2)$-dimensional scattering process, we denote the $i$-th, mass $m$, helicity/spin $s$ particle as $i^{s}_m$. 
If taking the shadow transform on the $i$-th particle, we denote it as $\shadow{i^{s}_m}$.
For example, the $2\to 2$ celestial amplitude of four massless scalars is denoted as
\begin{equation}
    \cA_{1^0_0+2^0_0\to 3^0_0+4^0_0}
    \; ,
\end{equation}
and if choosing the shadow basis for outgoing states, it is denoted as 
\begin{equation}
    \cA_{1^0_0+2^0_0\to \shadow{3^0_0}+\shadow{4^0_0}}
    \; .
\end{equation}

For a $d$-dimensional CFT, we denote the coordinate-dependence of the standard conformal three-point functions as
\begin{align}
    \vev{\op_{i}\op_{j}\op_{k}}
    =
    |x_{12}|^{\Delta_{3,12}}|x_{13}|^{\Delta_{2,13}}|x_{23}|^{\Delta_{1,23}}
    \; .
\end{align}
Here the default position and conformal dimension of the primary operator $\op_{i}$ are $x_{i}$ and $\Delta_{i}$ respectively, \ie, $\op_{i}\equiv\op_{\Delta_{i}}(x_{i}),\, \wave\op_{i}\equiv\op_{d-\Delta_{i}}(x_{i})$.
Moreover, we introduce the shadow and/or multiple conformal dimensions as
\begin{align}
    &
    \wave\Delta_{a}
    \equiv d-\Delta_{a}
    \; ,
    \\
    &
    \Delta_{a_{1}\cdots a_{n}}
    \equiv
    \sum_{i=1}^{n}\Delta_{a_{i}}
    \; ,
    \\
    &
    \Delta_{a_{1}\cdots a_{n},b_{1}\cdots b_{m}}
    \equiv
    \sum_{i=1}^{n}\Delta_{a_{i}}-
    \sum_{j=1}^{m}\Delta_{b_{j}}
    \; ,
\end{align}
and the multiple $\Gamma$-symbols as
\begin{align}
    &\Gamma[a_{1},\cdots,a_{n}]
    \equiv
    \prod_{i=1}^{n}\gm{a_{i}}
    \; ,
    \\
    &\multiGamma{a_{1},\cdots,a_{n} }{b_{1},\cdots,b_{m} }
    \equiv
    \left.
    \prod_{i=1}^{n}\gm{a_{i}}
    \middle/
    \prod_{j=1}^{m}\gm{b_{j}}
    \right.
    \; .
\end{align}
%

\subsection{Shadow formalism}

The shadow formalism is a widely used technique in CFT.
It's based on the representation theory and harmonic analysis of conformal groups, and allows us to conveniently compute conformal partial waves and conformal blocks \cite{Ferrara:1972uq,Dobrev:1976vr,Dobrev:1977qv,SimmonsDuffin:2012uy,Karateev:2018oml,Kravchuk:2018htv,Chen:2022cpx}.

In Euclidean CFT$_d$, the shadow transform of a (scalar) primary operator is defined as
\begin{equation}
    \shadow{\op_{\Delta}}(x)
    =
    N_{\Delta}
    \intt{d^{d}x'}|x-x'|^{-2\wave\Delta}
    \op_{\Delta}(x')
    \; ,
\end{equation}
where the normalization factor $N_{\Delta}$ is to ensure that the double shadow transform is equal to the identity $\shadow^{2}[\op_{\Delta}]=\op_{\Delta}$,
\begin{equation}
    \label{eq:shadowTransformNormalization}
    N_{\Delta}
    =
    \pi^{-\halfdim}\frac{\gm{\Delta}}{\gm{\Delta-\halfdim}}
    \; .
\end{equation}

For a local primary operator $\op_{\Delta}(x)$, the non-local operator $\shadow{\op_{\Delta}}(x)$ transforms as a primary operator with conformal dimension $\wave\Delta$. This implies that $\vev{\op_{1}\op_{2}\shadow{\op_{3}}}$ is proportional to $\vev{\op_{1}\op_{2}\wave\op_{3}}$: 
\begin{equation}
    \vev{\op_{1}\op_{2}\shadow{\op_{3}}}
    =
    N_{\Delta_{3}} 
    S^{\Delta_1,\Delta_2}_{\Delta_{3}}
    \vev{\op_{1}\op_{2}\wave\op_{3}}
    \; ,
\end{equation}
where the shadow coefficient $S^{\Delta_1,\Delta_2}_{\Delta_{3}}$ is
\begin{equation}\label{eq:shadow_coefficient}
    S^{\Delta_1,\Delta_2}_{\Delta_{3}}
    =
    \pi^{\frac{d}{2}}
    \multiGamma{
        \Delta_{3}-\halfdim,\frac{d+\Delta_{1,23}}{2},\frac{d+\Delta_{2,31}}{2}
    }{
        d-\Delta_{3},\frac{\Delta_{23,1}}{2},\frac{\Delta_{31,2}}{2}
    }
    \; .
\end{equation}
%

\subsection{Conformal primary wavefunctions and celestial amplitudes}\label{sec:ConformalBasis}

In a $d$-dimensional CCFT, the celestial amplitudes are defined by expanding the scattering amplitudes in the conformal basis \cite{Pasterski:2016qvg,Pasterski:2017kqt}, instead of in the usual plane-wave basis.
The conformal basis can be thought as a bulk-to-boundary propagator in $\RR^{d+1,1}$ since it connects a bulk point $X$ in the $(d+2)$-dimensional Minkowski spacetime and a boundary point $x$ on the $d$-dimensional celestial sphere \cite{Chang:2023ttm}.\footnotemark{}

\footnotetext{
Recently it was shown that the conformal basis can be obtained by taking a flat space limit of AdS$_{d+2}$ bulk-to-boundary propagators \cite{deGioia:2022fcn,deGioia:2024yne}.
}

Given a $n\to m$ amputated scattering amplitude $\mathcal{M}(X_i)$ of scalars in the position space, the celestial amplitude $\cA(x_i)$ is defined as
\begin{align}\label{eq:CA}
\cA(x_i)
=
\bigg(\prod_{i=1}^{n+m}\int d^{d+2}X_i\bigg)
\bigg(\prod_{i=1}^n\phi_{\Delta_i}^{-}(x_i;X_i)\bigg)
\bigg(\prod_{i=n+1}^{n+m}\phi^{+}_{\Delta_i}(x_i;X_{i})\bigg)
\mathcal{M}(X_i)
\; ,
\end{align}
where $\phi^{\pm}_{\Delta}(x;X)$-s are the incoming $(-)$ and outgoing $(+)$ scalar conformal primary wavefunctions with the conformal dimension $\Delta$. 
For example, the $2\to 2$ contact celestial amplitude of massless scalars is given by
\begin{align}
\cA^{\Delta_i}_{1^0_0+2^0_0\to 3^0_0+4^0_0}(x_i)=\int d^{d+2}X\,
\phi^{-}_{\Delta_1}(x_1,X)\phi^{-}_{\Delta_2}(x_2,X)\phi^{+}_{\Delta_3}(x_3,X)\phi^{+}_{\Delta_4}(x_4,X)
\; .
\end{align}

The conformal primary wavefunctions $\phi^{\pm}_{\Delta,m}(x;X)$ and $\phi^{\pm}_{\Delta}(x;X)$ for massive and massless scalars are
\begin{align}
    \label{eq:CPWFScalarMassive}
    &\phi^{\pm}_{\Delta,m}(x;X)
    =
    \intt{\frac{d^{d+1}\phat}{\phat^{0}}}
    (-\qhat(x)\cdot\phat)^{-\Delta} 
    e^{\pm im\phat\cdot X}
    \; ,
    \\
    \label{eq:CPWFScalarMassless}
    &
    \phi^{\pm}_{\Delta}(x;X)
    =
    \intrange{d\omega}{0}{+\infty}
    \omega^{\Delta-1}
    e^{\pm i\omega\qhat(x)\cdot X-\epsilon\omega}
    =
    (\mp i)^{\Delta}\Gamma[\Delta]
    (-\hat{q}\cdot X\mp i\epsilon)^{-\Delta}
    \; ,
\end{align}
%
where $m$ is the mass of the massive scalar,
and $p^\mu$ and $q^{\mu}$ are massive and massless on-shell momenta with the parameterizations 
\begin{align}
    &p^{\mu}(m,y,x)= m\hat{p}^{\mu}(y,x)=\frac{m}{2y}(1+y^2+x^2,2x,1-y^2-x^2)
    \; ,
    \\
    &q^{\mu}(\omega,x)=\omega\hat{q}^{\mu}(x)=\omega(1+x^2,2x,1-x^2)
    \; .
\end{align}

There is another set of conformal primary wavefunctions given by the shadow transform of the wavefunctions in \eqref{eq:CPWFScalarMassive} and \eqref{eq:CPWFScalarMassless}. 
As shown in \cite{Pasterski:2017kqt}, the shadow transform of the massive scalar conformal primary wavefunction $\phi^{\pm}_{\Delta,m}(x;X)$ is simply $\phi^{\pm}_{\wave\Delta,m}(x;X)$, \ie,
\begin{align}
    \label{eq:CPWFScalarMassiveShadow}
    \shadow{\phi^{\pm}_{\Delta,m}}(x;X)=\phi^{\pm}_{\widetilde{\Delta},m}(x;X)
    \; .
\end{align}
The shadow transform of the massless scalar conformal primary wavefunction $\phi^{\pm}_{\Delta}(x;X)$ is 
\begin{align}
    \label{eq:CPWFScalarMasslessShadow}
    \shadow{\phi_{\Delta}^{\pm}}(x;X)
    =
    N_{\Delta}
    \int d^dx^{\prime}
    |x-x^{\prime}|^{-2\wave\Delta}
    \phi_{\Delta}^{\pm}(x^{\prime};X)
    =
    \frac{
        (\mp i)^{\Delta}\Gamma[\Delta]
    }{
        (\mp i)^{\wave\Delta}\Gamma[\wave\Delta]
    }
    (-X^2)^{\frac{d}{2}-\Delta}\phi_{\wave\Delta}^{\pm}(x;X)
    \; .
\end{align}
where the normalization factor $N_{\Delta}$ is given in \eqref{eq:shadowTransformNormalization}.

Starting with the massive conformal primary wavefunctions $\phi^{\pm}_{\Delta,m}$, one can reproduce either the massless conformal primary wavefunctions $\phi^{\pm}_{\Delta}$ or the massless shadow conformal primary wavefunctions $\shadow{\phi^{\pm}_{\widetilde{\Delta}}}$ by taking the massless limit \cite{Chang:2022seh}:
\begin{align}
    \label{eq:CPWFScalarMasslessLimit}
    &\phi_{\Delta}^{\pm}
    =
    \lim_{m \to 0} 
    2^{-\Delta}\pi^{-\halfdim} m^{\Delta}\frac{\gm{\Delta}}{\gm{\Delta-\halfdim}}
    \phi_{\Delta, m}^{\pm}
    \; ,
    \\
    &\shadow{\phi_{\widetilde{\Delta}}^\pm}
    =
    \lim_{m\to 0}
    2^{-\wave\Delta}\pi^{-\halfdim} m^{\wave\Delta}\frac{\gm{\wave\Delta}}{\gm{\wave\Delta-\halfdim}}
    \phi_{\Delta, m}^{\pm}
    \; .
\end{align}
%

\subsection{Split representation of Feynman propagators}
The split representation of the Feynman propagator in the Minkowski spacetime has been derived in \cite{Chang:2023ttm}. It enables us to rewrite the Feynman propagator as a product of a pair of (tachyonic) massive conformal bases with conformal dimensions $\Delta$ and $\wave\Delta$, and with the common boundary point, the conformal dimension $\Delta$ and the (imaginary) mass integrated. For example, the scalar Feynman propagator in position space 
\begin{align}\label{eq:BtB}
\mathcal{K}_m(X_1,X_2)=\int\frac{d^{d+2}p}{(2\pi)^{d+2}}\frac{i}{p^2+m^2-i\epsilon}e^{-ip\cdot(X_1-X_2)}
\end{align}
can be rewritten as
\begingroup
%
\begin{align}
\label{eq:SplitRep}
\begin{split}
\mathcal{K}_{m}(X_1,X_2)
&=
\int_{0}^{+\infty}\frac{dM}{(2\pi)^{d+2}}\frac{iM^{d+1}}{-M^2+m^2}
\int_{\frac{d}{2}}^{\frac{d}{2}+i\infty}\frac{d\Delta}{2\pi i\, \mu_{\Delta}}
\int d^dx\, 
\left(
    \phi^{-}_{\Delta,M}\phi^{+}_{\wave\Delta,M}
    +
    \phi^{+}_{\Delta,M}\phi^{-}_{\wave\Delta,M}
\right)
\\
&\peq
+
\frac{1}{2}
\int_{0}^{\infty}\frac{dM}{(2\pi)^{d+2}}\frac{iM^{d+1}}{M^2+m^2}
\bigg(
\sum_{\epsilon=0,1}
\int_{\frac{d}{2}}^{\frac{d}{2}+i\infty}\frac{(-1)^{\epsilon}d\Delta}{2\pi i\, \mu_{\Delta}}
\int d^dx\, 
\phi_{\Delta,iM,\epsilon}\phi_{\wave\Delta,iM,\epsilon}
\\
&\peq
+
\constOfDiscreteSeries
\underset{(\Delta,\epsilon)\in D_{\text{dS}}}{\sum\Res}
\frac{(-1)^{\epsilon}}{\mu_{\Delta}}
\int d^dx\, 
\phi_{\Delta,iM,\epsilon}\phi_{d-\Delta,iM,\epsilon}
\bigg)
\; .
\end{split}
\end{align}
\endgroup
Here $\phi_{\Delta,iM,\epsilon}$-s are the conformal primary wavefunctions of tachyonic scalars,
$\constOfDiscreteSeries=1$ for odd $d$ and $\constOfDiscreteSeries=\half$ for even $d$, the summation range of the discrete series is
\begin{equation}
    D_{\text{dS}}=\set{(\Delta\in\ZZ,\epsilon):\,  \Delta>\halfdim \textInMath{and} \epsilon=1+d+\Delta\modd{\ZZ_2}}
    \; ,
\end{equation}
and $\mu^{-1}_{\Delta}$ is the Plancherel measure given by
\begin{align}\label{eq:plancherel_measure}
    \mu_{\Delta}=
    \pi^d \, 
    \multiGamma{
        \Delta -\frac{d}{2},
        \frac{d}{2}-\Delta
    }{
        \Delta,
        d-\Delta
    }
    \; .
\end{align}

The split representation is a powerful tool for efficiently calculating the conformal partial wave expansion of celestial amplitudes. For example, by employing the split representation, it has been demonstrated that in the case of tree-level scalar celestial amplitudes, the coefficient, apart from the Plancherel measure, in the conformal partial wave expansion can be factorized into a product of three-point coefficients involving massive and/or tachyonic scalars \cite{Chang:2023ttm}.

In this paper, we will utilize the split representation \eqref{eq:SplitRep} to compute the conformal block expansions of four- and higher-point celestial amplitudes. Since all the examples discussed in this paper are $s$-channel amplitudes, only the first line in \eqref{eq:SplitRep} contributes to the final results.

\section{Conformal partial wave expansions}\label{sec:CPW}
The conformal block expansions of four- and higher-point celestial amplitudes can be obtained from the conformal partial wave expansions of the corresponding celestial amplitudes. 
In this section, we will start with a brief review of the conformal partial wave expansion of four-point celestial amplitudes. Then, we will derive the relations between the five- and six-point conformal partial waves and the corresponding conformal blocks. Finally, we will establish the conformal partial wave expansions of five- and six-point celestial amplitudes, with a focus on the comb channel.
For further results of higher-point conformal blocks, see \cite{Alkalaev:2015fbw,Rosenhaus:2018zqn,Parikh:2019ygo,Goncalves:2019znr,Jepsen:2019svc,Parikh:2019dvm,Fortin:2019zkm,Fortin:2020yjz,Anous:2020vtw,Haehl:2021tft,Fortin:2020bfq,Hoback:2020pgj,fortin2020all,Buric:2020dyz,Hoback:2020syd,Poland:2021xjs,Buric:2021ywo,Buric:2021ttm,Buric:2021kgy,fortin2022feynman,Fortin:2023xqq}.

\subsection{Four-point conformal partial waves}

The (scalar) four-point conformal partial wave is 
\begin{equation}
    \label{eq:fourPointCPW}
    \Psi^{\Delta_{i}}_{\Delta_{a}}=
    \intt{d^dx_{a}}
    \vev{\op_{1}\op_{2}\op_{a}}
    \vev{\wave{\op}_{a}\op_{3}\op_{4}}
    \; ,
\end{equation}
and is related to the conformal block by
\begin{align}
\label{eq:fourPointCPW2Block}
\begin{split}
&\Psi^{\Delta_{i}}_{\Delta_{a}}
=
S^{\Delta_3,\Delta_4}_{\wave\Delta_{a}}G^{\Delta_{i}}_{\Delta_{a}}
+
S^{\Delta_1,\Delta_2}_{\Delta_{a}}G^{\Delta_{i}}_{\wave\Delta_{a}}
\; .
\end{split}
\end{align}
The shadow symmetry of the four-point conformal partial wave is
\begin{equation}
    \label{eq:fourPointCPWShadowSymmetry}
    \Psi_{\wave\Delta_{a}}^{\Delta_i}
    =
    \frac{S_{\Delta_{a}}^{\Delta_3, \Delta_4}}{S_{\Delta_{a}}^{\Delta_1, \Delta_2}} 
    \Psi_{\Delta_{a}}^{\Delta_i}
    \; .
\end{equation}
For a (scalar) four-point celestial amplitude $\cA^{\Delta_i}_4$, the conformal partial wave expansion is\footnotemark{}
\footnotetext{
    For generic celestial amplitudes, spinning conformal partial waves $\Psi^{\Delta_i}_{\Delta_a,J_a}$ with spin $J_a$ can appear in the conformal partial wave expansion. 
    However for the examples in this paper, it turns out that there appear only scalar exchanges with $J_a=0$. Thus we omit the spinning contributions and write the conformal partial wave expansion as \eqref{eq:conformal_partial_wave_expansion}.
}
\begin{align}\label{eq:conformal_partial_wave_expansion}
    \cA^{\Delta_i}_4
    =
    \int_{\frac{d}{2}}^{\frac{d}{2}+i\infty}\frac{d\Delta_a}{2\pi i\, \mu_{\Delta_a}}
    I_{\Delta_a}\Psi^{\Delta_{i}}_{\Delta_{a}}
    \; ,
\end{align}
where $\mu_{\Delta_a}^{-1}$ is the Plancherel measure, given in \eqref{eq:plancherel_measure}. 
The density function $I_{\Delta_a}$ contains all the theory-specific information in the four-point function $\cA^{\Delta_i}_4$, and can be obtained by the Euclidean inversion formula \cite{Karateev:2018oml}
\begin{equation}
    \label{eq:inversionFormula}
    I_{\Delta_a}
    =
    \frac{1}{2^{d}\volume{\sogroup(d)}}
    \int 
    \frac{
        d^{d}x_{1}
        d^{d}x_{2}
        d^{d}x_{3}
        d^{d}x_{4}
    }{
        \volume{\sogroup(1,d+1)}
    }
    \cA^{\Delta_i}_4(x_i)
    \left(
        \Psi^{\Delta_{i}}_{\Delta_{a}}(x_{i})
    \right)^{*}
    \; .
\end{equation}
From the inversion formula the density function satisfies the shadow symmetry automatically,
\begin{equation}\label{eq:I_symmetry}
    I_{\wave\Delta_{a}}=
    \frac{S_{\Delta_{a}}^{\Delta_1, \Delta_2}}{S_{\Delta_{a}}^{\Delta_3, \Delta_4}} 
    I_{\Delta_{a}}
    \; .
\end{equation}

Under the shadow symmetries \eqref{eq:fourPointCPWShadowSymmetry} and \eqref{eq:I_symmetry}, substituting the conformal partial wave \eqref{eq:fourPointCPW2Block} into the expansion \eqref{eq:conformal_partial_wave_expansion}, we can rewrite \eqref{eq:conformal_partial_wave_expansion} as
\begin{align}\label{eq:conformal_partial_wave_expansion_block}
    \cA^{\Delta_i}_4
    =
    \int_{\frac{d}{2}-i\infty}^{\frac{d}{2}+i\infty}\frac{d\Delta_a}{2\pi i \, \mu_{\Delta_a}}
    I_{\Delta_a}
    S^{\Delta_3,\Delta_4}_{\wave{\Delta}_a}
    G^{\Delta_{i}}_{\Delta_{a}}
    \; .
\end{align}
Then enclosing the contour into the right half $\Delta_a$-plane and picking up the $\Delta_{a}$-poles give the conformal block expansion of the celestial amplitude $\cA^{\Delta_i}_4$.

\subsection{Higher-point conformal partial waves}

In this subsection, we will establish the conformal partial wave expansions for the scalar five- and six-point celestial amplitudes with scalar exchanges in comb channel. 
Similar to \eqref{eq:fourPointCPW2Block}, we will first express the conformal partial waves as a linear combination of conformal blocks and shadow conformal blocks.

By the shadow formalism, the five-point comb channel conformal partial wave is
\begin{equation}
    \label{eq:fivePointCPW}
    \partialWaveDressed^{\Delta_{i}}_{\Delta_{a},\Delta_{b}}
    =
    \int d^dx_a
    \int d^dx_b
    \vev{\op_{1}\op_{2}\op_{a}}
    \vev{\wave{\op}_{a}\op_{3}\op_{b}}
    \vev{\wave{\op}_{b}\op_{4}\op_{5}}
    \; .
\end{equation}
The shadow symmetry of the four-point conformal partial wave \eqref{eq:fourPointCPWShadowSymmetry} implies that $\partialWaveDressed^{\Delta_{i}}_{\Delta_{a},\Delta_{b}}$ satisfies
\begin{align}
    \label{eq:fivePointCPWShadowSymmetry1}
    &\partialWaveDressed^{\Delta_{i}}_{\wave\Delta_{a},\Delta_{b}}
    =
    \frac{S_{\Delta_{a}}^{\Delta_3, \Delta_b}}{S_{\Delta_{a}}^{\Delta_1, \Delta_2}} 
    \partialWaveDressed^{\Delta_{i}}_{\Delta_{a},\Delta_{b}}
    \; ,
    \\
    \label{eq:fivePointCPWShadowSymmetry2}
    &\partialWaveDressed^{\Delta_{i}}_{\Delta_{a},\wave\Delta_{b}}
    =
    \frac{S_{\Delta_{b}}^{\Delta_4, \Delta_5}}{S_{\Delta_{b}}^{\wave\Delta_a, \Delta_3}} 
    \partialWaveDressed^{\Delta_{i}}_{\Delta_{a},\Delta_{b}}
    \; .
\end{align}

We note that the conformal partial wave \eqref{eq:fivePointCPW} can be written as a linear combination of four conformal blocks since they satisfy the same set of conformal Casimir equations. In other words, we have
\begin{align}\label{eq:5ptCPW_a}
\begin{split}
\Psi^{\Delta_{i}}_{\Delta_{a},\Delta_{b}}
&=
a_{1}
G^{\Delta_{i}}_{\Delta_{a},\Delta_{b}}
+
a_{2}
G^{\Delta_{i}}_{\Delta_{a},\wave\Delta_{b}}
+
a_{3}G^{\Delta_{i}}_{\wave\Delta_{a},\Delta_{b}}
+
a_{4}
G^{\Delta_{i}}_{\wave\Delta_{a},\wave\Delta_{b}}
\; .
\end{split}
\end{align}
Using the $\op_{1}\times\op_{2}$ OPE, we can fix the first two coefficients as
\begin{equation}
    a_{1}=
    S^{\Delta_3,\Delta_b}_{\wave\Delta_{a}}
    S^{\Delta_4,\Delta_5}_{\wave\Delta_{b}}
    \; ,
    \quad 
    a_{2}=
    S^{\Delta_3,\Delta_b}_{\wave\Delta_{a}}
    S^{\Delta_a,\Delta_3}_{\Delta_{b}}
    \; ,
\end{equation}
and similarly using $\op_{4}\times \op_{5}$ OPE, we have
\begin{equation}
    a_{4}
    =
    S^{\Delta_{1},\Delta_{2}}_{\Delta_{a}}
    S^{\wave\Delta_{a},\Delta_3}_{\Delta_{b}}
    \; .
\end{equation}
Finally, the coefficient $a_{3}$ can be fixed by comparing the coefficients on both sides of \eqref{eq:fivePointCPWShadowSymmetry1}, leading to
\begin{equation}
    a_{3}
    =
    \left(
        \frac{S_{\Delta_{a}}^{\Delta_3, \Delta_b}}{S_{\Delta_{a}}^{\Delta_1, \Delta_2}}
        a_{1}
    \right)
    \bigg|_{\Delta_{a}\to \wave\Delta_{a}}
    =
    S^{\Delta_1,\Delta_2}_{\Delta_{a}}
    S^{\Delta_4,\Delta_5}_{\wave\Delta_{b}}
    \; .
\end{equation}
This fixes all of coefficients in \eqref{eq:5ptCPW_a} and gives 
\begin{align}
\label{eq:fivePointCPW2Block}
\begin{split}
\Psi^{\Delta_{i}}_{\Delta_{a},\Delta_{b}}
&=
S^{\Delta_3,\Delta_b}_{\wave\Delta_{a}}
S^{\Delta_4,\Delta_5}_{\wave\Delta_{b}}
G^{\Delta_{i}}_{\Delta_{a},\Delta_{b}}
+
S^{\Delta_3,\Delta_b}_{\wave\Delta_{a}}
S^{\Delta_a,\Delta_3}_{\Delta_{b}}
G^{\Delta_{i}}_{\Delta_{a},\wave\Delta_{b}}
\\
&\peq
+
S^{\Delta_1,\Delta_2}_{\Delta_{a}}
S^{\Delta_4,\Delta_5}_{\wave\Delta_{b}}
G^{\Delta_{i}}_{\wave\Delta_{a},\Delta_{b}}
+
S^{\Delta_{1},\Delta_{2}}_{\Delta_{a}}
S^{\wave\Delta_{a},\Delta_3}_{\Delta_{b}}
G^{\Delta_{i}}_{\wave\Delta_{a},\wave\Delta_{b}}
\; .
\end{split}
\end{align}

The six-point comb channel conformal partial wave is 
\begin{equation}
    \label{eq:sixPointCPW}
    \partialWaveDressed^{\Delta_{i}}_{\Delta_{a},\Delta_{b},\Delta_{c}}
    =
    \int d^dx_a
    \int d^dx_b
    \int d^dx_c
    \vev{\op_{1}\op_{2}\op_{a}}
    \vev{\wave{\op}_{a}\op_{3}\op_{b}}
    \vev{\wave{\op}_{b}\op_{4}\op_{c}}
    \vev{\wave{\op}_{c}\op_{5}\op_{6}}
    \; ,
\end{equation}
which respects the following shadow symmetries,
\begin{align}
    \label{eq:sixPointCPWShadowSymmetry1}
    &\partialWaveDressed^{\Delta_{i}}_{\wave\Delta_{a},\Delta_{b},\Delta_{c}}
    =
    \frac{S_{\Delta_{a}}^{\Delta_3, \Delta_b}}{S_{\Delta_{a}}^{\Delta_1, \Delta_2}} 
    \partialWaveDressed^{\Delta_{i}}_{\Delta_{a},\Delta_{b},\Delta_{c}}
    \; ,
    \\
    \label{eq:sixPointCPWShadowSymmetry2}
    &\partialWaveDressed^{\Delta_{i}}_{\Delta_{a},\wave\Delta_{b},\Delta_{c}}
    =
    \frac{S_{\Delta_{b}}^{\Delta_4, \Delta_c}}{S_{\Delta_{b}}^{\wave\Delta_a, \Delta_3}} 
    \partialWaveDressed^{\Delta_{i}}_{\Delta_{a},\Delta_{b},\Delta_{c}}
    \; ,
    \\
    \label{eq:sixPointCPWShadowSymmetry3}
    &\partialWaveDressed^{\Delta_{i}}_{\Delta_{a},\Delta_{b},\wave\Delta_{c}}
    =
    \frac{S_{\Delta_{c}}^{\Delta_5, \Delta_6}}{S_{\Delta_{c}}^{\wave\Delta_b, \Delta_4}} 
    \partialWaveDressed^{\Delta_{i}}_{\Delta_{a},\Delta_{b},\Delta_{c}}
    \; .
\end{align}
Following the steps in the five-point case, we get the relation between the six-point conformal partial wave and the conformal blocks, given by
\begin{align}
\label{eq:sixPointCPW2Block}
\begin{split}
\Psi^{\Delta_{i}}_{\Delta_{a},\Delta_{b},\Delta_{c}}
&=
S_{\wave\Delta_c}^{\Delta_5,\Delta_6} S_{\wave\Delta_a}^{\Delta_3,\Delta_b} S_{\wave\Delta_b}^{\Delta_4,\Delta_c}
G^{\Delta_{i}}_{\Delta_{a},\Delta_{b},\Delta_{c}}
+
S_{\Delta_c}^{\Delta_b,\Delta_4} S_{\wave\Delta_a}^{\Delta_3,\Delta_b} S_{\wave\Delta_b}^{\Delta_4,\Delta_c}
G^{\Delta_{i}}_{\Delta_{a},\Delta_{b},\wave\Delta_{c}}
\\
&\peq
+
S_{\Delta_b}^{\Delta_a,\Delta_3} S_{\wave\Delta_c}^{\Delta_5,\Delta_6} S_{\wave\Delta_a}^{\Delta_3,\Delta_b}
G^{\Delta_{i}}_{\Delta_{a},\wave\Delta_{b},\Delta_{c}}
+
S_{\Delta_b}^{\Delta_a,\Delta_3} S_{\wave\Delta_a}^{\Delta_3,\Delta_b} S_{\Delta_c}^{\wave\Delta_b,\Delta_4}
G^{\Delta_{i}}_{\Delta_{a},\wave\Delta_{b},\wave\Delta_{c}}
\\
&\peq
+
S_{\Delta_a}^{\Delta_1,\Delta_2} S_{\wave\Delta_c}^{\Delta_5,\Delta_6} S_{\wave\Delta_b}^{\Delta_4,\Delta_c}
G^{\Delta_{i}}_{\wave\Delta_{a},\Delta_{b},\Delta_{c}}
+
S_{\Delta_a}^{\Delta_1,\Delta_2} S_{\wave\Delta_b}^{\Delta_4,\wave\Delta_c} S_{\Delta_c}^{\wave\Delta_b,\Delta_4}
G^{\Delta_{i}}_{\wave\Delta_{a},\Delta_{b},\wave\Delta_{c}}
\\
&\peq
+
S_{\Delta_a}^{\Delta_1,\Delta_2} S_{\wave\Delta_c}^{\Delta_5,\Delta_6} S_{\Delta_b}^{\Delta_3,\wave\Delta_a}
G^{\Delta_{i}}_{\wave\Delta_{a},\wave\Delta_{b},\Delta_{c}}
+
S_{\Delta_a}^{\Delta_1,\Delta_2} S_{\Delta_b}^{\wave\Delta_a,\Delta_3} S_{\Delta_c}^{\wave\Delta_b,\Delta_4}
G^{\Delta_{i}}_{\wave\Delta_{a},\wave\Delta_{b},\wave\Delta_{c}}
\; .
\end{split}
\end{align}

Similar to \eqref{eq:conformal_partial_wave_expansion}, five- and six-point celestial amplitudes can be expanded in term of the conformal partial waves:
\begin{align}\label{eq:conformal_partial_wave_expansion_5}
    \cA^{\Delta_i}_5
    =
    \int_{\frac{d}{2}}^{\frac{d}{2}+i\infty}\frac{d\Delta_a}{2\pi i\, \mu_{\Delta_a}}\int_{\frac{d}{2}}^{\frac{d}{2}+i\infty}\frac{d\Delta_b}{2\pi i\, \mu_{\Delta_b}}
    I_{\Delta_a,\Delta_b}\Psi^{\Delta_{i}}_{\Delta_{a},\Delta_b}
    \; ,
\end{align}
and
\begin{align}\label{eq:conformal_partial_wave_expansion_6}
    \cA^{\Delta_i}_6
    =
    \int_{\frac{d}{2}}^{\frac{d}{2}+i\infty}\frac{d\Delta_a}{2\pi i\, \mu_{\Delta_a}}\int_{\frac{d}{2}}^{\frac{d}{2}+i\infty}\frac{d\Delta_b}{2\pi i\, \mu_{\Delta_b}}\int_{\frac{d}{2}}^{\frac{d}{2}+i\infty}\frac{d\Delta_c}{2\pi i\, \mu_{\Delta_c}}
    I_{\Delta_a,\Delta_b,\Delta_c}\Psi^{\Delta_{i}}_{\Delta_{a},\Delta_b,\Delta_c}
    \; .
\end{align}
Together with the shadow symmetries of the conformal partial waves and the density functions, the integration contour of exchanged conformal dimensions can be extended to the whole principle series, leading to
\begin{align}\label{eq:conformal_partial_wave_expansion_block5}
    \cA^{\Delta_i}_5
    =
    \int_{\frac{d}{2}-i\infty}^{\frac{d}{2}+i\infty}\frac{d\Delta_a}{2\pi i\, \mu_{\Delta_a}}\int_{\frac{d}{2}-i\infty}^{\frac{d}{2}+i\infty}\frac{d\Delta_b}{2\pi i\, \mu_{\Delta_b}}
    I_{\Delta_a,\Delta_b}S^{\Delta_3,\Delta_b}_{\wave\Delta_{a}}
    S^{\Delta_4,\Delta_5}_{\wave\Delta_{b}}
    G^{\Delta_{i}}_{\Delta_{a},\Delta_{b}}
    \; ,
\end{align}
and
\begin{align}\label{eq:conformal_partial_wave_expansion_block6}
    \cA^{\Delta_i}_6
    =
    \int_{\frac{d}{2}-i\infty}^{\frac{d}{2}+i\infty}\frac{d\Delta_a}{2\pi i\, \mu_{\Delta_a}}\int_{\frac{d}{2}-i\infty}^{\frac{d}{2}+i\infty}\frac{d\Delta_b}{2\pi i\, \mu_{\Delta_b}}\int_{\frac{d}{2}-i\infty}^{\frac{d}{2}+i\infty}\frac{d\Delta_c}{2\pi i\, \mu_{\Delta_c}}
    I_{\Delta_a,\Delta_b,\Delta_c}
    S_{\wave\Delta_a}^{\Delta_3,\Delta_b} 
    S_{\wave\Delta_b}^{\Delta_4,\Delta_c}
    S_{\wave\Delta_c}^{\Delta_5,\Delta_6} 
    G^{\Delta_{i}}_{\Delta_{a},\Delta_{b},\Delta_{c}}
    \; .
\end{align}
Evaluating the integrals of exchanged conformal dimensions by enclosing the contours into the right half planes and picking up the poles, we obtain the conformal block expansions of the five- and six-point celestial amplitudes.

\section{Three-point celestial amplitudes}\label{sec:3pt}

To obtain the conformal block expansions of four- and higher-point celestial amplitudes, it is necessary to compute scalar three-point celestial amplitudes containing more than one massive scalar. 
The scalar three-point celestial amplitudes with one massive scalar have been obtained in \cite{Lam:2017ofc}. 
The scalar three-point celestial amplitudes of three massive scalars with mass $m$, $m$, and $2m(1+\epsilon)$ in CCFT$_2$ have been obtained in \cite{Pasterski:2016qvg} to the leading order of $\epsilon$.
In this section, we will compute scalar three-point celestial amplitudes containing more than one massive scalar. It turns out that the three-point coefficient of scalar three-point celestial amplitudes with two massive scalars contains a hypergeometric function \eqref{eq:threePointm0m}, while the three-point coefficient of celestial amplitudes with three massive scalars can be written as a triple Mellin-Barnes integral \eqref{eq:tripleIntegral}.

One of the obstacles to evaluate the celestial amplitudes with more than one massive scalar is to solve the $(d+2)$-dimensional momentum conservation $\delta$-function. To overcome this, we rewrite the integral over the on-shell hypersurface $\hat{p}^2=-1$ as an integral over the entire momentum space by virtue of the identity
\begin{align}
    \label{eq:onshell_integral_to_full_integral}
    \int\frac{d^{d+1}\hat{p}}{\hat{p}^0}
    =
    \int d^{d+2}\hat{p}\;
    2\delta(\hat{p}^2+1)\theta(\hat{p}^0)
    \; .
\end{align}
The $(d+2)$-dimensional momentum conservation $\delta$-function then can be removed by performing the integral over $\hat{p}$, leaving with a one-dimensional $\delta$-function, which is easier to be solved.

As an example, we use our strategy to re-compute the celestial amplitudes with one massive scalar. The celestial amplitude of massless-massless-massive scalars $1_{0}^{0}+2_{0}^{0}\to 3_{m_{3}}^{0}$ is given by
\begin{align}
    &
    \cA_{1_{0}^{0}+2_{0}^{0}\to 3_{m_{3}}^{0}}^{\Delta_{1},\Delta_{2},\Delta_{3}}
    =
    \intrange{d\omega_{1}}{0}{+\oo}
    \omega_{1}^{\Delta_{1}-1}
    \intrange{d\omega_{2}}{0}{+\oo}
    \omega_{2}^{\Delta_{2}-1}
    \int\frac{d^{d+1}\phat_{3'}}{\hat{p}^0_{3'}}\delta^{(d+2)}(q_1+q_2-p_{3'})
    \; .
\end{align}
Applying \eqref{eq:onshell_integral_to_full_integral} and performing the integral over $\hat{p}_{3'}$ lead to 
\begin{align}
    \cA_{1_{0}^{0}+2_{0}^{0}\to 3_{m_{3}}^{0}}^{\Delta_{1},\Delta_{2},\Delta_{3}}
    &=m_{3}^{-d-2}
    \intrange{d\omega_{1}}{0}{+\oo}
    \omega_{1}^{\Delta_{1}-1}
    \intrange{d\omega_{2}}{0}{+\oo}
    \omega_{2}^{\Delta_{2}-1}
    2
    \delta(1-\frac{4x_{12}^{2}\omega_{1}\omega_{2}}{m_{3}})
    \left(2\frac{\omega_{1}x_{13}^{2}+\omega_{2}x_{23}^{2}}{m_{3}} \right)^{-\Delta_{3}}
    \nn
    \\
    &=
    C_{1_{0}^{0}+2_{0}^{0}\to 3_{m_{3}}^{0}}^{\Delta_{1},\Delta_{2},\Delta_{3}}\;
    \vev{\op_{1}\op_{2}\op_{3}}
    \; ,
\end{align}
where the three-point coefficient $C_{1_{0}^{0}+2_{0}^{0}\to 3_{m_{3}}^{0}}^{\Delta_{1},\Delta_{2},\Delta_{3}}$ is
\begin{equation}
    \label{eq:threePoint00m}
    C_{1_{0}^{0}+2_{0}^{0}\to 3_{m_{3}}^{0}}^{\Delta_{1},\Delta_{2},\Delta_{3}}
    =
    2^{-\Delta_{12}}m_{3}^{\Delta_{12}-d-2}
    \multiGamma{
        \frac{\Delta_{13,2}}{2},\frac{\Delta_{23,1}}{2}
    }{
        \Delta_3
    }
    \; .
\end{equation}
This matches the result of \cite[Section 3.1]{Lam:2017ofc}. Now, we are ready to compute the scalar three-point celestial amplitudes with more than one massive scalar by the same strategy.

\subsection{Massive-massless-massive scalars}

The celestial three-point amplitude of massive-massless-massive scalars $1_{m_{1}}^{0}+2_{0}^{0}\to 3_{m_{3}}^{0}$ takes the form as
\begin{align}
    \cA_{1_{m_{1}}^{0}+2_{0}^{0}\to 3_{m_{3}}^{0}}^{\Delta_{1},\Delta_{2},\Delta_{3}}
    &=
    \intrange{d\omega_{2}}{0}{+\oo}
    \omega_{2}^{\Delta_{2}-1}
    \intt{\frac{d^{d+1}\phat_{1'}}{\phat_{1'}^{0}}}
    \intt{\frac{d^{d+1}\phat_{3'}}{\hat{p}^0_{3'}}}
    \\
    &\peq
    \times
    (-\qhat_{1}\cdot\phat_{1'})^{-\Delta_{1}} 
    (-\qhat_{3}\cdot\phat_{3'})^{-\Delta_{3}}
    \delta^{(d+2)}(p_{1'}+q_2-p_{3'})
    \; .
    \nn
    \\
    &=
    C_{1_{m_{1}}^{0}+2_{0}^{0}\to 3_{m_{3}}^{0}}^{\Delta_{1},\Delta_{2},\Delta_{3}}
    \vev{\op_{1}\op_{2}\op_{3}}
    \; ,
    \nn
\end{align}
and in the following we will compute the three-point coefficient $C_{1_{m_{1}}^{0}+2_{0}^{0}\to 3_{m_{3}}^{0}}^{\Delta_{1},\Delta_{2},\Delta_{3}}$.

Applying \eqref{eq:onshell_integral_to_full_integral} and performing the integral over $\hat{p}_{3'}$ then lead to
\begin{align}
    \cA_{1_{m_{1}}^{0}+2_{0}^{0}\to 3_{m_{3}}^{0}}^{\Delta_{1},\Delta_{2},\Delta_{3}}
    &=
    m_{3}^{-d-2}\intrange{d\omega_{2}}{0}{+\oo}
    \omega_{2}^{\Delta_{2}-1}
    \intt{\frac{d^{d+1}\phat_{1'}}{\phat_{1'}^{0}}}
    2
    \delta\left( 
        \frac{m_{3}^{2}-m_{1}^{2}+2m_{1}\omega_{2}\phat_{1'}\cdot \qhat_{2}}{m_{3}^{2}}
    \right)
    \\
    &\peq
    \times
    (-\qhat_{1}\cdot\phat_{1'})^{-\Delta_{1}} 
    \left(-\frac{m_{1}\qhat_{3}\cdot \phat_{1'}+\omega_{2}\qhat_{3}\cdot \qhat_{2}}{m_{3}}\right)^{-\Delta_{3}}
    \; .
    \nn
\end{align}
Integrating out $\omega_{2}$ leads to
\begin{align}
    \cA_{1_{m_{1}}^{0}+2_{0}^{0}\to 3_{m_{3}}^{0}}^{\Delta_{1},\Delta_{2},\Delta_{3}}
    &=
    2^{\Delta_{3}-\Delta_{2}+1}
    m_{3}^{\Delta_{3}-d}
    m_{1}^{\Delta_{3}-\Delta_{2}}
    (m_{3}^{2}-m_{1}^{2})^{\Delta_{2}-1}
    \intt{\frac{d^{d+1}\phat_{1'}}{\phat_{1'}^{0}}}
    \\
    &\peq
    \times
    (-\qhat_{1}\cdot\phat_{1'})^{-\Delta_{1}}
    (-\qhat_{2}\cdot\phat_{1'})^{\Delta_{3}-\Delta_{2}} 
    \left( 
        2m_{1}^{2}\qhat_{2}\cdot\phat_{1'}\qhat_{3}\cdot\phat_{1'}
        -
        (m_{3}^{2}-m_{1}^{2})\qhat_{2}\cdot\qhat_{3} 
    \right)^{-\Delta_{3}}
    \; .
    \nn
\end{align}
Since $m_{1}>0,\, \omega_2\geq 0$, the $\delta$-function can be solved by $\omega_{2}=\frac{m_{3}^{2}-m_{1}^{2}}{-2m_{1}\phat_{1'}\cdot \qhat_{2}}$ if and only if $m_3\geq m_1$, which manifests the mass threshold.

Now we use the Mellin-Barnes relation \eqref{eq:MBRelation} to separate the factor $(\cdots)^{-\Delta_{3}}$ into a product $(-\qhat_{2}\cdot\phat_{1'})^{-s-\Delta_{3}}(-\qhat_{3}\cdot\phat_{1'})^{-s-\Delta_{3}}(-\qhat_{2}\cdot\qhat_{3} )^{s}$. Then the integral over $\phat_{1'}$ is the tree-level three-point Witten diagram on EAdS \eqref{eq:threePointWittenDiagram}, and we can read off the three-point coefficient as
\begin{align}
    C_{1_{m_{1}}^{0}+2_{0}^{0}\to 3_{m_{3}}^{0}}^{\Delta_{1},\Delta_{2},\Delta_{3}}
    &=
    d_{\Delta_{1},\Delta_{2},\Delta_{3}}
    2^{1-\Delta_{2}}m_{1}^{-\Delta_{23}}m_{3}^{-d+\Delta_{3}}
    (m_{3}^{2}-m_{1}^{2})^{\Delta_{2}-1}
    \\
    \label{eq:threePointm0mTemp}
    &\peq\xx
    \intrange{\frac{ds}{2\pi i}}{-i\oo}{+i\oo}
    \multiGamma{
        -s,
        \Delta_2,
        \frac{\Delta_{123}-d}{2}+s,
        \frac{\Delta_{23,1}}{2}+s
    }{
        \Delta_2+s,
        \frac{\Delta_{123}-d}{2},
        \frac{\Delta_{23,1}}{2}
    }
    \left( 
        -\frac{m_{1}^{2}-m_{3}^{2}}{m_{1}^{2}}
     \right)^{s}
    \; ,
\end{align}
where $d_{\Delta_{1},\Delta_{2},\Delta_{3}}$ is the coefficient of the three-point Witten diagram on EAdS:
\begin{equation}
    d_{\Delta_{1},\Delta_{2},\Delta_{3}}
    =
    \half \pi^{d/2}
    \multiGamma{
        \frac{\Delta_{12,3}}{2},
        \frac{\Delta_{23,1}}{2},
        \frac{\Delta_{31,2}}{2},
        \frac{\Delta_{123}-d}{2}
    }{
        \Delta_{1},
        \Delta_{2},
        \Delta_{3}
    }
    \; .
\end{equation}
The $s$-integral in \eqref{eq:threePointm0mTemp} is exactly the Mellin-Barnes representation of the hypergeometric function \eqref{eq:F21MB}, hence the three-point coefficient is
\begin{align}
    \label{eq:threePointm0m}
    C_{1_{m_{1}}^{0}+2_{0}^{0}\to 3_{m_{3}}^{0}}^{\Delta_{1},\Delta_{2},\Delta_{3}}
    &=
    d_{\Delta_{1},\Delta_{2},\Delta_{3}}
    2^{1-\Delta_{2}}m_{1}^{-\Delta_{23}}m_{3}^{-d+\Delta_{3}}
    (m_{3}^{2}-m_{1}^{2})^{\Delta_{2}-1}
    \Fpq{2}{1}{
        \frac{\Delta_{23,1}}{2},\frac{\Delta_{123}-d}{2}
    }{\Delta_{2}}{\frac{m_{1}^{2}-m_{3}^{2}}{m_{1}^{2}}}
    \; .
\end{align}
We stress that if the masses leave from the physical region $m_{3}\geq m_{1}>0$, there are no solutions for the $\delta$-function, and then the three-point coefficient vanishes.

\subsubsection{Sanity checks}\label{sec:threePointm0mCheck}

We check the result \eqref{eq:threePointm0m} by the shadow symmetry and the massless limit.
From the shadow symmetry between the massive conformal primary wavefunction and its shadow \eqref{eq:CPWFScalarMassive}, 
the three-point coefficient \eqref{eq:threePointm0m} should respect the following shadow symmetries:
\begin{equation}
    C_{\shadow{1_{m_{1}}^{0}}+2_{0}^{0}\to 3_{m_{3}}^{0}}^{\wave\Delta_{1},\Delta_{2},\Delta_{3}}
    =
    C_{1_{m_{1}}^{0}+2_{0}^{0}\to 3_{m_{3}}^{0}}^{\wave\Delta_{1},\Delta_{2},\Delta_{3}}
    \; ,
    \qquad 
    C_{1_{m_{1}}^{0}+2_{0}^{0}\to \shadow{3_{m_{3}}^{0}}}^{\Delta_{1},\Delta_{2},\wave\Delta_{3}}
    =
    C_{1_{m_{1}}^{0}+2_{0}^{0}\to {3}_{m_{3}}^{0}}^{\Delta_{1},\Delta_{2},\wave\Delta_{3}}
    \; .
\end{equation}
This implies that 
\begin{align}
    \label{eq:threePointm0mShadow}
    &C_{1_{m_{1}}^{0}+2_{0}^{0}\to 3_{m_{3}}^{0}}^{\wave\Delta_{1},\Delta_{2},\Delta_{3}}
    =
    C_{1_{m_{1}}^{0}+2_{0}^{0}\to 3_{m_{3}}^{0}}^{\Delta_{1},\Delta_{2},\Delta_{3}}
    S^{\Delta_{2},\Delta_{3}}_{\Delta_{1}}N_{\Delta_{1}}
    \; ,
    \\
    &C_{1_{m_{1}}^{0}+2_{0}^{0}\to 3_{m_{3}}^{0}}^{\Delta_{1},\Delta_{2},\wave\Delta_{3}}
    =
    C_{1_{m_{1}}^{0}+2_{0}^{0}\to 3_{m_{3}}^{0}}^{\Delta_{1},\Delta_{2},\Delta_{3}}
    S^{\Delta_{1},\Delta_{2}}_{\Delta_{3}}N_{\Delta_{3}}
    \; ,
\end{align}
which can be directly verified by the property of the hypergeometric function \eqref{eq:F21aToc-a}.

To analyze the $m_{1}\to 0$ limit of the three-point coefficient $C_{1_{m_{1}}^{0}+2_{0}^{0}\to 3_{m_{3}}^{0}}^{\Delta_{1},\Delta_{2},\Delta_{3}}$, by the property of the hypergeometric function \eqref{eq:F21zTo1/z} we rewrite it as
\begin{align}
    \frac{C_{1_{m_{1}}^{0}+2_{0}^{0}\to 3_{m_{3}}^{0}}^{\Delta_{1},\Delta_{2},\Delta_{3}}}{d_{\Delta_{1},\Delta_{2},\Delta_{3}}}
    &=
    2^{1-\Delta_2} 
    m_1^{-\Delta_1} 
    m_3^{\Delta_3-d} 
    (m_3^2-m_1^2)^{\frac{\Delta_{12,3}}{2}-1}
    \multiGamma{
        \Delta_2,
        \Delta_1-\frac{d}{2}
    }{
        \frac{\Delta_{12,3}}{2},
        \frac{\Delta_{123}-d}{2}
    }
    \Fpq{2}{1}{
        \frac{\Delta_{3,12}+2}{2},
        \frac{\Delta_{23,1}}{2}
    }{
        \Delta_{1}-\halfdim+1
    }{\frac{m_{1}^{2}}{m_{1}^{2}-m_{3}^{2}}}
    \nn
    \\
    &\peq 
    +
    (\Delta_{1}\to \wave\Delta_{1})
    \; .
    \label{eq:threePointm0mTwoTerms}
\end{align}
After multiplied by the factor in \eqref{eq:CPWFScalarMasslessLimit}, 
the first term is convergent as $m_{1}\to 0$; 
the second term is proportional to $m_{1}^{2\Delta_{1}-d}$, and is convergent to $0$ (divergent to $\oo$) if $\Re{\Delta_{1}}>\halfdim$ ($\Re{\Delta_{1}}<\halfdim$). 
Hence under the condition $\Re{\Delta_{1}}>\halfdim$, we have 
\begin{equation}
    C_{1_{0}^{0}+2_{0}^{0}\to 3_{m_{3}}^{0}}^{\Delta_{1},\Delta_{2},\Delta_{3}}
    =
    \lim_{m_{1}\to 0}
    2^{-\Delta_{1}}\pi^{-\halfdim} m_{1}^{\Delta_{1}}\frac{\gm{\Delta_{1}}}{\gm{\Delta_{1}-\halfdim}}
    C_{1_{m_{1}}^{0}+2_{0}^{0}\to 3_{m_{3}}^{0}}^{\Delta_{1},\Delta_{2},\Delta_{3}}
    \; .
\end{equation}
Similar analysis shows that if $\Re{\Delta_{1}}<\halfdim$, the $m_{1}\to 0$ limit of $C_{1_{m_{1}}^{0}+2_{0}^{0}\to 3_{m_{3}}^{0}}^{\Delta_{1},\Delta_{2},\Delta_{3}}$ recovers the three-point coefficient under the shadow conformal basis $C_{\shadow{1_{0}^{0}}+2_{0}^{0}\to 3_{m_{3}}^{0}}^{\wave\Delta_{1},\Delta_{2},\Delta_{3
}}$.

\subsection{Massive-massive-massive scalars}

In this section we evaluate the three-point coefficient $C_{1_{m_{1}}^{0}+2_{m_{2}}^{0}\to 3_{m_{3}}^{0}}^{\Delta_{1},\Delta_{2},\Delta_{3}}$ of massive-massive-massive scalars. The celestial amplitude is 
\begin{align}
    \cA_{1_{m_{1}}^{0}+2_{m_{2}}^{0}\to 3_{m_{3}}^{0}}^{\Delta_{1},\Delta_{2},\Delta_{3}}
    &=
    \intt{\frac{d^{d+1}\phat_{1'}}{\phat_{1'}^{0}}}
    \intt{\frac{d^{d+1}\phat_{2'}}{\phat_{2'}^{0}}}
    \intt{\frac{d^{d+1}\phat_{3'}}{\phat_{3'}^{0}}}
    \delta^{(d+2)}(p_{1'}+p_{2'}-p_{3'})
    \\
    &\peq
    \times
    (-\qhat_{1}\cdot\phat_{1'})^{-\Delta_{1}} 
    (-\qhat_{2}\cdot\phat_{2'})^{-\Delta_{2}} 
    (-\qhat_{3}\cdot\phat_{3'})^{-\Delta_{3}} 
    \nn
    \\
    &=
    C_{1_{m_{1}}^{0}+2_{m_{2}}^{0}\to 3_{m_{3}}^{0}}^{\Delta_{1},\Delta_{2},\Delta_{3}}
    \vev{\op_{1}\op_{2}\op_{3}}
    \; .
\end{align}

Applying \eqref{eq:onshell_integral_to_full_integral} and performing the $\hat{p}_{3'}$-integral, we get
\begin{align}
    \cA_{1_{m_{1}}^{0}+2_{m_{2}}^{0}\to 3_{m_{3}}^{0}}^{\Delta_{1},\Delta_{2},\Delta_{3}}
    &=
    m_{1}^{-\Delta_{3}-1}
    m_{2}^{-1}
    m_{3}^{-d+\Delta_{3}}
    \intt{\frac{d^{d+1}\phat_{1'}}{\phat_{1'}^{0}}}
    \intt{\frac{d^{d+1}\phat_{2'}}{\phat_{2'}^{0}}}
    \delta\left( 
        \frac{m_{3}^{2}-(m_{1}+m_{2})^{2}}{2 m_{1}m_{2}}+1+\phat_{1'}\cdot \phat_{2'}
    \right)
    \nn
    \\
    &\peq
    \times
    (-\qhat_{1}\cdot\phat_{1'})^{-\Delta_{1}} 
    (-\qhat_{2}\cdot\phat_{2'})^{-\Delta_{2}} 
    \left(
        -\qhat_{3}\cdot \phat_{1'}-\frac{m_{2}}{m_{1}}\qhat_{3}\cdot \phat_{2'}
    \right)^{-\Delta_{3}}
    \; .
\end{align}
If the masses do not satisfy the physical condition $m_{3}\geq m_{1}+m_{2}$, there are no solutions for the $\delta$-function, and then the three-point coefficient vanishes.
To manifest this mass threshold, we introduce the new mass parameter $A=\frac{m_{3}^{2}-(m_{1}+m_{2})^{2}}{2 m_{1}m_{2}}\geq 0$ in the physical region $m_{1}>0\, ,m_{2}>0,\, m_{3}\geq m_{1}+m_{2}$.
Inserting an identity of the Mellin transform 
\begin{equation}
    f(A)
    =
    \intrange{\frac{ds}{2\pi i}}{-i\oo}{+i\oo}A^{-s}
    \intrange{dA'}{0}{+\oo}A'^{s-1}f(A')
    \; ,
\end{equation}
we have
\begin{align}
    \cA_{1_{m_{1}}^{0}+2_{m_{2}}^{0}\to 3_{m_{3}}^{0}}^{\Delta_{1},\Delta_{2},\Delta_{3}}
    &=
    m_{1}^{-\Delta_{3}-1}
    m_{2}^{-1}
    m_{3}^{-d+\Delta_{3}}
    \intrange{\frac{ds}{2\pi i}}{-i\oo}{+i\oo}A^{-s}
    \intt{\frac{d^{d+1}\phat_{1'}}{\phat_{1'}^{0}}}
    \intt{\frac{d^{d+1}\phat_{2'}}{\phat_{2'}^{0}}}
    \intrange{dA'}{0}{+\oo}A'^{s-1}
    \\
    \nn
    &\peq
    \times
    \delta(A'+1+\phat_{1'}\cdot \phat_{2'})
    (-\qhat_{1}\cdot\phat_{1'})^{-\Delta_{1}} 
    (-\qhat_{2}\cdot\phat_{2'})^{-\Delta_{2}} 
    \left(
        -\qhat_{3}\cdot \phat_{1'}
        -\frac{m_{2}}{m_{1}}\qhat_{3}\cdot \phat_{2'}
    \right)^{-\Delta_{3}}
 \; .
\end{align}
After integrating out $A'$, we use the Mellin-Barnes relation \eqref{eq:MBRelation} to separate the sum $(-\qhat_{3}\cdot \phat_{1'}-\frac{m_{2}}{m_{1}}\qhat_{3}\cdot \phat_{2'})$ into a product, then the integrand is proportional to 
\begin{equation}
    (\cdots)
    (-1-\phat_{1'}\cdot \phat_{2'})^{s-1}
    (-\qhat_{1}\cdot\phat_{1'})^{-\Delta_{1}} 
    (-\qhat_{2}\cdot\phat_{2'})^{-\Delta_{2}} 
    (-\qhat_{3}\cdot \phat_{1'})^{-\Delta_{3}-t}
    (-\qhat_{3}\cdot \phat_{2'})^{t}
    \; .
\end{equation}

The integral over $\hat{p}_{1'}$ can be computed as follows.
By the Feynman-Schwinger parameterization \eqref{eq:FSParameterization} and the EAdS integral \eqref{eq:EAdSIntegral}, we have 
\begin{align}
    \label{eq:p1Integral}
    &\peq
    \intt{\frac{d^{d+1}\phat_{1'}}{\phat_{1'}^{0}}}
    (-1-\phat_{1'}\cdot \phat_{2'})^{s-1}
    (-\qhat_{1}\cdot\phat_{1'})^{-\Delta_{1}} 
    (-\qhat_{3}\cdot \phat_{1'})^{-t-\Delta_{3}}
    \\
    &=
    (\cdots)
    \intrange{d\alpha d\beta}{0}{+\oo}
    \alpha^{\Delta_{1}-1}
    \beta^{t+\Delta_{3}-1}
    \intt{\frac{d^{d+1}\phat_{1'}}{\phat_{1'}^{0}}}
    \left(
        -1-\phat_{1'}\cdot (\phat_{2'}+\alpha\qhat_{1}+\beta\qhat_{3})
    \right)^{
        -\Delta_{13}+s-t-1
    }
    \nn
    \\
    &=
    (\cdots)
    \intrange{\frac{du}{2\pi i}}{-i\oo}{+i\oo}
    \intrange{d\alpha d\beta}{0}{+\oo}
    \alpha^{\Delta_{1}-1}
    \beta^{t+\Delta_{3}-1}
    (
        1
        -2\alpha \phat_{2'}\cdot \qhat_{1}
        -2\beta \phat_{2'}\cdot \qhat_{3}
        -2\alpha\beta \qhat_{1}\cdot \qhat_{3}
    )^{\frac{-\Delta_{13}+s-t-u-1}{2}}
    \nn
    \; .
\end{align}
The $\alpha$-integral gives 
\begin{equation}
    (\cdots)
    \beta^{t+\Delta_{3}-1}
    (
        -\phat_{2'}\cdot \qhat_{1}
        -\beta \qhat_{1}\cdot \qhat_{3}
    )^{-\Delta_1} 
    (
        1-2 \beta  \phat_{2'}\cdot \qhat_{3}
    )^{\frac{\Delta_{1,3}+s-t-u-1}{2}}
    \; ,
\end{equation}
then using the Mellin-Barnes relation \eqref{eq:MBRelation} we separate the two sums $(-\phat_{2'}\cdot \qhat_{1}-\beta \qhat_{1}\cdot \qhat_{3})$ and $(1-2 \beta  \phat_{2'}\cdot \qhat_{3})$ into products with two new Mellin parameters $\gamma$ and $v$. 
After performing the $\beta$- and $\gamma$-integrals, we find that the $\hat{p}_{1'}$ -integral \eqref{eq:p1Integral} is written as a double Mellin-Barnes integral over $u$ and $v$, and the integrand is proportional to 
\begin{equation}
    (\cdots)
    (-\phat_{2'}\cdot \qhat_{1})^{-v-\Delta_{1}}
    (-\phat_{2'}\cdot \qhat_{3})^{-v-\Delta_{3}}
    (-\qhat_{1}\cdot \qhat_{3})^{v}
    \; .
\end{equation}

The remaining integral over $\hat{p}_{2'}$ is the tree-level three-point Witten diagram on EAdS \eqref{eq:threePointWittenDiagram}. Then we can read off the three-point coefficient as 
\begin{align}
    C_{1_{m_{1}}^{0}+2_{m_{2}}^{0}\to 3_{m_{3}}^{0}}^{\Delta_{1},\Delta_{2},\Delta_{3}}
    &=
    \intrange{\frac{dsdtdv}{(2\pi i)^{3}}}{-i\oo}{+i\oo}
    \pi^{d/2} 
    m_1^{-\Delta_3+s-t-1}
    m_2^{s+t-1} 
    m_3^{\Delta_3-d} 
    (m_3^2-(m_1+m_2)^2)^{-s}
    \\
    &\peq
    \xx
    \intrange{\frac{du}{2\pi i}}{-i\oo}{+i\oo}
    (-2)^u 
    \Gamma\left[\textstyle{
        -u,
        \frac{-\Delta_{13}-s-t+u-2 v+1}{2},
        \frac{\Delta_{13}-d-s+t+u+1}{2}
    }\right]
    \nn
    \\
    &\peq
    \xx
    d_{\Delta_{1},\Delta_{2},\Delta_{3}}
    \multiGamma{
        -t,
        -v,
        \Delta_3+t+v,
        \frac{\Delta_{13,2}+2 v}{2},
        \frac{\Delta_{123}-d+2 v}{2}
    }{
        1-s,
        \Delta_3+v,
        \frac{\Delta_{13,2}}{2},
        \frac{\Delta_{123}-d}{2}
    }
    \; .
    \nn
\end{align}
The $u$-integral can be performed as follows. We first rewrite $(-2)^{u}\gm{-u}$ as $\frac{(-1)^u}{2 \sqrt{\pi }} \gm{\frac{1-u}{2}} \gm{-\frac{u}{2}}$, which gives two series of $u$-poles in the right half plane. For each series of $u$-poles, the residues can be resummed as a hypergeometric function with argument unity \eqref{eq:F21ArgumentUnity}, leading to 
\begin{align}
    \label{eq:tripleIntegral}
    C_{1_{m_{1}}^{0}+2_{m_{2}}^{0}\to 3_{m_{3}}^{0}}^{\Delta_{1},\Delta_{2},\Delta_{3}}
    &=
    \intrange{\frac{dsdtdv}{(2\pi i)^{3}}}{-i\oo}{+i\oo}
    \pi^{\frac{d+1}{2}} 
    m_1^{-\Delta_3+s-t-1} 
    m_2^{s+t-1} 
    m_3^{\Delta_3-d} 
    (m_3^2-(m_{1}+m_{2})^{2})^{-s} 
    \\
    &\peq
    \xx
    d_{\Delta_{1},\Delta_{2},\Delta_{3}}
    \left(
        C_{1}+C_{2}
    \right)
    \; ,
    \nn
\end{align}
where 
\begin{align}
    &C_{1}
    =
    \multiGamma{
        -t,
        -v,
        \frac{d}{2}+s+v-\frac{1}{2},
        \frac{-\Delta_{13}-s-t-2 v+1}{2},
        \Delta_3+t+v,
        \frac{\Delta_{13}-d-s+t+1}{2},
        \frac{\Delta_{13,2}+2 v}{2},
        \frac{\Delta_{123}-d+2 v}{2}
    }{
        1-s,
        \frac{-\Delta_{13}+d+s-t}{2},
        \Delta_3+v,
        \frac{\Delta_{13}+s+t+2 v}{2},
        \frac{\Delta_{13,2}}{2},
        \frac{\Delta_{123}-d}{2}
    }
    \; ,
    \\[0.2cm]
    &C_{2}
    =
    \multiGamma{
        -t,
        -v,
        \frac{d}{2}+s+v-\frac{1}{2},
        \frac{-\Delta_{13}-s-t-2 v+2}{2},
        \Delta_3+t+v,
        \frac{\Delta_{13}-d-s+t+2}{2},
        \frac{\Delta_{13,2}+2 v}{2},
        \frac{\Delta_{123}-d+2 v}{2}
    }{
        1-s,
        \frac{-\Delta_{13}+d+s-t+1}{2},
        \Delta_3+v,
        \frac{\Delta_{13}+s+t+2 v+1}{2},
        \frac{\Delta_{13,2}}{2},
        \frac{\Delta_{123}-d}{2}
    }
    \; .
\end{align}
We stress that if the masses leave from the physical region $m_{1},m_{2}>0$ and $m_{3}\geq m_{1}+m_{2}>0$, there are no solutions for the $\delta$-function, and then the three-point coefficient vanishes.

\subsubsection{Mass threshold expansion}

Near the mass threshold $m_3=m_1+m_2$, we can enclose the $s$-contour to the left half plane. After performing the $s$-integral, we enclose the $v$- and $t$-contours to the right half planes successively. In each of the steps there is only one series of poles at $s=\frac{1-d-2v}{2}-n_{1}$, $t=n_{2}$ and $v=n_{3}$. 
It turns out that the contributions from $C_{1}$ and $C_{2}$ in \eqref{eq:tripleIntegral} are equal, which can be further simplified to a double series of the Wilson polynomial \eqref{eq:Wilson_polynomial}.
This leads to the following series expansion of $C_{1_{m_{1}}^{0}+2_{m_{2}}^{0}\to 3_{m_{3}}^{0}}^{\Delta_{1},\Delta_{2},\Delta_{3}}$ near the mass threshold,
\begin{align}
    \label{eq:threePointCoefficientMMMMassThresholdExpansion}
    &
    C_{1_{m_{1}}^{0}+2_{m_{2}}^{0}\to 3_{m_{3}}^{0}}^{\Delta_{1},\Delta_{2},\Delta_{3}}
    =
    \sum_{n_{1},n_{2}=0}^{+\oo}
    \pi^{\frac{d+1}{2}} 
    m_1^{-\frac{d+1}{2}-\Delta_3-n_{1}-n_{2}} 
    m_2^{-\frac{d+1}{2}-n_1+n_2} 
    m_3^{\Delta_3-d} 
    (m_3^2-(m_{1}+m_{2})^{2})^{\frac{d-1}{2}+n_{1}} 
    \\
    &\hspace{-1.5ex}
    \xx
    \frac{2(-1)^{n_2}d_{\Delta_{1},\Delta_{2},\Delta_{3}}}{n_{1}!n_{2}!}
    \multiGamma{
        \Delta_3+n_2
    }{
        \frac{d+1}{2}+n_1,
        \Delta_3+n_1
    }
    \wilsonPolynomial_{n_{1}}({\textstyle
        -\frac{(d-2 \Delta_2)^2}{16};
        \frac{2 \Delta_{13}-d}{4},
        \frac{-2 n_1+2 n_2+1}{4},
        \frac{-2 n_1+2 n_2+3}{4},
        \frac{2\Delta_{3,1}+d}{4}
    })
    \; ,
    \nn
\end{align}
%


To compare with the result in \cite[Section 3]{Pasterski:2016qvg}, we set $m_{3}=(1+\epsilon)(m_{1}+m_{2})$ with $\epsilon\rightarrow 0^{+}$, and in this limit we have the estimate
\begin{equation}
    (m_3^2-(m_{1}+m_{2})^{2})^{\frac{d-1}{2}+n_{1}}
    \sim
    \epsilon^{\frac{d-1}{2}+n_{1}}
    \; .
\end{equation}
This implies the leading term of \eqref{eq:threePointCoefficientMMMMassThresholdExpansion} can be obtained by setting $n_1=0$, then the remaining sum of $n_{2}$ can be performed, which leads to
\begin{equation}
    C_{1_{m_{1}}^{0}+2_{m_{2}}^{0}\to 3_{m_{3}}^{0}}^{\Delta_{1},\Delta_{2},\Delta_{3}}
    =
    d_{\Delta_{1},\Delta_{2},\Delta_{3}}
    \frac{
        (2\pi)^{\frac{d+1}{2}} 
        (m_1 m_2)^{-\frac{d+1}{2}} 
    }{
        (m_1+m_2) 
        \gm{\frac{d+1}{2}}
    }
    \epsilon^{\frac{d-1}{2}}
    +
    O(\epsilon^{\frac{d+1}{2}})
    \; .
\end{equation}
Notice that except the factor $d_{\Delta_{1},\Delta_{2},\Delta_{3}}$ the leading term does not depend on the external conformal dimensions $\Delta_{i}$. Setting $d=2$ and $m_{1}=m_{2}=m$ then reproduce the result in \cite[Section 3]{Pasterski:2016qvg} upto a factor $(2\pi)^4i$ due to the different conventions of the scattering amplitude.

\subsubsection{Massless limit}

We further check our results by taking the massless limit \eqref{eq:CPWFScalarMasslessLimit}. 
To take the $m_2\rightarrow 0$ limit, we should enclose the $s$-contour into the right half plane after shifting $s$ by $s\rightarrow s-t$. 
Using the method of pole pinching, one can check that the $C_2$-term in \eqref{eq:tripleIntegral} does not survive in the limit $m_{2}\to 0$. The regular term can be obtained from the $C_{1}$-term by picking up poles at $t=n_1$, $ v=\frac{-\Delta_{13}-s+1}{2}+n_2$, and $s=1-\Delta_2+2n_2+2n_3$ with $n_1,n_2,n_3\in \NN$, successively. 
For each triplet $(n_1,n_2,n_3)$, the residue is proportional to the factor $m_2^{2n_2+2n_3}$, and  taking $m_1\rightarrow0$ forces $n_2=n_3=0$. This leads to
\begin{align}\label{eq:mmm_m2_0}
    &\peq
    \lim_{m_2\rightarrow 0}
    2^{-\Delta_{2}}\pi^{-\halfdim} m_{2}^{\Delta_{2}}\frac{\gm{\Delta_{2}}}{\gm{\Delta_{2}-\halfdim}}
    C_{1_{m_{1}}^{0}+2_{m_{2}}^{0}\to 3_{m_{3}}^{0}}^{\Delta_{1},\Delta_{2},\Delta_{3}}
    \\
    &=
    \sum_{n_{1}=0}^{+\oo}
    (-1)^{n_1} 
    2^{1-\Delta_2}
    m_1^{-\Delta_{23}-2 n_1} 
    m_3^{\Delta_3-d} 
    (m_3^2-m_1^2)^{\Delta_2+n_1-1} 
    \multiGamma{
        \Delta_2,
        \frac{\Delta_{23,1}+2 n_1}{2},
        \frac{\Delta_{123}-d+2 n_1}{2}
    }{
        n_1+1,
        \Delta_2+n_1,
        \frac{\Delta_{23,1}}{2},
        \frac{\Delta_{123}-d}{2}
    }
    \nn
    \\
    &=
    C_{1_{m_{1}}^{0}+2_{0}^{0}\to 3_{m_{3}}^{0}}^{\Delta_{1},\Delta_{2},\Delta_{3}}
    \nn
    \; ,
\end{align}
which reproduces the result in \eqref{eq:threePointm0m}. 
We stress that to get \eqref{eq:mmm_m2_0}, we have ignored the terms that may not converge in the limit $m_{2}\to 0$. The situation here is similar to that in Section \ref{sec:threePointm0mCheck}, see the discussions after \eqref{eq:threePointm0mTwoTerms}.

Similarly, to take the $m_1\rightarrow 0$ limit, we should enclose the $s$-contour into the right half plane after shifting $s$ by $s\rightarrow s+t$. 
Again, the $C_2$-term in \eqref{eq:tripleIntegral} does not survive in the limit $m_{1}\to 0$. The regular term can be obtained from the $C_{1}$-term by picking up poles at $t=-\Delta_3-v-n_1$, $ s=-\Delta_{1,3}+1+2n_1+2n_2$, and $v=n_3$ with $n_1,n_2,n_3\in \NN$, successively. 
For each triplet $(n_1,n_2,n_3)$, the residue is proportional to the factor $m_1^{2n_1+2n_2}$, and  taking $m_1\rightarrow0$ forces $n_2=n_3=0$. This leads to
\begin{align}\label{eq:mmm_m1_0}
    &\peq
    \lim_{m_1\rightarrow 0}
    2^{-\Delta_{1}}\pi^{-\halfdim} m_{1}^{\Delta_{1}}\frac{\gm{\Delta_{1}}}{\gm{\Delta_{1}-\halfdim}}
    C_{1_{m_{1}}^{0}+2_{m_{2}}^{0}\to 3_{m_{3}}^{0}}^{\Delta_{1},\Delta_{2},\Delta_{3}}
    \\
    &=
    \sum_{n_{3}=0}^{+\oo}
    (-1)^{n_3} 
    2^{1-\Delta_1} 
    m_2^{-\Delta_{13}-2 n_3} 
    m_3^{\Delta_3-d} 
    (m_3^2-m_2^2)^{\Delta_1+n_3-1} 
    \multiGamma{
        \Delta_1,
        \frac{\Delta_{13,2}+2 n_3}{2},
        \frac{\Delta_{123}-d+2 n_3}{2}
    }{
        n_3+1,
        \Delta_1+n_3,
        \frac{\Delta_{13,2}}{2},
        \frac{\Delta_{123}-d}{2}
    }
    \nn
    \\
    &=
    C_{1_0^{0}+2_{m_2}^{0}\to 3_{m_{3}}^{0}}^{\Delta_{1},\Delta_{2},\Delta_{3}}
    \nn
    \; .
\end{align}
%

\section{Four-point celestial amplitudes}\label{sec:4pt}
In this section, we will utilize the split representation \eqref{eq:SplitRep} to analyze four-point celestial amplitudes involving massive scalars. 
The split representation is a useful technique that allows us to easily obtain the conformal partial wave expansion and the conformal block expansion of celestial amplitudes.
To illustrate this, we will warm up with the simple case of massless four-point celestial amplitudes.

\subsection{Warm up: four massless scalars with massless exchange}\label{sec:fourPoint1}

Let us consider the \schannel tree-level scattering amplitude 
\begin{align}
    \mathcal{M}_{1^0_0+2^0_0\to 3^0_0+4^0_0}=
    \frac{i}{(q_1+q_2)^2}\delta^{(d+2)}(q_1+q_2-q_3-q_4)
    \; ,
\end{align}
where the external and the exchanged particles are massless scalars.
In this example, the internal momentum $(q_1+q_2)$ is timelike or lightlike, thus only the first line in the split representation \eqref{eq:SplitRep} contributes. Then the celestial amplitude is 
\begin{equation}\label{eq:AsMassless=LR}
\begin{split}
\cA^{\Delta_i}_{1_0^0+2_0^0\rightarrow3_0^0+4_0^0}
&=
\int_{0}^{\infty}\frac{dm_a}{(2\pi)^{d+2}}\frac{i m_a^{d+1}}{-m_a^2}
\int_{\frac{d}{2}}^{\frac{d}{2}+i\infty}\frac{d\Delta_a}{2 \pi i\,  \mu_{\Delta_a}}
\int d^dx_a\;
\cA^{\Delta_1,\Delta_2,\Delta_a}_{1_0^0+2_0^0\rightarrow a_{m_a}^0}\cA^{\wave{\Delta}_a,\Delta_3,\Delta_4}_{a_{m_a}^0\rightarrow3_0^0+4_0^0}
\; ,
\end{split}
\end{equation}
where $\cA^{\Delta_1,\Delta_2,\Delta_a}_{1_0^0+2_0^0\rightarrow a_{m_a}^0}$ and $\cA^{\wave{\Delta}_a,\Delta_3,\Delta_4}_{a_{m_a}^0\rightarrow3_0^0+4_0^0}$ are three-point celestial amplitudes containing one massive scalar. Using \eqref{eq:fourPointCPW} we obtain the conformal partial wave expansion of the celestial amplitude,
\begin{align}\label{eq:CPWAs}
\begin{split}
\cA^{\Delta_i}_{1_0^0+2_0^0\rightarrow3_0^0+4_0^0}
&=
\int_{0}^{\infty}\frac{dm_a}{(2\pi)^{d+2}}\frac{im_a^{d+1}}{-m_{a}^2}
\int_{\frac{d}{2}}^{\frac{d}{2}+i\infty}\frac{d\Delta_a}{2\pi i\, \mu_{\Delta_a}}
I_{\Delta_{a}}\Psi^{\Delta_i}_{\Delta_{a}}
\; .
\end{split}
\end{align}
where the density function is
\begin{equation}
    I_{\Delta_{a}}
    =
    C^{\Delta_1,\Delta_2,\Delta_a}_{1^0_0+2^0_0\rightarrow a_{m_a}^0}C^{\wave{\Delta}_a,\Delta_3,\Delta_4}_{a^0_{m_a}\rightarrow3_0^0+4_0^0}
    \; .
\end{equation}

Armed with the shadow symmetry of the three-point coefficient \eqref{eq:threePointm0mShadow}, it is easy to check that the density function satisfies
\begin{equation}
    I_{\wave\Delta_{a}}=
    \frac{S_{\Delta_{a}}^{\Delta_1, \Delta_2}}{S_{\Delta_{a}}^{\Delta_3, \Delta_4}} 
    I_{\Delta_{a}}
    \; .
\end{equation}
Together with the shadow symmetry of the conformal partial wave \eqref{eq:fourPointCPWShadowSymmetry} and the relation to the conformal blocks \eqref{eq:fourPointCPW2Block}, we can rewrite \eqref{eq:CPWAs} as
\begin{equation}
\label{eq:fourPointContourIntegral}
\cA_{1^0_0+2^0_0\to 3^0_0+4^0_0}^{\Delta_i}
=
\int_{\frac{d}{2}-i\infty}^{\frac{d}{2}+i\infty}\frac{d\Delta_{a}}{2\pi i}\, 
\rho_{\Delta_{a}}
G^{\Delta_{i}}_{\Delta_{a}}
\; .
\end{equation}
Here $\rho_{\Delta_{a}}$ takes the form as
\begin{align}
    \rho_{\Delta_{a}}
    &=
    \int_{m}^{+\infty}\frac{dm_{a}}{(2\pi)^{d+2}}\frac{i m_{a}^{d+1}}{-m_{a}^2}
    \mu_{\Delta_{a}}^{-1}
    I_{\Delta_{a}}
    S^{\Delta_3,\Delta_4}_{\wave\Delta_{a}}
    \\
    &=
    -i\,  \pi^{-\frac{3 d}{2}-2} 2^{-\Delta_{1234}-d-2} \complexDelta{\Delta_{1234}-d-4}
    \multiGamma{
        \frac{\Delta_{1a,2}}{2},
        \frac{\Delta_{2a,1}}{2},
        \frac{\Delta_{3a,4}}{2},
        \frac{\Delta_{4a,3}}{2}
    }{
        \Delta_a,
        \Delta_{a}-\frac{d}{2}
    }
    \; ,
\end{align}
where in the second line we have performed the integral over $m_a$ by the $\delta$-function with complex argument \eqref{eq:complexDeltaFunction}.

We find that $\rho_{\Delta_{a}}$ does not have $\Delta_{a}$-poles to the right of the principal series. This implies if the conformal cross-ratio $\chi$ satisfies $|\chi|< 1$, enclosing the contour to the right half $\Delta_a$-plane leads to a vanishing conformal block expansion. While if $|\chi|>1$ the integrand in \eqref{eq:fourPointContourIntegral} does not decay to zero as $\Re(\Delta_a)\to +\oo$, and we cannot enclose the contour. 
These two aspects reflect the fact that four-point massless celestial amplitudes contain a nonanalytic factor $\theta(|\chi|-1)$, and the celestial amplitude $\cA^{\Delta_i}_{1^0_0+2^0_0\rightarrow 3^0_0+4^0_0}$ does not have a proper conformal block expansion \cite{Chang:2022jut}.

To fix this issue, we will use the shadow conformal basis for outgoing states \cite{Chang:2022jut,Furugori:2023hgv}. Following the same steps, we find that the celestial amplitude in this basis takes the form as
\begin{equation}
\label{eq:shadow_four_point_contour_integral}
\cA_{1^0_0+2^0_0\to \shadow{3^0_0}+\shadow{4^0_0}}^{\Delta_1,\Delta_2,\wave{\Delta}_3,\wave{\Delta}_4}
=
\int_{\frac{d}{2}-i\infty}^{\frac{d}{2}+i\infty}\frac{d\Delta_{a}}{2\pi i}\, 
\sigma_{\Delta_{a}}
G^{\Delta_{1},\Delta_{2},\wave\Delta_{3},\wave\Delta_{4}}_{\Delta_{a}}
\; ,
\end{equation}
where $\sigma_{\Delta_{a}}$ is given by
\begin{align}
    \sigma_{\Delta_{a}}
    &=
    -i\,  \pi^{-\frac{3 d}{2}-2} 2^{-\Delta_{1234}-d-2} \complexDelta{\Delta_{1234}-d-4}
    \\
    &\peq
    \times
    \multiGamma{
        \Delta_3,
        \Delta_4,
        \mathred{\frac{2 d-\Delta_{34a}}{2}},
        \frac{\Delta_{1a,2}}{2},
        \frac{\Delta_{2a,1}}{2},
        \frac{\Delta_{a,34}+d}{2},
        \frac{\Delta_{3a,4}}{2},
        \frac{\Delta_{4a,3}}{2}
    }{
        d-\Delta_3,
        d-\Delta_4,
        \mathred{\frac{\Delta_{34,a}}{2}},
        \Delta_a,
        \Delta_{a}-\frac{d}{2},
        \frac{\Delta_{34a}-d}{2}
    }
    \; .
    \nn
\end{align}
Then there are double-trace $\Delta_{a}$-poles from the factor $\gm{\frac{2 d-\Delta_{34a}}{2}}$ at 
\begin{align}
    &\Delta_{a}=\wave\Delta_{3}+\wave\Delta_{4}+2N
    \; ,
    \quad 
    N\in\NN
    \; .
\end{align}

In the rest of this paper, unless specified otherwise we will use the shadow conformal basis for outgoing states to discuss the conformal block expansion of celestial amplitudes.

\subsection{Two massive and two massless scalars with massive exchange}

We consider a theory containing the interaction vertex $\frac{1}{2}\phi\Phi^2$, where $\phi$ is a massless scalar and $\Phi$ is a massive scalar with mass $m$. In this theory, the $s$-channel tree-level scattering amplitude of $\Phi+\phi\to\Phi+\phi$ is given by
\begin{align}
    \label{eq:fourPointM}
    \mathcal{M}_{1^0_m+2^0_0\to 3^0_m+4^0_0}=
    \frac{i}{(p_1+q_2)^2+m^2}\delta^{(d+2)}(p_1+q_2-p_3-q_4)
    \; .
\end{align}
Using the split representation \eqref{eq:SplitRep} and the integral expression of the conformal partial wave \eqref{eq:fourPointCPW}, the corresponding celestial amplitude of \eqref{eq:fourPointM} can be written as
\begin{align}
\label{eq:fourPointSplit}
\begin{split}
\cA_{1^0_m+2^0_0\to \shadow{3^0_m}+\shadow{4^0_0}}^{\Delta_{1},\Delta_{2},\wave\Delta_{3},\wave\Delta_{4}}
&=
\int_{m}^{+\infty}\frac{d m_{a}}{(2\pi)^{d+2}}\frac{i m_{a}^{d+1}}{-m_{a}^2+m^2}
\int_{\frac{d}{2}}^{\frac{d}{2}+i\infty}\frac{d\Delta_{a}}{2\pi i\, \mu_{\Delta_{a}}}
I_{\Delta_{a}}
\Psi^{\Delta_{1},\Delta_{2},\wave\Delta_{3},\wave\Delta_{4}}_{\Delta_{a}}
\; ,
\end{split}
\end{align}
where the density function is
\begin{equation}
    I_{\Delta_{a}}
    =
    C_{1^0_m+2^0_0\to  a^0_{m_{a}}}^{\Delta_1,\Delta_2,\Delta_{a}}
    C_{a^0_{m_{a}}\to \shadow{3^0_m}+\shadow{4^0_0}}^{\wave\Delta_{a},\wave\Delta_3,\wave{\Delta}_4}
    \; .
\end{equation}

Following the same steps as in the previous subsection, we can rewrite \eqref{eq:fourPointSplit} as
\begin{equation}
\cA_{1^0_m+2^0_0\to \shadow{3^0_m}+\shadow{4^0_0}}^{\Delta_{1},\Delta_{2},\wave\Delta_{3},\wave\Delta_{4}}
=
\int_{\frac{d}{2}-i\infty}^{\frac{d}{2}+i\infty}\frac{d\Delta_{a}}{2\pi i}\, 
\sigma_{\Delta_{a}}
G^{\Delta_{1},\Delta_{2},\wave\Delta_{3},\wave\Delta_{4}}_{\Delta_{a}}
\; ,
\end{equation}
where $\sigma_{\Delta_{a}}$ takes the form as
\begin{align}
\sigma_{\Delta_{a}}
&=
\int_{m}^{+\infty}\frac{d m_{a}}{(2\pi)^{d+2}}\frac{i m_{a}^{d+1}}{-m_{a}^2+m^2}
\mu_{\Delta_{a}}^{-1}
I^{\Delta_i}_{\Delta_{a}}
S_{\wave\Delta_{a}}^{\wave\Delta_{3},\wave\Delta_{4}}
\\
&=
(\cdots)
\int_{m}^{+\infty}dm_{a}\, m_{a}(m_{a}^2-m^2)^{\Delta_{24}-3}
\Fpq{2}{1}{\frac{\Delta_{2a,1}}{2},\frac{\Delta_{12a}-d}{2}}{\Delta_{2}}{1-\frac{m_{a}^{2}}{m^{2}}}
\Fpq{2}{1}{\frac{d+\Delta_{4,3a}}{2},\frac{\Delta_{34,a}}{2}}{\Delta_{4}}{1-\frac{m_{a}^{2}}{m^{2}}}
\; .
\nn
\end{align}
Using the Mellin-Barnes representation of the hypergeometric function \eqref{eq:F21MB} with Mellin parameters $s$ and $t$, and changing the integration variable from $m_{a}$ to $\alpha=m_{a}^2-m^2$, the mass integral can be separated as the $\delta$-function with complex argument \eqref{eq:complexDeltaFunction}, $\int_{0}^{+\infty}d\alpha\, \alpha^{s+t+\Delta_{24}-3}$.
Then we can evaluate the integral over $s$, leading to
\begin{align}
    \sigma_{\Delta_{a}}
    &=
    -i\,  \pi^{-\frac{d}{2}-2} 2^{-\Delta_{24}-d-3} m^{\Delta_{24}-d-4}
    \multiGamma{
        \Delta_4,
        \mathred{\frac{\Delta_{12,a}}{2}},
        \mathred{\frac{2 d-\Delta_{34a}}{2}},
        \frac{\Delta_{1a,2}}{2},
        \frac{\Delta_{a,34}+d}{2},
        \frac{\Delta_{3a,4}}{2},
        \frac{\Delta_{4a,3}}{2}
    }{
        \Delta_1,
        d-\Delta_3,
        d-\Delta_4,
        \mathred{\frac{\Delta_{4,3a}+d}{2}},
        \mathred{\frac{\Delta_{34,a}}{2}},
        \Delta_a,
        \Delta_{a}-\frac{d}{2}
    }
    \nn
    \\
    &\peq
    \times\int_{-i\infty}^{+i\infty}\frac{dt}{2\pi i}
    \multiGamma{
        -t,
        \Delta_{24}+t-2,
        \mathred{\frac{\Delta_{4,3a}+d}{2}+t},
        \mathred{\frac{\Delta_{34,a}}{2}+t},
        \frac{\Delta_{a,12}-2 \Delta_4+4}{2}-t,
        \frac{\Delta_{1a,2}-d-2 \Delta_4+4}{2}-t
    }{
        -\Delta_4-t+2,
        \Delta_4+t
    }
    \; .
    \nn
\end{align}

By the method of pole pinching, we note that performing the $t$-integral does not produce $\Delta_a$-poles. 
The $\Delta_{a}$-poles from the factors $\gm{-t}$ and $\gm{\frac{\Delta_{4,3a}+d}{2}+t}$ get canceled by $\gm{\frac{\Delta_{4,3a}+d}{2}}$ in the denominator, and the $\Delta_{a}$-poles from the factors $\gm{-t}$ and $\gm{\frac{\Delta_{34,a}}{2}+t}$ get canceled by $\gm{\frac{\Delta_{34,a}}{2}}$ in the denominator. 
Thus, only the factors $\gm{\frac{\Delta_{12,a}}{2}}$ and $\gm{\frac{2d-\Delta_{34a}}{2}}$ provide two series of double-trace $\Delta_{a}$-poles at 
\begin{align}\label{eq:four-point_0m0m_poles}
    &\Delta_{a}=\Delta_{1}+\Delta_{2}+2N
    \; ,
    \quad 
    N\in\NN
    \; .
    \\
    &\Delta_{a}=\wave\Delta_{3}+\wave\Delta_{4}+2N
    \; ,
    \quad 
    N\in\NN
    \; .
\end{align}
We note that the feature of double-trace exchange is similar to that of AdS, see \eg \cite{Liu:1998th}. But unlike AdS/CFT, the conformal dimension here is not related to the mass in the bulk.

\section{Higher-point celestial amplitudes}\label{sec:5&6pt}

Following the previous section, we will use the split representation \eqref{eq:SplitRep} to analyze higher-point celestial amplitudes. 
Compared to the four-point celestial amplitudes, we observe that there are new exchanged operator appearing in the conformal block expansions of higher-point ones. Specifically, in the $2d$ five-point case, the conformal dimensions of the dominant new operators are given by $(\Delta_{12}-2)$ and $(\Delta_{12}+2\Delta_{3}-4)$, which supports the findings of two-particle operators in \cite[Section 4]{Ball:2023sdz}.

For convenience, we introduce the following naming of poles:
\begin{equation}
    \label{eq:namingOfPoles}
    \begin{tabular}{l@{{}\qquad{}} L}
     type-1: &   \Delta=\Delta_{i_{1}i_{2}\cdots}+\text{integer}
     \; ,
     \\
     type-2: &   \Delta=\Delta_{i_{1}i_{2}\cdots}+2\Delta_{j}+\text{integer}
     \; ,
     \\
     type-3: &   \Delta=\Delta_{i_{1}i_{2}\cdots}+2\Delta_{j}+2\Delta_{k}+\text{integer}
     \; ,
     \\
     $\cdots$ &     
    \end{tabular}
\end{equation}

\subsection{Five massless scalars with massless exchange}\label{eq:00000}

\begin{figure}[htbp]
\centering

\tikzset{every picture/.style={line width=0.75pt}} 

\begin{tikzpicture}[x=0.75pt,y=0.75pt,yscale=-1,xscale=1]

\draw    (200,120) -- (200,200) ;
\draw [shift={(200,165)}, rotate = 270] [fill={rgb, 255:red, 0; green, 0; blue, 0 }  ][line width=0.08]  [draw opacity=0] (8.93,-4.29) -- (0,0) -- (8.93,4.29) -- cycle    ;
\draw    (200,200) -- (120,200) ;
\draw [shift={(166.5,200)}, rotate = 180] [fill={rgb, 255:red, 0; green, 0; blue, 0 }  ][line width=0.08]  [draw opacity=0] (8.93,-4.29) -- (0,0) -- (8.93,4.29) -- cycle    ;
\draw    (200,200) -- (360,200) ;
\draw    (280,120) -- (280,200) ;
\draw [shift={(280,165)}, rotate = 270] [fill={rgb, 255:red, 0; green, 0; blue, 0 }  ][line width=0.08]  [draw opacity=0] (8.93,-4.29) -- (0,0) -- (8.93,4.29) -- cycle    ;
\draw    (360,120) -- (360,200) ;
\draw [shift={(360,153.5)}, rotate = 90] [fill={rgb, 255:red, 0; green, 0; blue, 0 }  ][line width=0.08]  [draw opacity=0] (8.93,-4.29) -- (0,0) -- (8.93,4.29) -- cycle    ;
\draw    (360,200) -- (440,200) ;
\draw [shift={(405,200)}, rotate = 180] [fill={rgb, 255:red, 0; green, 0; blue, 0 }  ][line width=0.08]  [draw opacity=0] (8.93,-4.29) -- (0,0) -- (8.93,4.29) -- cycle    ;

\draw (92,192.4) node [anchor=north west][inner sep=0.75pt]  [xscale=0.8,yscale=0.8]  {$1_{0}^{0}$};
\draw (192,92.4) node [anchor=north west][inner sep=0.75pt]  [xscale=0.8,yscale=0.8]  {$2_{0}^{0}$};
\draw (272,92.4) node [anchor=north west][inner sep=0.75pt]  [xscale=0.8,yscale=0.8]  {$3_{0}^{0}$};
\draw (352,92.4) node [anchor=north west][inner sep=0.75pt]  [xscale=0.8,yscale=0.8]  {$4_{0}^{0}$};
\draw (452,192.4) node [anchor=north west][inner sep=0.75pt]  [xscale=0.8,yscale=0.8]  {$5_{0}^{0}$};

\end{tikzpicture}

\caption{A tree-level diagram for $1_{0}^{0}+2_{0}^{0}+3_{0}^{0}\to 4_{0}^{0}+5_{0}^{0}$. The external and internal particles are all massless scalars.}
\label{fig:fivePointFeynmanDiagram}
\end{figure}
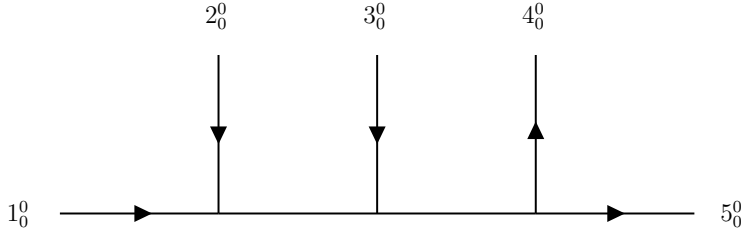

In the $\phi^{3}$ theory, the contribution of the tree-level diagram \ref{fig:fivePointFeynmanDiagram} to the scattering amplitude is 
\begin{align}
    \label{eq:fivePointM}
    \mathcal{M}_{1_{0}^{0}+2_{0}^{0}+3_{0}^{0}\to 4_{0}^{0}+5_{0}^{0}}
    =
    i^{2}\frac{\delta^{(d+2)}(q_1+q_2+q_3-q_4-q_5)}{(q_1+q_2)^2(q_1+q_2+q_3)^2}
    \; .
\end{align}
Using the split representation \eqref{eq:SplitRep} and the integral expression of the five-point conformal partial wave \eqref{eq:fivePointCPW}, we get the celestial amplitude of \eqref{eq:fivePointM} as
\begin{align}
    \label{eq:fivePointSplit}
    \cA_{1_{0}^{0}+2_{0}^{0}+3_{0}^{0}\to 4_{0}^{0}+5_{0}^{0}}^{\Delta_i}
    &=
    \int_{0}^{+\infty}\frac{dm_{a}}{(2\pi)^{d+2}}(-i m_{a}^{d-1})
    \int_{m_{a}}^{+\infty}\frac{dm_{b}}{(2\pi)^{d+2}}(-i m_{b}^{d-1})
    \\
    &\peq 
    \times
    \int_{\frac{d}{2}}^{\frac{d}{2}+i\infty}\frac{d\Delta_a}{2\pi i \, \mu_{\Delta_a}}
    \int_{\frac{d}{2}}^{\frac{d}{2}+i\infty}\frac{d\Delta_b}{2\pi i \, \mu_{\Delta_b}}
    I_{\Delta_{a},\Delta_{b}}
    \partialWaveDressed^{\Delta_{i}}_{\Delta_{a},\Delta_{b}}
    \; ,
\end{align}
where the density function is
\begin{equation}
    I_{\Delta_{a},\Delta_{b}}
    =
    C_{1^0_0+2^0_0\to a^0_{m_{a}}}^{\Delta_1,\Delta_2,\Delta_a}
    C_{a^0_{m_{a}}+3^0_0\to b^0_{m_{b}}}^{\wave\Delta_a,\Delta_3,\Delta_b}
    C_{b^0_{m_{b}}\to 4^0_0+5^0_0}^{\wave\Delta_b,\Delta_4,\Delta_5}
    \; .
\end{equation}

From \eqref{eq:threePointm0mShadow}, the density function satisfies
\begin{equation}
    I_{\wave\Delta_{a},\Delta_{b}}
    =
    \frac{S_{\Delta_{a}}^{\Delta_1, \Delta_2}}{S_{\Delta_{a}}^{\Delta_3, \Delta_b}}
    I_{\Delta_{a},\Delta_{b}}
    \; ,
    \quad 
    I_{\Delta_{a},\wave\Delta_{b}}
    =
    \frac{S_{\Delta_{b}}^{\wave\Delta_a, \Delta_3}}{S_{\Delta_{b}}^{\Delta_4, \Delta_5}}
    I_{\Delta_{a},\Delta_{b}}
    \; .
\end{equation}
The by the shadow symmetry of the conformal partial wave \eqref{eq:fivePointCPWShadowSymmetry1} and the relation to the conformal blocks \eqref{eq:fivePointCPW2Block}, 
we can rewrite \eqref{eq:fivePointSplit} as a double contour integral,
\begin{align}
\label{eq:fivePointContourIntegral}
\begin{split}
\cA_{1_{0}^{0}+2_{0}^{0}+3_{0}^{0}\to 4_{0}^{0}+5_{0}^{0}}^{\Delta_i}
&=
\int_{\frac{d}{2}-i\infty}^{\frac{d}{2}+i\infty}\frac{d\Delta_a}{2\pi i}
\int_{\frac{d}{2}-i\infty}^{\frac{d}{2}+i\infty}\frac{d\Delta_b}{2\pi i}
\rho_{\Delta_a,\Delta_{b}}
G^{\Delta_{i}}_{\Delta_a,\Delta_{b}}
\; ,
\end{split}
\end{align}
where
\begin{align}
\begin{split}
    &\peq
    \rho_{\Delta_a,\Delta_{b}}
    =
    \int_{0}^{+\infty}\frac{dm_{a}}{(2\pi)^{d+2}}(-i m_{a}^{d-1})
    \int_{m_{a}}^{+\infty}\frac{dm_{b}}{(2\pi)^{d+2}}(-i m_{b}^{d-1})
    \mu^{-1}_{\Delta_a}
    \mu^{-1}_{\Delta_b}
    I_{\Delta_{a},\Delta_{b}}
    S^{\Delta_3,\Delta_b}_{\wave\Delta_{a}}
    S^{\Delta_4,\Delta_5}_{\wave\Delta_{b}}
    \; .
\end{split}
\end{align}
We first use the Mellin-Barnes representation of the hypergeometric function \eqref{eq:F21MB} with the Mellin parameter $s$, then integrate out $\alpha$ by the change of variable $\alpha=m_{b}^{2}-m_{a}^{2}$, 
\begin{align}
    \rho_{\Delta_a,\Delta_{b}}
    &=
    (\cdots)
    \int_{0}^{+\infty}dm_{a} m_{a}^{\Delta_{12345}-d-7}
    \intrange{\frac{ds}{2\pi i}}{-i\infty}{+i\infty}
    \\
    &\peq
    \times
    \Gamma\left[{
        -s,
        -s-\Delta_{3}+\frac{-\Delta_{45b}+d+4}{2},
        s+\frac{\Delta_{3b,a}}{2},
        s+\frac{\Delta_{3ab}-d}{2}
    }\right]
    \nn
    \; .
\end{align}
The $m_{a}$-integral gives a $\delta$-function with complex argument \eqref{eq:complexDeltaFunction}, and the $s$-integral can be done by the first Barnes lemma \eqref{eq:first_mellin_barnes_lemma}, leading to
\begin{align}
    \rho_{\Delta_a,\Delta_{b}}
    &=
    -\pi^{-\frac{5 d}{2}-4} 2^{-\Delta_{12345}-2 d-5} \complexDelta{\Delta_{12345}-d-6}
    \\
    &\peq
    \times
    \multiGamma{
        \mathred{\frac{-\Delta_{345a}+d+4}{2}},
        \frac{\Delta_{1a,2}}{2},
        \frac{\Delta_{2a,1}}{2},
        \frac{\Delta_{a,345}+4}{2},
        \mathblue{\frac{\Delta_{3a,b}}{2}},
        \frac{\Delta_{4b,5}}{2},
        \frac{\Delta_{5b,4}}{2},
        \mathred{\frac{\Delta_{3b,a}}{2}},
        \frac{\Delta_{ab,3}}{2},
        \frac{\Delta_{3ab}-d}{2}
    }{
        \Delta_a,
        \Delta_{a}-\frac{d}{2},
        \mathblue{\frac{-\Delta_{45b}+d+4}{2}},
        \Delta_b,
        \Delta_{b}-\frac{d}{2},
        \frac{\Delta_{b,45}+4}{2}
    }
    \nn
    \; .
\end{align}

If we first perform the $\Delta_{a}$-integral, there are two series of $\Delta_{a}$-poles to the right of the principal series.
But in any case we cannot enclose the contour of $\Delta_{b}$, and the conformal block expansion can be schematically written as
\begin{align}
    \cA_{1_{0}^{0}+2_{0}^{0}+3_{0}^{0}\to 4_{0}^{0}+5_{0}^{0}}^{\Delta_i}
    &=
    \sum_{N=0}^{+\infty}
    \int_{\frac{d}{2}-i\infty}^{\frac{d}{2}+i\infty}\frac{d\Delta_b}{2\pi i}
    \bigl(
        \cdots
        G^{\Delta_{i}}_{\Delta_{12}-2+2N,\Delta_{b}}
        +
        \cdots
        G^{\Delta_{i}}_{\Delta_{3b}+2N,\Delta_{b}}
    \bigr)
    \; .
\end{align}
Similarly, if firstly performing the $\Delta_{b}$-integral, we cannot enclose the contour of $\Delta_{a}$ to the right of the principal series, and the conformal block expansion is schematically
\begin{align}
    \cA_{1_{0}^{0}+2_{0}^{0}+3_{0}^{0}\to 4_{0}^{0}+5_{0}^{0}}^{\Delta_i}
    &=
    \sum_{N=0}^{+\infty}
    \int_{\frac{d}{2}-i\infty}^{\frac{d}{2}+i\infty}\frac{d\Delta_a}{2\pi i}
    \cdots
    G^{\Delta_{i}}_{\Delta_{a},\Delta_{3a}+2N}
    \; .
\end{align}

The fact that the double contour integral \eqref{eq:fivePointContourIntegral} cannot be jointly enclosed can be understood that the origin in the cross-ratio space is disallowed by the momentum conservation.
As discussed in Section \ref{sec:fourPoint1}, this has already happened in the case of four-point celestial amplitudes of four massless scalars: the momentum conservation contributes a non-analytic factor $\theta(|\chi|-1)$ to $\cA_{1_{0}^{0}+2_{0}^{0}\to 4_{0}^{0}+5_{0}^{0}}^{\Delta_i}$ that obstructs the \schannel conformal block expansion.

Instead, we choose the shadow conformal basis for outgoing states.
Comparing with \eqref{eq:fivePointContourIntegral}, the new double contour integral is
\begin{align}
    \label{eq:fivePointContourIntegralNew}
    \cA_{1_{0}^{0}+2_{0}^{0}+3_{0}^{0}\to \shadow{4_{0}^{0}}+\shadow{5_{0}^{0}}}^{\Delta_1,\Delta_{2},\Delta_{3},\wave\Delta_{4},\wave\Delta_{5}}
    &=
    \int_{\frac{d}{2}-i\infty}^{\frac{d}{2}+i\infty}\frac{d\Delta_a}{2\pi i}
    \int_{\frac{d}{2}-i\infty}^{\frac{d}{2}+i\infty}\frac{d\Delta_b}{2\pi i}
    \sigma_{\Delta_a,\Delta_{b}}
    G_{\Delta_a,\Delta_{b}}^{\Delta_1,\Delta_{2},\Delta_{3},\wave\Delta_{4},\wave\Delta_{5}}
    \; ,
\end{align}
where
\begin{align}
    \label{eq:sigma00000}
    \sigma_{\Delta_a,\Delta_{b}}
    &=
    \rho_{\Delta_a,\Delta_{b}}
    N_{\Delta_{4}}
    S_{\Delta_{4}}^{\Delta_{5},\wave\Delta_{b}}
    N_{\Delta_{5}}
    S_{\Delta_{5}}^{\wave\Delta_{4},\wave\Delta_{b}}
    S_{\wave\Delta_{b}}^{\wave\Delta_{4},\wave\Delta_{5}}/S_{\wave\Delta_{b}}^{\Delta_{4},\Delta_{5}}
    \\
    &=
    -\pi^{-\frac{5 d}{2}-4} 2^{-\Delta_{12345}-2 d-5} \complexDelta{\Delta_{12345}-d-6}
    \nn
    \\
    &\peq
    \times
    \multiGamma{
        \Delta_4,
        \Delta_5,
        \frac{\Delta_{1a,2}}{2},
        \frac{\Delta_{2a,1}}{2},
        \mathblue{\frac{\Delta_{3a,b}}{2}},
        \frac{\Delta_{4b,5}}{2},
        \frac{\Delta_{5b,4}}{2},
        \mathred{\frac{\Delta_{3b,a}}{2}},
        \frac{\Delta_{ab,3}}{2}
    }{
        d-\Delta_4,
        d-\Delta_5,
        \Delta_a,
        \Delta_{a}-\frac{d}{2},
        \mathblue{\frac{\Delta_{45,b}}{2}},
        \Delta_b,
        \Delta_{b}-\frac{d}{2}
    }
    \nn
    \\
    &\peq
    \times
    \multiGamma{
        \mathred{\frac{-\Delta_{345a}+d+4}{2}},
        \frac{\Delta_{a,345}+4}{2},
        \mathblue{\frac{2 d-\Delta_{45b}}{2}},
        \frac{\Delta_{b,45}+d}{2},
        \frac{\Delta_{3ab}-d}{2}
    }{
        \mathblue{\frac{-\Delta_{45b}+d+4}{2}},
        \frac{\Delta_{b,45}+4}{2},
        \frac{\Delta_{45b}-d}{2}
    }
    \nn
    \; .
\end{align}

We perform the integrals in the order $\Delta_{a}\to \Delta_{b}$. There are two series of $\Delta_{a}$-poles (to the right of the principal series).
\begin{enumerate}
    \item The factor $\gm{\frac{-\Delta_{345a}+d+4}{2}}$ provides double-trace $\Delta_{a}$-poles at 
    \begin{equation}
        \label{eq:fivePointDoubleTracePolesA}
        \Delta_{a}
        =
        \Delta_{12}-2+2N
        \; ,
        \quad  
        N\in\NN
        \; .
    \end{equation}
    After integrating out $\Delta_{a}$, the $\Delta_{b}$-poles are captured by 
    \begin{equation}
        \label{eq:fivePointDoubleTracePolesB}
        \multiGamma{
            \frac{\Delta_{123,b}+d-6}{2},
            \frac{\Delta_{123,b}+2N-2}{2}
        }{
            \frac{\Delta_{123,b}-2}{2}
        }
        \; .
    \end{equation}
    \item The factor $\gm{\frac{\Delta_{3b,a}}{2}}$ provides ``running'' $\Delta_{a}$-poles at 
    \begin{equation}
        \label{eq:fivePointTwoParticlePolesA}
        \Delta_{a}
        =
        \Delta_{3b}+2N
        \; ,
        \quad  
        N\in\NN
        \; .
    \end{equation}
    After integrating out $\Delta_{a}$, the $\Delta_{b}$-poles are captured by 
    \begin{equation}
        \label{eq:fivePointTwoParticlePolesB}
        \multiGamma{
            \frac{\Delta_{123,b}+d-6}{2}
        }{
            \frac{\Delta_{123,b}-2}{2}
        }
        \; .
    \end{equation}
\end{enumerate}
Due to the factors \eqref{eq:fivePointDoubleTracePolesB} and \eqref{eq:fivePointTwoParticlePolesB}, the pole structure depends on the dimension $d$, and we summarize the result in Table \ref{tab:fivePointPoleStructure}.
The type-1 $\Delta_{a}$-poles \eqref{eq:fivePointDoubleTracePolesA} accompanied with the type-1 $\Delta_{b}$-poles \eqref{eq:fivePointDoubleTracePolesB} are present in any dimension,
but the type-2 $\Delta_{a}$-poles \eqref{eq:fivePointTwoParticlePolesA} disappear in even dimensions except $d=2$.

Now we can compare with the results in \cite{Ball:2023sdz}. When $d=2$, the leading type-1 $\Delta_{a}$-pole is $\Delta_{a}=\Delta_{12}-2$, matching with the conformal dimension of the single-particle operator in \cite[Equation 4.1]{Ball:2023sdz}; the leading type-2 $\Delta_{a}$-pole is $\Delta_{a}=\Delta_{12}+2\Delta_{3}-4$, matching with the conformal dimension of the two-particle operator in \cite[Equation 4.1]{Ball:2023sdz}. 

At this stage, it is unclear whether the subleading $\Delta_{a}$-poles correspond to single- or two particle exchanged operators, so we will retain the current naming convention \eqref{eq:namingOfPoles}. We leave this question in a future work.
Another concern arises in even dimensions, the first bulk dimension $(d+2)$ at which the type-2 poles disappear coincides with the critical dimension of the $\phi^{3}$ theory. Is this merely a coincidence, or are there more informations of the bulk theory hidden behind the tree-level celestial amplitudes?

\subsection{One massive and four massless scalars}

To test the stability of the type-2 poles, in the tree-level diagram \ref{fig:fivePointFeynmanDiagram}, we replace the first incoming massless scalar with a massive one, and choose the first propagator as massless or massive.
The computation is similar to that in Section \ref{eq:00000}, and in this section, we will only present the celestial amplitudes in the shadow conformal basis for outgoing states. As before, the celestial amplitudes are represented as
\begin{equation}
    \label{eq:m0000}
    \cA_{1_{m}^{0}+2_{0}^{0}+3_{0}^{0}\to \shadow{4_{0}^{0}}+\shadow{5_{0}^{0}}}^{\Delta_1,\Delta_{2},\Delta_{3},\wave\Delta_{4},\wave\Delta_{5}}
    =
    \int_{\frac{d}{2}-i\infty}^{\frac{d}{2}+i\infty}\frac{d\Delta_a}{2\pi i}
    \int_{\frac{d}{2}-i\infty}^{\frac{d}{2}+i\infty}\frac{d\Delta_b}{2\pi i}
    \sigma_{\Delta_a,\Delta_{b}}
    G_{\Delta_a,\Delta_{b}}^{\Delta_1,\Delta_{2},\Delta_{3},\wave\Delta_{4},\wave\Delta_{5}}
    \; .
\end{equation}

\subsubsection{Massless exchanges}

The scattering amplitude is
\begin{align}
    \mathcal{M}_{1_{m}^{0}+2_{0}^{0}+3_{0}^{0}\to 4_{0}^{0}+5_{0}^{0}}
    =
    i^{2}\frac{\delta^{(d+2)}(q_1+q_2+q_3-q_4-q_5)}{(q_1+q_2)^2(q_1+q_2+q_3)^2}
    \; ,
\end{align}
and the corresponding celestial amplitude \eqref{eq:m0000} is characterized by
\begin{align}
    \label{eq:sigmam0000}
    \sigma_{\Delta_a,\Delta_{b}}
    &=
    -\pi^{-2 (d+2)}2^{-\Delta_{2345}-2 d-6} m^{\Delta_{2345}-d-6}
    \\
    &\peq
    \times
    \multiGamma{
        \Delta_4,
        \Delta_5,
        \mathred{\frac{\Delta_{12,a}}{2}},
        \frac{\Delta_{1a,2}}{2},
        \frac{\Delta_{2a,1}}{2},
        \mathblue{\frac{\Delta_{3a,b}}{2}},
        \frac{\Delta_{4b,5}}{2},
        \frac{\Delta_{5b,4}}{2},
        \mathred{\frac{\Delta_{3b,a}}{2}},
        \frac{\Delta_{ab,3}}{2}
    }{
        \Delta_1,
        d-\Delta_4,
        d-\Delta_5,
        \Delta_a,
        \Delta_{a}-\frac{d}{2},
        \mathblue{\frac{\Delta_{45,b}}{2}},
        \Delta_b,
        \Delta_{b}-\frac{d}{2}
    }
    \nn
    \\
    &\peq
    \times
    \multiGamma{
        \frac{-\Delta_{12345}+d+6}{2},
        \frac{\Delta_{1,2345}+6}{2},
        \mathred{\frac{-\Delta_{345a}+d+4}{2}},
        \frac{\Delta_{12a}-d}{2},
        \frac{\Delta_{a,345}+4}{2},
        \mathblue{\frac{2 d-\Delta_{45b}}{2}},
        \frac{\Delta_{b,45}+d}{2},
        \frac{\Delta_{3ab}-d}{2}
    }{
        \mathred{\frac{-\Delta_{345a}+d+6}{2}},
        \frac{\Delta_{a,345}+6}{2},
        \mathblue{\frac{-\Delta_{45b}+d+4}{2}},
        \frac{\Delta_{b,45}+4}{2},
        \frac{\Delta_{45b}-d}{2}
    }
    \; .
    \nn
\end{align}

We first check the massless limit $m\to 0$.
After taking into account the prefactor in \eqref{eq:CPWFScalarMasslessLimit}, the factor $m^{\Delta_{12345}-d-6} \gm{\frac{-\Delta_{12345}+d+6}{2}}$ recovers $\complexDelta{\Delta_{12345}-d-6}$ in \eqref{eq:sigma00000} by the approximation of the $\delta$-function with complex argument \eqref{eq:complexDeltaFunctionAsLimit}, leading to 
\begin{align}
    &\peq
    \lim_{m\rightarrow 0}
    2^{-\Delta_{1}}\pi^{-\halfdim} m^{\Delta_{1}}\frac{\gm{\Delta_{1}}}{\gm{\Delta_{1}-\halfdim}}
    \cA_{1_{m}^{0}+2_{0}^{0}+3_{0}^{0}\to \shadow{4_{0}^{0}}+\shadow{5_{0}^{0}}}^{\Delta_1,\Delta_{2},\Delta_{3},\wave\Delta_{4},\wave\Delta_{5}}
    =
    \cA_{1_{0}^{0}+2_{0}^{0}+3_{0}^{0}\to \shadow{4_{0}^{0}}+\shadow{5_{0}^{0}}}^{\Delta_1,\Delta_{2},\Delta_{3},\wave\Delta_{4},\wave\Delta_{5}}
    \; .
\end{align}

The pole structure of \eqref{eq:sigmam0000} is summarized in Table \ref{tab:fivePointPoleStructure2}. 
The names of poles are chosen by comparing with the massless limit $m\to 0$.
For example, the $\Delta_{a}$-poles at $\Delta_{a}=\Delta_{3,45}+4+2N$ in Table \ref{tab:fivePointPoleStructure2} correspond to the type-2 poles in Table \ref{tab:fivePointPoleStructure}. 
We notice that after turning on the mass of the first particle, the type-1 poles $\Delta_{a}=\Delta_{12}-2+2N$ in the massless case split into two parts:
$\Delta_{a}=\Delta_{12}+2N$ and $\Delta_{a}=-\Delta_{345}+d+4$.

\subsubsection{One massive exchange}

The scattering amplitude is
\begin{align}
    \mathcal{M}_{1_{m}^{0}+2_{0}^{0}+3_{0}^{0}\to 4_{0}^{0}+5_{0}^{0}}
    =
    i^{2}\frac{\delta^{(d+2)}(q_1+q_2+q_3-q_4-q_5)}{\left((q_1+q_2)^2+m^{2}\right)(q_1+q_2+q_3)^2}
    \; .
\end{align}
and the corresponding celestial amplitude \eqref{eq:m0000} is characterized by
\begin{align}
    \label{eq:sigmam00002}
    \sigma_{\Delta_a,\Delta_{b}}
    &=
    -\pi^{-2 (d+2)} 2^{-\Delta_{2345}-2 d-6} m^{\Delta_{2345}-d-6}
    \multiGamma{
        \frac{\Delta_{a,345}+4}{2},
        \mathblue{\frac{2 d-\Delta_{45b}}{2}},
        \frac{\Delta_{b,45}+d}{2},
        \frac{\Delta_{3ab}-d}{2}
    }{
        \mathblue{\frac{-\Delta_{45b}+d+4}{2}},
        \frac{\Delta_{b,45}+4}{2},
        \frac{\Delta_{45b}-d}{2}
    }
    \\
    &\peq
    \times
    \multiGamma{
        \Delta_4,
        \Delta_5,
        \mathred{\frac{\Delta_{12,a}}{2}},
        \frac{\Delta_{1a,2}}{2},
        \mathblue{\frac{\Delta_{3a,b}}{2}},
        \frac{\Delta_{4b,5}}{2},
        \frac{\Delta_{5b,4}}{2},
        \mathred{\frac{\Delta_{3b,a}}{2}},
        \frac{\Delta_{ab,3}}{2}
    }{
        \Delta_1,
        d-\Delta_4,
        d-\Delta_5,
        \Delta_a,
        \Delta_{a}-\frac{d}{2},
        \mathblue{\frac{\Delta_{45,b}}{2}},
        \Delta_b,
        \Delta_{b}-\frac{d}{2}
    }
    \nn
    \\
    &\peq
    \times
    \int_{-i\infty}^{+i\infty}\frac{ds}{2\pi i}
    \multiGamma{
        -s,
        \Delta_2+s-1,
        \mathred{\frac{-\Delta_{345a}+d-2 \Delta_2+6}{2}-s},
        \frac{\Delta_{2a,1}}{2}+s,
        \frac{\Delta_{12a}-d}{2}+s
    }{
        \Delta_2+s
    }
    \; .
    \nn
\end{align}

We first check the massless limit $m\to 0$.
By the method of pole pinching, during the $s$-integral the factors $\gm{\frac{-\Delta_{345a}+d-2 \Delta_2+6}{2}-s}$ and $\gm{\frac{\Delta_{2a,1}}{2}+s}$ provide a simple pole at $\Delta_{1}=d+6-\Delta_{2345}$. Then the factor $\frac{1}{\Delta_{12345}-d-6}m^{\Delta_{12345}-d-6}$ recovers $\complexDelta{\Delta_{12345}-d-6}$ in \eqref{eq:sigma00000}. After performing the $s$-integral and reading off the residue at $s=\frac{1}{2}(-\Delta_{345a}-2 \Delta_2+d+6)$, we obtain the expected massless limit,
\begin{align}
    &\peq
    \lim_{m\rightarrow 0}
    2^{-\Delta_{1}}\pi^{-\halfdim} m^{\Delta_{1}}\frac{\gm{\Delta_{1}}}{\gm{\Delta_{1}-\halfdim}}
    \cA_{1_{m}^{0}+2_{0}^{0}+3_{0}^{0}\to \shadow{4_{0}^{0}}+\shadow{5_{0}^{0}}}^{\Delta_1,\Delta_{2},\Delta_{3},\wave\Delta_{4},\wave\Delta_{5}}
    =
    \cA_{1_{0}^{0}+2_{0}^{0}+3_{0}^{0}\to \shadow{4_{0}^{0}}+\shadow{5_{0}^{0}}}^{\Delta_1,\Delta_{2},\Delta_{3},\wave\Delta_{4},\wave\Delta_{5}}
    \; .
\end{align}

The pole structure of \eqref{eq:sigmam00002} is summarized in Table \ref{tab:fivePointPoleStructure3}. 
The names of poles are chosen by comparing with the massless limit $m\to 0$.
We notice that after turning on the mass of the first propagator, 
the type-1 poles $\Delta_{a}=-\Delta_{345}+d+4$ in the massive case extend to a series poles at $\Delta_{a}=-\Delta_{345}+d+4+2N$. This can be understood by the small mass expansion of the propagator,
\begin{equation}
    \frac{1}{p^{2}+m^{2}}=\sum_{n=0}^{+\oo}(-1)^{n}\frac{m^{2n}}{(p^{2})^{n+1}}
    \; ,
\end{equation}
and the new $\Delta_{a}$-poles correspond to the higher order terms in this expansion.

\subsection{Six massless scalars with massless exchange}

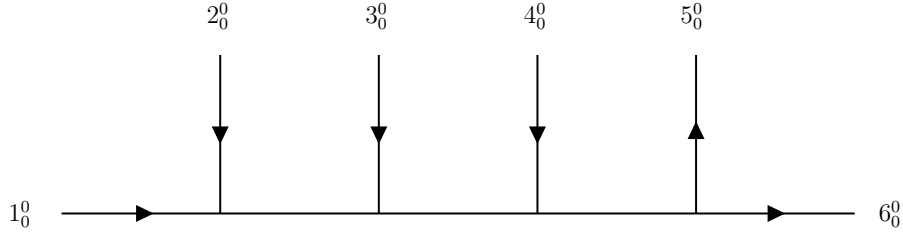
\begin{figure}[htbp]
\centering
\tikzset{every picture/.style={line width=0.75pt}} 

\begin{tikzpicture}[x=0.75pt,y=0.75pt,yscale=-1,xscale=1]
    
    \draw    (220,120) -- (220,200) ;
    \draw [shift={(220,165)}, rotate = 270] [fill={rgb, 255:red, 0; green, 0; blue, 0 }  ][line width=0.08]  [draw opacity=0] (8.93,-4.29) -- (0,0) -- (8.93,4.29) -- cycle    ;
    \draw    (220,200) -- (140,200) ;
    \draw [shift={(186.5,200)}, rotate = 180] [fill={rgb, 255:red, 0; green, 0; blue, 0 }  ][line width=0.08]  [draw opacity=0] (8.93,-4.29) -- (0,0) -- (8.93,4.29) -- cycle    ;
    \draw    (220,200) -- (460,200) ;
    \draw    (300,120) -- (300,200) ;
    \draw [shift={(300,165)}, rotate = 270] [fill={rgb, 255:red, 0; green, 0; blue, 0 }  ][line width=0.08]  [draw opacity=0] (8.93,-4.29) -- (0,0) -- (8.93,4.29) -- cycle    ;
    \draw    (380,120) -- (380,200) ;
    \draw [shift={(380,165)}, rotate = 270] [fill={rgb, 255:red, 0; green, 0; blue, 0 }  ][line width=0.08]  [draw opacity=0] (8.93,-4.29) -- (0,0) -- (8.93,4.29) -- cycle    ;
    \draw    (460,200) -- (540,200) ;
    \draw [shift={(505,200)}, rotate = 180] [fill={rgb, 255:red, 0; green, 0; blue, 0 }  ][line width=0.08]  [draw opacity=0] (8.93,-4.29) -- (0,0) -- (8.93,4.29) -- cycle    ;
    \draw    (460,120) -- (460,200) ;
    \draw [shift={(460,153.5)}, rotate = 90] [fill={rgb, 255:red, 0; green, 0; blue, 0 }  ][line width=0.08]  [draw opacity=0] (8.93,-4.29) -- (0,0) -- (8.93,4.29) -- cycle    ;
    
    \draw (112,192.4) node [anchor=north west][inner sep=0.75pt]  [xscale=0.8,yscale=0.8]  {$1_{0}^{0}$};
    \draw (212,92.4) node [anchor=north west][inner sep=0.75pt]  [xscale=0.8,yscale=0.8]  {$2_{0}^{0}$};
    \draw (292,92.4) node [anchor=north west][inner sep=0.75pt]  [xscale=0.8,yscale=0.8]  {$3_{0}^{0}$};
    \draw (372,92.4) node [anchor=north west][inner sep=0.75pt]  [xscale=0.8,yscale=0.8]  {$4_{0}^{0}$};
    \draw (451,92.4) node [anchor=north west][inner sep=0.75pt]  [xscale=0.8,yscale=0.8]  {$5_{0}^{0}$};
    \draw (551,192.4) node [anchor=north west][inner sep=0.75pt]  [xscale=0.8,yscale=0.8]  {$6_{0}^{0}$};

\end{tikzpicture}

\caption{
    A tree-level diagram for $1_{0}^{0}+2_{0}^{0}+3_{0}^{0}+4_{0}^{0}\to 5_{0}^{0}+6_{0}^{0}$. The external and internal particles are all massless scalars.}
\label{fig:sixPointFeynmanDiagram}
\end{figure}

In this subsection we explore whether there are further new exchanged operators in the tree-level celestial amplitudes in the $\phi^{3}$ theory.
We consider the scattering amplitude of the tree-level diagram \ref{fig:sixPointFeynmanDiagram},
\begin{align}
    \label{eq:sixPointM}
    \mathcal{M}_{1_{0}^{0}+2_{0}^{0}+3_{0}^{0}\to 4_{0}^{0}+5_{0}^{0}}
    =
    i^{3}\frac{\delta^{(d+2)}(q_1+q_2+q_3+q_4-q_5-q_6)}{(q_1+q_2)^2(q_1+q_2+q_3)^2(q_1+q_2+q_3+q_4)^2}
    \; .
\end{align}
Using the split representation \eqref{eq:SplitRep} and the integral expression of the six-point conformal partial wave \eqref{eq:sixPointCPW}, the celestial amplitude of \eqref{eq:sixPointM} is
\begin{align}
    \cA_{1_{0}^{0}+2_{0}^{0}+3_{0}^{0}+4_{0}^{0}\to \shadow{5_{0}^{0}}+\shadow{6_{0}^{0}}}^{\Delta_1,\Delta_{2},\Delta_{3},\Delta_{4},\wave\Delta_{5},\wave\Delta_{6}}
    &=
    \int_{0}^{+\infty}dm_{a}
    \int_{m_{a}}^{+\infty}dm_{b}
    \int_{m_{b}}^{+\infty}dm_{c}
    \left(
        \prod_{n\in\set{a,b,c}}
        \frac{-i m_{n}^{d-1}}{(2\pi)^{d+2}}
        \int_{\frac{d}{2}}^{\frac{d}{2}+i\infty}\frac{d\Delta_n}{2\pi i \, \mu_{\Delta_n}}
    \right)
    \nn
    \\
    \peq&
    \times
    C_{1^0_0+2^0_0\to a^0_{m_{a}}}^{\Delta_1,\Delta_2,\Delta_a}
    C_{a^0_{m_{a}}+3^0_0\to b^0_{m_{b}}}^{\wave\Delta_a,\Delta_3,\Delta_b}
    C_{b^0_{m_{b}}+4^0_0\to c^0_{m_{c}}}^{\wave\Delta_b,\Delta_4,\Delta_{c}}
    C_{c^0_{m_{c}}\to \shadow{5_{0}^{0}}+\shadow{6_{0}^{0}}}^{\wave\Delta_c,\wave\Delta_5,\wave\Delta_6}
    \partialWaveDressed^{\Delta_{1},\Delta_{2},\Delta_{3},\Delta_{4},\wave\Delta_{5},\wave\Delta_{6}}_{\Delta_{a},\Delta_{b},\Delta_{c}}
    \; .
    \label{eq:sixPointSplit}
\end{align}
There are two hypergeometric functions in the integrand, and by \eqref{eq:F21MB} we rewrite them as contour integrals of $s$ and $t$. Then the integrals of $m_{b}$ and $m_{c}$ can be done by the change of variables $\alpha=m_{b}^{2}-m_{a}^{2},\, \beta=m_{c}^{2}-m_{b}^{2}$, and the $m_{a}$-integral gives a $\delta$-function with complex argument \eqref{eq:complexDeltaFunction}. Finally we integrate out $s$ and $t$ by the first Barnes lemma \eqref{eq:first_mellin_barnes_lemma}. The result is 
\begin{align}
    \label{eq:sixPointContourIntegralShadow}
    \cA_{1_{0}^{0}+2_{0}^{0}+3_{0}^{0}+4_{0}^{0}\to \shadow{5_{0}^{0}}+\shadow{6_{0}^{0}}}^{\Delta_1,\Delta_{2},\Delta_{3},\Delta_{4},\wave\Delta_{5},\wave\Delta_{6}}
    &=
    \left(
        \prod_{n\in\set{a,b,c}}\int_{\frac{d}{2}-i\infty}^{\frac{d}{2}+i\infty}\frac{d\Delta_n}{2\pi i}
    \right)
    \sigma_{\Delta_a,\Delta_{b},\Delta_{c}}
    G_{\Delta_a,\Delta_{b},\Delta_{c}}^{\Delta_1,\Delta_{2},\Delta_{3},\Delta_{4},\wave\Delta_{5},\wave\Delta_{6}}
    \; ,
\end{align}
where
\begin{align}
    \sigma_{\Delta_a,\Delta_{b},\Delta_{c}}
    &=
    -i\,  \pi^{-\frac{7 d}{2}-6} 2^{-\Delta_{123456}-3 d-8} \complexDelta{\Delta_{123456}-d-8}
    \nn
    \\
    &\peq
    \times
    \multiGamma{
        \Delta_5,
        \Delta_6,
        \frac{\Delta_{1a,2}}{2},
        \frac{\Delta_{2a,1}}{2},
        \mathblue{\frac{\Delta_{3a,b}}{2}},
        \mathred{\frac{\Delta_{3b,a}}{2}},
        \frac{\Delta_{ab,3}}{2},
        \mathgreen{\frac{\Delta_{4b,c}}{2}},
        \frac{\Delta_{5c,6}}{2},
        \frac{\Delta_{6c,5}}{2},
        \mathblue{\frac{\Delta_{4c,b}}{2}},
        \frac{\Delta_{bc,4}}{2}
    }{
        d-\Delta_5,
        d-\Delta_6,
        \Delta_a,
        \Delta_{a}-\frac{d}{2},
        \Delta_b,
        \Delta_{b}-\frac{d}{2},
        \mathgreen{\frac{\Delta_{56,c}}{2}},
        \Delta_c,
        \Delta_{c}-\frac{d}{2}
    }
    \\
    &\peq 
    \times 
    \multiGamma{
        \mathred{\frac{-\Delta_{3456a}+d+6}{2}},
        \frac{\Delta_{a,3456}+6}{2},
        \mathblue{\frac{-\Delta_{456b}+d+4}{2}},
        \frac{\Delta_{b,456}+4}{2},
        \frac{\Delta_{3ab}-d}{2},
        \mathgreen{\frac{2 d-\Delta_{56c}}{2}},
        \frac{\Delta_{c,56}+d}{2},
        \frac{\Delta_{4bc}-d}{2}
    }{
        \mathblue{\frac{-\Delta_{456b}+d+6}{2}},
        \frac{\Delta_{b,456}+6}{2},
        \mathgreen{\frac{-\Delta_{56c}+d+4}{2}},
        \frac{\Delta_{c,56}+4}{2},
        \frac{\Delta_{56c}-d}{2}
    }
    \nn
    \; .
\end{align}

We perform the triple contour integrals in \eqref{eq:sixPointContourIntegralShadow} in the order $\Delta_{a} \to \Delta_{b} \to \Delta_{c}$.
Firstly there are two series of $\Delta_{a}$-poles (to the right of the principal series).
\begin{enumerate}
    \item The factor $\gm{\frac{-\Delta_{3456a}+d+6}{2}}$ provides the single-particle $\Delta_{a}$-poles at
    \begin{equation}
        \Delta_{a}=\Delta_{12}-2+2N
        \; ,
        \quad 
        N\in\NN
        \; .
    \end{equation}
    After integrating out $\Delta_{a}$, there are three types of $\Delta_{b}$-poles. 
    \begin{enumerate}
        \item 
        The first type is an alone single-particle $\Delta_{b}$-pole at 
        \begin{equation}
            \label{eq:sixPointBPoleAlone}
            \Delta_{b}=\Delta_{123}-4
            \; .
        \end{equation}
        Then after integrating out $\Delta_{b}$, the $\Delta_{c}$-poles are captured by 
        \begin{equation}
            \label{eq:sixPointCPole2}
            \Gamma\left[\frac{\Delta_{1234,c}+d-8}{2}\right]
            \; .
        \end{equation}
        \item 
        During the $\Delta_{a}$-integral, the factors $\gm{\frac{-\Delta_{3456a}+d+6}{2}}$ and $\gm{\frac{\Delta_{3a,b}}{2}}$ provide single-particle $\Delta_{b}$-poles at
        \begin{equation}
            \Delta_{b}=\Delta_{123}-2+2N+2M
            \; ,
            \quad 
            M\in\NN
            \; .
        \end{equation}
        Then after integrating out $\Delta_{b}$, the $\Delta_{c}$-poles are captured by 
        \begin{equation}
            \multiGamma{
                \frac{\Delta_{1234,c}+d-8}{2},
                \frac{\Delta_{1234,c}+2N+2 M-2}{2}
            }{
                \frac{\Delta_{1234,c}-4}{2}
            }
            \; .
        \end{equation}
        \item The factor $\gm{\frac{\Delta_{4c,b}}{2}}$ provides a series of ``running'' $\Delta_{b}$-poles at 
        \begin{equation}
            \label{eq:sixPointBPoleRunning1}
            \Delta_{b}=\Delta_{4c}+2M
            \; ,
            \quad  
            M\in\NN
            \; ,
        \end{equation}
        Then after integrating out $\Delta_{b}$, the $\Delta_{c}$-poles are captured by 
        \begin{equation}
            \label{eq:sixPointCPole1}
            \multiGamma{
                \frac{\Delta_{1234,c}+d-8}{2}
            }{
                \frac{\Delta_{1234,c}-4}{2}
            }
            \; .
        \end{equation}
    \end{enumerate}
    \item The factor $\gm{\frac{\Delta_{3b,a}}{2}}$ provides a series of ``running'' $\Delta_{a}$-poles at 
    \begin{equation}
        \Delta_{a}=\Delta_{3b}+2N
        \; ,
        \quad  
        N\in\NN
        \; .
    \end{equation}
    After integrating out $\Delta_{a}$, there are two types of $\Delta_{b}$-poles. 
    \begin{enumerate}
        \item
        The first type of poles of $\Delta_{b}$ and $\Delta_{c}$ are the same as \eqref{eq:sixPointBPoleAlone} and \eqref{eq:sixPointCPole2} respectively.
        \item 
        The second type of poles of $\Delta_{b}$ and $\Delta_{c}$ are the same as \eqref{eq:sixPointBPoleRunning1} and \eqref{eq:sixPointCPole1} respectively.
    \end{enumerate}
\end{enumerate}
The pole structure is summarized in Table \ref{tab:sixPointPoleStructure}. We notice that there is a new series type-3 poles appearing in the conformal partial wave expansion \eqref{eq:sixPointContourIntegralShadow} comparing to the five-point case.

\section{Conclusion}
\label{sec:conclusion}

In this paper, we computed scalar three-point celestial amplitudes involving both two and three massive scalars. We found that the three-point coefficient of celestial amplitudes with two massive scalars contains a hypergeometric function \eqref{eq:threePointm0m}, while the three-point coefficient of celestial amplitudes with three massive scalars can be written as a triple Mellin-Barnes integral \eqref{eq:tripleIntegral}. 

Based on these results and using the split representation \eqref{eq:SplitRep}, we studied the conformal block expansion of four-point celestial amplitudes involving two massive scalars. By choosing the conformal basis and shadow conformal basis for the incoming and outgoing particles respectively, we found that in the conformal block expansion of four-point celestial amplitudes $\cA_{1^0_m+2^0_0\to \shadow{3^0_m}+\shadow{4^0_0}}^{\Delta_{1},\Delta_{2},\wave\Delta_{3},\wave\Delta_{4}}$, there exist two series of double-trace poles \eqref{eq:four-point_0m0m_poles}, which is similar to the spectrum of four-point contact diagrams in AdS.

In addition, we obtained the conformal block expansion of five- and six-point massless celestial amplitudes. We found that when choosing the usual conformal basis for all massless particles, it is impossible to enclose the contour in the conformal partial wave expansion. To address this issue, we chose the shadow conformal basis for outgoing particles, while still using the usual conformal basis for the incoming particles, and the exchanged operators in the conformal block expansion are summarized in Table \ref{tab:fivePointPoleStructure} and \ref{tab:sixPointPoleStructure}. It is interesting that the operator spectrum depends on the spacetime dimension $d$. For $d=2$, we found that the $\op_{1}^{+}\op_{2}^+$ OPE contains three series of exchanged operators:
\begin{align}
\op_{1}^+\op_{2}^+
&\sim
\mathcal{C}_{\Delta_{12}-2+2N}\op_{\Delta_{12}-2+2N}^{+}
+\mathcal{C}_{\Delta_{12}+2\Delta_3-4+2N}\op_{\Delta_{12}+2\Delta_3-4+2N}^{+}
\\
\nonumber
&\phantom{{}\sim{}}
+\mathcal{C}_{\Delta_{12}+2\Delta_{34}-6+2N}\op_{\Delta_{12}+2\Delta_{34}-6+2N}^{+}
\; .
\end{align}
The existence of the first two series of operators in the conformal block expansion of both five- and six-point celestial amplitudes confirms the analysis in \cite{Ball:2023sdz,Guevara:2024ixn}. The leading terms $\op_{\Delta_{12}-2}^{+}$ and $\op_{\Delta_{12}+2\Delta_3-4}^{+}$ in these two series of operators have been interpreted as single- and two-particle operators, respectively. 
The third series of operators is new and only appears in the conformal block expansion of six-point celestial amplitudes. Following the same spirit of \cite{Ball:2023sdz,Guevara:2024ixn}, 
we propose that the leading term $\op_{\Delta_{12}+2\Delta_{34}-6}^{+}$ in this series corresponds to a three-particle operator. 
It is now natural to conjecture that there would be $n$-particle operators $\op_{\Delta_{12}+2\Delta_{34\cdots n+1}-2(n-1)}^+$ in the $\op_{1}^{+}\op_{2}^+$ OPE. This conjecture can be confirmed by examining $(n+3)$-point celestial amplitudes. Although we currently do not have the conformal block expansion of $(n+3)$-point celestial amplitudes, the computation can be done without much effort using the split representations.

We also extended our computations to five-point celestial amplitudes involving one massive scalar. The results are summarized in Table \ref{tab:fivePointPoleStructure2} and \ref{tab:fivePointPoleStructure3}. For $d=2$, we found that the OPE between one massive scalar and one massless scalar contains three types of operators:
\begin{align}
\op_{1,m}^+\op_{2}^+
&\sim
\mathcal{C}_{\Delta_{12}+2N}\op_{\Delta_{12}+2N}^{+}
+\mathcal{C}_{-\Delta_{345}+6}\op_{-\Delta_{345}+6}^{+}
\\
&\phantom{{}\sim{}}
+\mathcal{C}_{\Delta_{3,45}+4+2N}\op_{\Delta_{3,45}+4+2N}^{+}
\; .
\nn
\end{align}
The first series consists of double-trace operators, which also appear in the conformal block expansion of four-point celestial amplitudes \eqref{eq:four-point_0m0m_poles}. 
The second and the third types appear only in the conformal block expansion of five-point celestial amplitudes. 
In the massless limit $m\rightarrow0$, due to the emergence of $\complexDelta{\Delta_{12345}-8}$, $\op_{-\Delta_{345}+6}^{+}$ and $\op_{\Delta_{3,45}+4+2N}^{+}$ are reduced to $\op_{\Delta_{12}-2}^{+}$ and $\op_{\Delta_{12}+2\Delta_3-4+2N}^{+}$ respectively.
Hence we conjecture that the leading operator $\op_{\Delta_{3,45}+4}^+$ can be interpreted as a two-particle operator in the $\op_{1,m}^+\op_{2}^+$ OPE. 
Furthermore, if turning on the mass of the first propagator, the OPE is corrected as:
\begin{align}
\op_{1,m}^+\op_{2}^+
&\sim
\mathcal{C}_{\Delta_{12}+2N}\op_{\Delta_{12}+2N}^{+}
+\mathcal{C}_{-\Delta_{345}+6+2N}\op_{-\Delta_{345}+6+2N}^{+}
\\
&\phantom{{}\sim{}}
+\mathcal{C}_{\Delta_{3,45}+4+2N}\op_{\Delta_{3,45}+4+2N}^{+}
\; .
\nn
\end{align}
We explained that the extension of the series $\op_{-\Delta_{345}+6+2N}^{+}$ can be understood from the small mass expansion of the propagator.

Our work leads to many directions for future research. 
Firstly, it would be intriguing to study the conformal block expansion of six-point celestial amplitudes in other topologies and to deduce the corresponding operator spectra. 
It is known that at six-point there are two topologies: the comb topology and the snowflake topology \cite{Fortin:2020yjz}. We have shown that the comb channel conformal block expansion of six-point celestial amplitudes contains three-particle operators. 
To compute the one in the snowflake channel, it is necessary to use the three-point coefficients of three massive scalars \eqref{eq:tripleIntegral}. This computation is significantly more intricate and we leave it to the future work. 

Another avenue is to extend our methods to calculate three-point celestial amplitudes of massive spinning particles. 
These three-point celestial amplitudes play a crucial role in obtaining the conformal block expansion of higher-point celestial MHV amplitudes.
The authors of \cite{Ball:2023sdz} and \cite{Guevara:2024ixn} have shown that two-particle operators also appear in the OPEs of gluons.
It would be important to verify this at the level of conformal block expansion and to investigate if there are additional operators appearing in the OPE of gluons.

Finally, it would be of great interest to examine the consistency conditions of OPEs. 
Using our explicit results of conformal block expansions, it is important to study whether the coefficients can be factorized into a product of three-point coefficients, as expected from our understanding of CFTs.

\section*{Acknowledgements}

We thank Chi-Ming Chang and Yangrui Hu for the useful discussions. The work of WJM is supported by the National Natural Science Foundation of China No. 12405082.

\vspace{1cm}
\appendix
\renewcommand{\appendixname}{Appendix~\Alph{section}}

\section{Useful formulas}\label{sec:formulas}

In this appendix we list some useful formulas.
The Mellin-Barnes relation is
\begin{equation}
    \label{eq:MBRelation}
    (A+B)^{-\Delta}=
    \intrange{\frac{ds}{2\pi i}}{-i\oo}{+i\oo}
    \multiGamma{
        \Delta+s,-s
    }{
        \Delta
    }
    A^{s}B^{-s-\Delta}
    \; .
\end{equation}
The first Barnes lemma is
\begin{align}\label{eq:first_mellin_barnes_lemma}
    &\int_{-i\infty}^{+i\infty}\frac{ds}{2\pi i}
    \Gamma[
	a_1+s,
	a_2+s,
	b_1-s,
	b_2-s
    ]
    =
    \multiGamma{
        a_1+b_1,
	a_1+b_2,
        a_2+b_1,
        a_2+b_2
    }{
        a_1+a_2+b_1+b_2
    }
    \; .
\end{align}
For $\Re(\Delta_i)>0$ and $d=\sum_{i=1}^{n} \Delta_i$, the Feynman-Schwinger parameterization is
\begin{align}
    \label{eq:FSParameterization}
    \prod_{i=1}^{n} A_i^{-\Delta_i}
    &=
    \frac{\gm{d}}{\prod_{i=1}^{n} \gm{\Delta_i}}
    \int_0^{\infty} \prod_{i=2}^n d s_i s_i^{\Delta_i-1}\,
    (A_1+\sum_{i=2}^n s_i A_i)^{-d}
    \; .
\end{align}

The Mellin-Barnes representation of the hypergeometric function is 
\begin{equation}
    \label{eq:F21MB}
    \Fba(a,b,c,z)=
    \intrange{ds}{-i\oo}{+i\oo}
    (-z)^s 
    \multiGamma{
        c,
        -s,
        a+s,
        b+s
    }{
        a,
        b,
        c+s
    }
    \; .
\end{equation}
If $\arg(1-z)<\pi$, the hypergeometric function satisfies
\begin{equation}
    \label{eq:F21aToc-a}
    \Fba(a,b,c,z)=(1-z)^{c-a-b}\Fba(c-a,c-b,c,z)
    \; .
\end{equation}
If $\arg(-z)<\pi$, the hypergeometric function near $z=0$ is related to that near $z=\oo$ by 
\begin{align}
    \label{eq:F21zTo1/z}
    \Fba(a,b,c,z)
    &=
    (-z)^{-a}
    \multiGamma{
        b-a,
        c
    }{
        b,
        c-a
    }
    \Fba(a,a-c+1,a-b+1,\frac{1}{z}) 
    \\
    &\peq
    +
    (-z)^{-b} 
    \multiGamma{
        a-b,
        c
    }{
        a,
        c-b
    }
    \Fba(b,b-c+1,-a+b+1,\frac{1}{z})
    \; .
    \nn
\end{align}
If $\Re(c-a-b)>0$, the hypergeometric function with argument unity is 
\begin{equation}
    \label{eq:F21ArgumentUnity}
    \Fba(a,b,c,1)
    =
    \multiGamma{
        c,c-a-b
    }{
        c-a,
        c-b
    }
    \; .
\end{equation}

The Wilson polynomial is a special hypergeometric function ${}_{4}F_{3}(1)$, see \eg \cite{wilson1980some} and \cite[Section 3.8]{andrews1999special}, defined as
\begin{align}
    \label{eq:Wilson_polynomial}
    W_n(x^2 ; a, b, c, d)
    &= 
    \multiGamma{
        a+b+n,
        a+c+n,
        a+d+n
    }{
        a+b,
        a+c,
        a+d
    }
    \\
    &\peq
    \xx
    \Fpq{4}{3}{-n, a+b+c+d+n-1, a-i x, a+i x}{a+b, a+c, a+d}{1}
    \; ,
    \nn
\end{align}
It also appears in the crossing kernel of the Euclidean conformal group, see \eg \cite{groenevelt2003wilson,groenevelt2006wilson,Hogervorst:2017sfd,Sleight:2018ryu,Liu:2018jhs}.

\subsection{EAdS integrals}

The tree-level three-point contact Witten diagram is
\begin{equation}
    \label{eq:threePointWittenDiagram}
    D_{\Delta_{1},\Delta_{2},\Delta_{3}}
    =
    \intt{\frac{d^{d+1}\phat}{\phat^{0}}}
    (-\qhat_{1}\cdot\phat)^{-\Delta_{1}} 
    (-\qhat_{2}\cdot\phat)^{-\Delta_{2}} 
    (-\qhat_{3}\cdot\phat)^{-\Delta_{3}} 
    =
    d_{\Delta_{1},\Delta_{2},\Delta_{3}}\vev{\op_{1}\op_{2}\op_{3}}
    \; ,
\end{equation}
where 
\begin{equation}
    d_{\Delta_{1},\Delta_{2},\Delta_{3}}=
    \frac{1}{2} \pi^{d/2} 
    \multiGamma{
        \frac{\Delta_{12,3}}{2},
        \frac{\Delta_{13,2}}{2},
        \frac{\Delta_{23,1}}{2},
        \frac{\Delta_{123}-d}{2}
    }{
        \Delta_1,
        \Delta_2,
        \Delta_3
    }
    \; .
\end{equation}

When evaluating the massive-massive-massive three-point function we need the following EAdS integral,
\begin{align}
    \label{eq:EAdSIntegral}
    &I_{d,\Delta}(p)
    =
    \intt{\frac{d^{d+1}\phat_{1}}{\phat_{1}^{0}}}
    (-\phat_{1}\cdot p)^{-\Delta}
    \; ,
    \\
    &I_{d,\Delta}(p,a)
    =
    \intt{\frac{d^{d+1}\phat_{1}}{\phat_{1}^{0}}}
    (a-\phat_{1}\cdot p)^{-\Delta}
    \; ,
\end{align}
where $a\leq 0$ and $p\in\RR^{1,d+1}$ is a timelike vector. 
The first integral can be computed as follows: using the Lorentz symmetry and the scaling behavior of $p$, it takes the form as 
\begin{equation}
    I_{d,\Delta}(p)
    =
    f(\Delta)(-p^2)^{-\frac{\Delta}{2}}
    \; .
\end{equation}
We choose the rest frame of $p$, \ie $p^{\mu}=(1,0,\cdots,0)$ to determine the $p$-independent function $f(\Delta)$,
and the above integral becomes
\begin{align}
    f(\Delta)
    =
    \intt{d^{d+1}\phat_{1}}
    (\phat_{1}^0)^{-\Delta-1}
    =
    \int_0^{+\infty}dr
    \int_{\sphere{d}} d\Omega\, 
    r^d(1+r^2)^{-\frac{\Delta+1}{2}}
    =
    \pi^{\frac{d+1}{2}}
    \multiGamma{\frac{\Delta-d}{2}}{\frac{\Delta+1}{2}}
    \; .
\end{align}
This implies 
\begin{equation}
    I_{d,\Delta}(p)
    =
    \pi^{\frac{d+1}{2}}
    \multiGamma{\frac{\Delta-d}{2}}{\frac{\Delta+1}{2}}
    (-p^2)^{-\frac{\Delta}{2}}
    \; .
\end{equation}
By the Mellin-Barnes relation \eqref{eq:MBRelation} and the integral \eqref{eq:EAdSIntegral}, we have 
\begin{align}
    I_{d,\Delta}(p,a)
    &=
    \intt{\frac{d^{d+1}\phat_{1}}{\phat_{1}^{0}}}
    \intrange{\frac{ds}{2\pi i}}{-i\oo}{+i\oo}
    \multiGamma{
        \Delta+s,-s
    }{
        \Delta
    }
    (-\phat_{1}\cdot p)^{-s-\Delta}
    a^{s}
    \\
    &=
    \pi^{\frac{d+1}{2}}
    \intrange{\frac{ds}{2\pi i}}{-i\oo}{+i\oo}
    \multiGamma{
        \Delta+s,-s,\frac{\Delta+s-d}{2}
    }{
        \Delta,\frac{\Delta+s+1}{2}
    }
    (-p^2)^{-\frac{\Delta+s}{2}}
    a^{s}
    \nn
    \; .
\end{align}
%

\subsection{Analytic functional}

The $\delta$-function with complex argument is defined by the following properties,
\begin{align}
    \label{eq:complexDeltaFunction}
    &\complexDelta{z-w}
    =
    \int_0^{+\infty}d\alpha\;\alpha^{z-w-1}
    \; ,
    \\
    \label{eq:complexDeltaFunction2}
    &\intrange{\frac{dz}{2\pi i}}{a-i\infty}{a+i\infty}\complexDelta{z-w}f(z)=f(w)
    \; ,
\end{align}
and when the argument locates on the integration contour, it reduces to the usual $\delta$-function.
This is an analytic functional belonging to the Gelfand–Shilov space $Z'$, see e.g. \cite{Gelfand1,Gelfand2} and \cite[Section 4]{Donnay:2020guq}.

The $\delta$-function with complex argument admits the following approximations in the sense of weak limit,
\begin{equation}
    \label{eq:complexDeltaFunctionAsLimit}
    \lim_{m\to 0}\frac{1}{2}\gm{-\frac{\Delta}{2}} m^{\Delta}=\complexDelta{\Delta}
    \; .
\end{equation}
To prove this, we integrate the left side with an analytic test function $f(\Delta)$, then enclose the contour to the right half $\Delta$-plane and pick up poles at $\Delta=2n,\, n\in\NN$, leading to
\begin{align}
    &\peq
    \lim_{m\to 0}
    \intrange{\frac{d\Delta}{2\pi i}}{a-i\oo}{a+i\oo}
    \frac{1}{2}\gm{-\frac{\Delta}{2}} m^{\Delta}f(\Delta)
    =
    \lim_{m\to 0}\sum_{n=0}^{+\oo}\frac{(-1)^{n}}{n!}m^{2n}f(2n)
    =
    f(0)
    \; ,
\end{align}
which justifies the defining property \eqref{eq:complexDeltaFunction2}.

\newpage
\section{Tables of operator spectra}\label{sec:tables}

\begin{table}[H]
    \renewcommand{\arraystretch}{1.2}
    \begin{tabularx}{\textwidth}{|L|L|L|}
    \cline{1-3}
    d 
    & 
    \begin{array}{l}
        \text{type-1 $\Delta_{a}$}\\
        \text{type-1 $\Delta_{b}$}
    \end{array} 
    & 
    \begin{array}{l}
        \text{type-2 $\Delta_{a}$}\\
        \text{type-1 $\Delta_{b}$}
    \end{array} 
    \\
    \cline{1-3}
    2
    &
    \begin{array}{l}
        \Delta_{a}=\Delta_{12}-2+2N\\
        \Delta_{b}=\Delta_{123}-4 \textInMath{or}\\
        \phantom{\Delta_{c}={}}\Delta_{123}-2+2N+2M
    \end{array} 
    &
    \begin{array}{l}
        \Delta_{a}=\Delta_{12}+2\Delta_{3}-4+2N\\
        \Delta_{b}=\Delta_{123}-4\\
        \phantom{\Delta_{c}={}}
    \end{array} 
    \\
    \cline{1-3}
    4,6,\dots 
    &
    \begin{array}{l}
        \Delta_{a}=\Delta_{12}-2+2N\\
        \Delta_{b}=\Delta_{123}-2+2N+2M
    \end{array} 
    &
    \textInMath{/}
    \\
    \cline{1-3}
    1,3,\dots 
    &
    \begin{array}{l}
        \Delta_{a}=\Delta_{12}-2+2N\\
        \Delta_{b}=\Delta_{123}+d-6+2M
    \end{array} 
    &  
    \begin{array}{l}
        \Delta_{a}=\Delta_{12}+2\Delta_{3}+d-6+2N+2M\\
        \Delta_{b}=\Delta_{123}+d-6+2M
    \end{array} 
    \\
    \cline{1-3}
    \end{tabularx}

    \caption{
    The conformal dimensions of exchanged operators in the \schannel conformal block expansion of $\cA_{1_{0}^{0}+2_{0}^{0}+3_{0}^{0}\to \shadow{4_{0}^{0}}+\shadow{5_{0}^{0}}}^{\Delta_1,\Delta_{2},\Delta_{3},\wave\Delta_{4},\wave\Delta_{5}}$, where $N,M\in\NN$.}
    \label{tab:fivePointPoleStructure}
\end{table}

\begin{table}[H]
    \renewcommand{\arraystretch}{1.2}
    \begin{tabularx}{\textwidth}{|L|L|L|L|}
    \cline{1-3}
    d 
    & 
    \begin{array}{l}
        \text{type-1 $\Delta_{a}$}\\
        \text{type-1 $\Delta_{b}$}
    \end{array} 
    & 
    \begin{array}{l}
        \text{type-2 $\Delta_{a}$}\\
        \text{type-1 $\Delta_{b}$}
    \end{array} 
    \\
    \cline{1-3}
    2
    &
    \begin{array}{l}
        \Delta_{a}=\Delta_{12}+2N\\
        \Delta_{b}=-\Delta_{45}+4 \textInMath{or}\\
        \phantom{\Delta_{b}={}}\Delta_{123}+2N+2M
    \end{array} 
    &
    \begin{array}{l}
        \Delta_{a}=\Delta_{3,45}+4+2N\\
        \Delta_{b}=-\Delta_{45}+4\\
        \phantom{\Delta_{b}={}}
    \end{array} 
    \\
    \cline{1-3}
    4,6,\dots 
    &
    \begin{array}{l}
        \Delta_{a}=\Delta_{12}+2N\\
        \Delta_{b}=\Delta_{123}+2N+2M
    \end{array} 
    &
    \textInMath{/}
    \\
    \cline{1-3}
    1,3,\dots 
    &
    \begin{array}{l}
        \Delta_{a}=\Delta_{12}+2N\\
        \Delta_{b}=-\Delta_{45}+2d+2M\textInMath{or}\\
        \phantom{\Delta_{b}={}}\Delta_{123}+2N+2M
    \end{array} 
    &  
    \begin{array}{l}
        \Delta_{a}=\Delta_{3,45}+2d+2N+2M\\
        \Delta_{b}=-\Delta_{45}+2d+2M\\
        \phantom{\Delta_{b}={}}
    \end{array} 
    \\
    \cline{1-3}
    \multicolumn{1}{c}{}
    &
    \multicolumn{1}{c}{}
    &
    \multicolumn{1}{c}{}
    \\[-1.3em]
    \cline{1-2}
    d
    & 
    \begin{array}{l}
        \text{type-1 $\Delta_{a}$}\\
        \text{type-1 $\Delta_{b}$}
    \end{array} 
    \\
    \cline{1-2}
    1,2,\dots
    &
    \begin{array}{l}
        \Delta_{a}=-\Delta_{345}+d+4\\
        \Delta_{b}=-\Delta_{45}+2d+2M
    \end{array}
    \\
    \cline{1-2}
    \end{tabularx}

    \caption{
    The conformal dimensions of exchanged operators in the \schannel conformal block expansion of $\cA_{1_{m}^{0}+2_{0}^{0}+3_{0}^{0}\to \shadow{4_{0}^{0}}+\shadow{5_{0}^{0}}}^{\Delta_1,\Delta_{2},\Delta_{3},\wave\Delta_{4},\wave\Delta_{5}}$ with massless exchange, where $N,M\in\NN$.
    }
    \label{tab:fivePointPoleStructure2}
\end{table}

\newpage

\begin{table}[H]
    \renewcommand{\arraystretch}{1.2}
    \begin{tabularx}{\textwidth}{|L|L|L|L|}
    \cline{1-3}
    d 
    & 
    \begin{array}{l}
        \text{type-1 $\Delta_{a}$}\\
        \text{type-1 $\Delta_{b}$}
    \end{array} 
    & 
    \begin{array}{l}
        \text{type-2 $\Delta_{a}$}\\
        \text{type-1 $\Delta_{b}$}
    \end{array} 
    \\
    \cline{1-3}
    2
    &
    \begin{array}{l}
        \Delta_{a}=\Delta_{12}+2N\\
        \Delta_{b}=-\Delta_{45}+4 \textInMath{or}\\
        \phantom{\Delta_{b}={}}\Delta_{123}+2N+2M
    \end{array} 
    &
    \begin{array}{l}
        \Delta_{a}=\Delta_{3,45}+4+2N\\
        \Delta_{b}=-\Delta_{45}+4\\
        \phantom{\Delta_{b}={}}
    \end{array} 
    \\
    \cline{1-3}
    4,6,\dots 
    &
    \begin{array}{l}
        \Delta_{a}=\Delta_{12}+2N\\
        \Delta_{b}=\Delta_{123}+2N+2M
    \end{array} 
    &
    \textInMath{/}
    \\
    \cline{1-3}
    1,3,\dots 
    &
    \begin{array}{l}
        \Delta_{a}=\Delta_{12}+2N\\
        \Delta_{b}=-\Delta_{45}+2d+2M\textInMath{or}\\
        \phantom{\Delta_{b}={}}\Delta_{123}+2N+2M
    \end{array} 
    &  
    \begin{array}{l}
        \Delta_{a}=\Delta_{3,45}+2d+2N+2M\\
        \Delta_{b}=-\Delta_{45}+2d+2M\\
        \phantom{\Delta_{b}={}}
    \end{array} 
    \\
    \cline{1-3}
    \multicolumn{1}{c}{}
    &
    \multicolumn{1}{c}{}
    &
    \multicolumn{1}{c}{}
    \\[-1.3em]
    \cline{1-2}
    d 
    & 
    \begin{array}{l}
        \text{type-1 $\Delta_{a}$}\\
        \text{type-1 $\Delta_{b}$}
    \end{array} 
    & 
    \multicolumn{1}{c}{}
    \\
    \cline{1-2}
    2
    &
    \begin{array}{l}
        \Delta_{a}=-\Delta_{345}+6+2N\\
        \Delta_{b}=-\Delta_{45}+4\textInMath{or}\\
        \phantom{\Delta_{b}={}}-\Delta_{45}+6+2N+2M
    \end{array}
    &
    \multicolumn{1}{c}{}
    \\
    \cline{1-2}
    4,6,\dots 
    &
    \begin{array}{l}
        \Delta_{a}=-\Delta_{345}+d+4+2N\\
        \Delta_{b}=-\Delta_{45}+d+4+2N+2M
    \end{array}
    &
    \multicolumn{1}{c}{}
    \\
    \cline{1-2}
    1,3,\dots 
    &
    \begin{array}{l}
        \Delta_{a}=-\Delta_{345}+d+4+2N\\
        \Delta_{b}=-\Delta_{45}+2d+2M
    \end{array}
    &
    \multicolumn{1}{c}{}
    \\
    \cline{1-2}
    \end{tabularx}

    \caption{
    The conformal dimensions of exchanged operators in the \schannel conformal block expansion of $\cA_{1_{m}^{0}+2_{0}^{0}+3_{0}^{0}\to \shadow{4_{0}^{0}}+\shadow{5_{0}^{0}}}^{\Delta_1,\Delta_{2},\Delta_{3},\wave\Delta_{4},\wave\Delta_{5}}$ with massive exchange, where $N,M\in\NN$.
    }
    \label{tab:fivePointPoleStructure3}
\end{table}

\newpage

\begin{table}[H]
    \renewcommand{\arraystretch}{1.2}
    \begin{tabularx}{\textwidth}{|L|L|L|}
    \cline{1-3}
    d 
    & 
    \begin{array}{l}
        \text{type-1 $\Delta_{a}$}\\
        \text{type-1 $\Delta_{b}$}\\
        \text{type-1 $\Delta_{c}$}
    \end{array} 
    &
    \begin{array}{l}
        \text{type-1,2 $\Delta_{a}$}\\
        \text{type-1 $\Delta_{b}$}\\
        \text{type-1 $\Delta_{c}$}
    \end{array} 
    \\
    \cline{1-3}
    2 
    &
    \begin{array}{l}
        \Delta_{a}=\Delta_{12}-2+2N\\
        \Delta_{b}=\Delta_{123}-2+2N+2M\\
        \Delta_{c}=\Delta_{1234}-6 
        \textInMath{or}\\
        \phantom{\Delta_{c}={}}\Delta_{1234}-2+2N+2M+2K 
    \end{array} 
    &
    \multirow[b]{3}{*}{
    $\!\begin{array}{l}
        \Delta_{a}=\Delta_{12}-2+2N \textInMath{or}\\
        \phantom{\Delta_{a}={}}\Delta_{12}+2\Delta_{3}-4+2N\\
        \Delta_{b}=\Delta_{123}-4\\
        \Delta_{c}=\Delta_{1234}+d-8+2K
    \end{array} $
    }
    \\
    \cline{1-2}
    4,6,\dots 
    &
    \begin{array}{l}
        \Delta_{a}=\Delta_{12}-2+2N\\
        \Delta_{b}=\Delta_{123}-2+2N+2M\\
        \Delta_{c}=\Delta_{1234}-2+2N+2M+2K 
    \end{array} 
    &
    \\
    \cline{1-2}
    1,3,\dots 
    &
    \begin{array}{l}
        \Delta_{a}=\Delta_{12}-2+2N\\
        \Delta_{b}=\Delta_{123}-2+2N+2M\\
        \Delta_{c}=\Delta_{1234}+d-8+2K 
    \end{array}
    &
    \\
    \cline{1-3}
    \multicolumn{1}{c}{}
    &
    \multicolumn{1}{c}{}
    &
    \multicolumn{1}{c}{}
    \\[-1.3em]
    \cline{1-2}
    d 
    & 
    \begin{array}{l}
        \text{type-1,3 $\Delta_{a}$}\\
        \text{type-2 $\Delta_{b}$}\\
        \text{type-1 $\Delta_{c}$}
    \end{array} 
    &
    \multicolumn{1}{c}{}
    \\
    \cline{1-2}
    2 
    &
    \begin{array}{l}
        \Delta_{a}=\Delta_{12}-2+2N \textInMath{or}\\
        \phantom{\Delta_{a}={}}\Delta_{12}+2\Delta_{34}-6+2N+2M\\
        \Delta_{b}=\Delta_{123}+2\Delta_{4}-6+2M\\
        \Delta_{c}=\Delta_{1234}-6
    \end{array} 
    &
    \multicolumn{1}{c}{}
    \\
    \cline{1-2}
    4,6,\dots 
    &
    \textInMath{/}
    &
    \multicolumn{1}{c}{}
    \\
    \cline{1-2}
    1,3,\dots 
    &  
    \begin{array}{l}
        \Delta_{a}=\Delta_{12}-2+2N \textInMath{or}\\
        \phantom{\Delta_{a}={}}\Delta_{12}+2\Delta_{34}+d-8+2N+2M+2K\\
        \Delta_{b}=\Delta_{123}+2\Delta_{4}+d-8+2M+2K\\
        \Delta_{c}=\Delta_{1234}+d-8+2K
    \end{array} 
    &
    \multicolumn{1}{c}{}
    \\
    \cline{1-2}
    \end{tabularx}

    \caption{
    The conformal dimensions of exchanged operators in the \schannel conformal block expansion of $\cA_{1_{0}^{0}+2_{0}^{0}+3_{0}^{0}+4_{0}^{0}\to \shadow{5_{0}^{0}}+\shadow{6_{0}^{0}}}^{\Delta_1,\Delta_{2},\Delta_{3},\Delta_{4},\wave\Delta_{5},\wave\Delta_{6}}$, where $N,M,K\in\NN$.}
    \label{tab:sixPointPoleStructure}
\end{table}

\newpage
\printbibliography

@article{Melton:2024gyu,
    title         = {{A Celestial Dual for MHV Amplitudes}},
    author        = {Melton, Walker and Sharma, Atul and Strominger, Andrew and Wang, Tianli},
    year          = {2024},
    eprint        = {2403.18896},
    archiveprefix = {arXiv},
    month         = {3},
    primaryclass  = {hep-th}
}

@article{deGioia:2024yne,
    title         = {{Celestial amplitudes from conformal correlators with bulk-point kinematics}},
    author        = {de Gioia, Leonardo Pipolo and Raclariu, Ana-Maria},
    year          = {2024},
    eprint        = {2405.07972},
    archiveprefix = {arXiv},
    month         = {5},
    primaryclass  = {hep-th}
}

@article{Fan:2023lky,
    title         = {{Celestial conformal blocks of massless scalars and analytic continuation of the Appell function F$_{1}$}},
    author        = {Fan, Wei},
    year          = {2024},
    eprint        = {2311.11345},
    archiveprefix = {arXiv},
    doi           = {10.1007/JHEP01(2024)145},
    journal       = {JHEP},
    pages         = {145},
    primaryclass  = {hep-th},
    volume        = {01}
}

@article{Iacobacci:2024nhw,
    title         = {{Celestial holography revisited. Part II. Correlators and K\"all\'en-Lehmann}},
    author        = {Iacobacci, Lorenzo and Sleight, Charlotte and Taronna, Massimo},
    year          = {2024},
    eprint        = {2401.16591},
    archiveprefix = {arXiv},
    doi           = {10.1007/JHEP08(2024)033},
    journal       = {JHEP},
    pages         = {033},
    primaryclass  = {hep-th},
    volume        = {08}
}

@article{Furugori:2023hgv,
    title         = {{Celestial two-point functions and rectified dictionary}},
    author        = {Furugori, Hideo and Ogawa, Naoki and Sugishita, Sotaro and Waki, Takahiro},
    year          = {2024},
    eprint        = {2312.07057},
    archiveprefix = {arXiv},
    doi           = {10.1007/JHEP02(2024)063},
    journal       = {JHEP},
    pages         = {063},
    primaryclass  = {hep-th},
    reportnumber  = {KUNS-2988, YITP-23-160},
    volume        = {02}
}

@article{Ball:2023sdz,
    title         = {{Multicollinear singularities in celestial CFT}},
    author        = {Ball, Adam and Hu, Yangrui and Pasterski, Sabrina},
    year          = {2024},
    eprint        = {2309.16602},
    archiveprefix = {arXiv},
    doi           = {10.1007/JHEP02(2024)219},
    journal       = {JHEP},
    pages         = {219},
    primaryclass  = {hep-th},
    volume        = {02}
}

@article{Guevara:2024ixn,
    title         = {{Multiparticle Contributions to the Celestial OPE}},
    author        = {Guevara, Alfredo and Hu, Yangrui and Pasterski, Sabrina},
    year          = {2024},
    eprint        = {2402.18798},
    archiveprefix = {arXiv},
    month         = {2},
    primaryclass  = {hep-th}
}

@article{Fortin:2023xqq,
    title         = {{One- and two-dimensional higher-point conformal blocks as free-particle wavefunctions in $ {\textrm{AdS}}_3^{\otimes m} $}},
    author        = {Fortin, Jean-Fran\c{c}ois and Ma, Wen-Jie and Parikh, Sarthak and Quintavalle, Lorenzo and Skiba, Witold},
    year          = {2024},
    eprint        = {2310.08632},
    archiveprefix = {arXiv},
    doi           = {10.1007/JHEP01(2024)031},
    journal       = {JHEP},
    pages         = {031},
    primaryclass  = {hep-th},
    volume        = {01}
}

@article{Himwich:2023njb,
    title         = {{${\rm w}_{1+\infty}$ in 4D Gravitational Scattering}},
    author        = {Himwich, Elizabeth and Pate, Monica},
    year          = {2023},
    eprint        = {2312.08597},
    archiveprefix = {arXiv},
    month         = {12},
    primaryclass  = {hep-th}
}

@article{Sleight:2023ojm,
    title         = {{Celestial Holography Revisited}},
    author        = {Sleight, Charlotte and Taronna, Massimo},
    year          = {2023},
    eprint        = {2301.01810},
    archiveprefix = {arXiv},
    month         = {1},
    primaryclass  = {hep-th}
}

@article{Melton:2023bjw,
    title         = {{Celestial Leaf Amplitudes}},
    author        = {Melton, Walker and Sharma, Atul and Strominger, Andrew},
    year          = {2023},
    eprint        = {2312.07820},
    archiveprefix = {arXiv},
    month         = {12},
    primaryclass  = {hep-th}
}

@article{deGioia:2022fcn,
    title         = {{Eikonal approximation in celestial CFT}},
    author        = {de Gioia, Leonardo Pipolo and Raclariu, Ana-Maria},
    year          = {2023},
    eprint        = {2206.10547},
    archiveprefix = {arXiv},
    doi           = {10.1007/JHEP03(2023)030},
    journal       = {JHEP},
    pages         = {030},
    primaryclass  = {hep-th},
    volume        = {03}
}

@article{Iacobacci:2022yjo,
    title         = {{From celestial correlators to AdS, and back}},
    author        = {Iacobacci, Lorenzo and Sleight, Charlotte and Taronna, Massimo},
    year          = {2023},
    eprint        = {2208.01629},
    archiveprefix = {arXiv},
    doi           = {10.1007/JHEP06(2023)053},
    journal       = {JHEP},
    pages         = {053},
    primaryclass  = {hep-th},
    volume        = {06}
}

@article{Chang:2022seh,
    title         = {{Missing corner in the sky: massless three-point celestial amplitudes}},
    author        = {Chang, Chi-Ming and Ma, Wen-Jie},
    year          = {2023},
    eprint        = {2212.07025},
    archiveprefix = {arXiv},
    doi           = {10.1007/JHEP04(2023)051},
    journal       = {JHEP},
    pages         = {051},
    primaryclass  = {hep-th},
    volume        = {04}
}

@article{Chang:2022jut,
    title         = {{Shadow celestial amplitudes}},
    author        = {Chang, Chi-Ming and Cui, Wei and Ma, Wen-Jie and Shu, Hongfei and Zou, Hao},
    year          = {2023},
    eprint        = {2210.04725},
    archiveprefix = {arXiv},
    doi           = {10.1007/JHEP02(2023)017},
    journal       = {JHEP},
    pages         = {017},
    primaryclass  = {hep-th},
    volume        = {02}
}

@article{Chang:2023ttm,
    title         = {{Split representation in celestial holography}},
    author        = {Chang, Chi-Ming and Liu, Reiko and Ma, Wen-Jie},
    year          = {2023},
    eprint        = {2311.08736},
    archiveprefix = {arXiv},
    month         = {11},
    primaryclass  = {hep-th}
}

@article{Chen:2022cpx,
    title         = {{The shadow formalism of Galilean CFT$_{2}$}},
    author        = {Chen, Bin and Liu, Reiko},
    year          = {2023},
    eprint        = {2203.10490},
    archiveprefix = {arXiv},
    doi           = {10.1007/JHEP05(2023)224},
    journal       = {JHEP},
    pages         = {224},
    primaryclass  = {hep-th},
    volume        = {05}
}

@article{Chang:2021wvv,
    title         = {{Bulk locality from the celestial amplitude}},
    author        = {Chang, Chi-Ming and Huang, Yu-tin and Huang, Zi-Xun and Li, Wei},
    year          = {2022},
    eprint        = {2106.11948},
    archiveprefix = {arXiv},
    doi           = {10.21468/SciPostPhys.12.5.176},
    journal       = {SciPost Phys.},
    number        = {5},
    pages         = {176},
    primaryclass  = {hep-th},
    volume        = {12}
}

@article{Fan:2022kpp,
    title         = {{Celestial Yang-Mills amplitudes and D = 4 conformal blocks}},
    author        = {Fan, Wei and Fotopoulos, Angelos and Stieberger, Stephan and Taylor, Tomasz R. and Zhu, Bin},
    year          = {2022},
    eprint        = {2206.08979},
    archiveprefix = {arXiv},
    doi           = {10.1007/JHEP09(2022)182},
    journal       = {JHEP},
    pages         = {182},
    primaryclass  = {hep-th},
    volume        = {09}
}

@article{fortin2022feynman,
    title   = {Feynman Rules for Scalar Conformal Blocks},
    author  = {Fortin, Jean-Fran{\c{c}}ois and Hoback, Sarah and Ma, Wen-Jie and Parikh, Sarthak and Skiba, Witold},
    year    = {2022},
    eprint  = {2204.08909},
    journal = { }
}

@article{Buric:2021kgy,
    title         = {{Gaudin models and multipoint conformal blocks III: comb channel coordinates and OPE factorisation}},
    author        = {Buric, Ilija and Lacroix, Sylvain and Mann, Jeremy A. and Quintavalle, Lorenzo and Schomerus, Volker},
    year          = {2022},
    eprint        = {2112.10827},
    archiveprefix = {arXiv},
    doi           = {10.1007/JHEP06(2022)144},
    journal       = {JHEP},
    pages         = {144},
    primaryclass  = {hep-th},
    volume        = {06}
}

@article{Garcia-Sepulveda:2022lga,
    title         = {{Notes on resonances and unitarity from celestial amplitudes}},
    author        = {García-Sepúlveda, Diego and Guevara, Alfredo and Kulp, Justin and Wu, Jingxiang},
    year          = {2022},
    eprint        = {2205.14633},
    archiveprefix = {arXiv},
    doi           = {10.1007/JHEP09(2022)245},
    journal       = {JHEP},
    pages         = {245},
    primaryclass  = {hep-th},
    volume        = {09}
}

@article{Atanasov:2021cje,
    title         = {{Conformal block expansion in celestial CFT}},
    author        = {Atanasov, Alexander and Melton, Walker and Raclariu, Ana-Maria and Strominger, Andrew},
    year          = {2021},
    eprint        = {2104.13432},
    archiveprefix = {arXiv},
    doi           = {10.1103/PhysRevD.104.126033},
    journal       = {Phys. Rev. D},
    number        = {12},
    pages         = {126033},
    primaryclass  = {hep-th},
    volume        = {104}
}

@article{Fan:2021isc,
    title         = {{Conformal blocks from celestial gluon amplitudes}},
    author        = {Fan, Wei and Fotopoulos, Angelos and Stieberger, Stephan and Taylor, Tomasz R. and Zhu, Bin},
    year          = {2021},
    eprint        = {2103.04420},
    archiveprefix = {arXiv},
    doi           = {10.1007/JHEP05(2021)170},
    journal       = {JHEP},
    pages         = {170},
    primaryclass  = {hep-th},
    volume        = {05}
}

@article{Fan:2021pbp,
    title         = {{Conformal blocks from celestial gluon amplitudes. Part II. Single-valued correlators}},
    author        = {Fan, Wei and Fotopoulos, Angelos and Stieberger, Stephan and Taylor, Tomasz R. and Zhu, Bin},
    year          = {2021},
    eprint        = {2108.10337},
    archiveprefix = {arXiv},
    doi           = {10.1007/JHEP11(2021)179},
    journal       = {JHEP},
    pages         = {179},
    primaryclass  = {hep-th},
    volume        = {11}
}

@article{Hoback:2020syd,
    title         = {{Dimensional reduction of higher-point conformal blocks}},
    author        = {Hoback, Sarah and Parikh, Sarthak},
    year          = {2021},
    eprint        = {2009.12904},
    archiveprefix = {arXiv},
    doi           = {10.1007/JHEP03(2021)187},
    journal       = {JHEP},
    pages         = {187},
    primaryclass  = {hep-th},
    volume        = {03}
}

@article{Buric:2020dyz,
    title         = {{From Gaudin Integrable Models to $d$-dimensional Multipoint Conformal Blocks}},
    author        = {Buric, Ilija and Lacroix, Sylvain and Mann, Jeremy A. and Quintavalle, Lorenzo and Schomerus, Volker},
    year          = {2021},
    eprint        = {2009.11882},
    archiveprefix = {arXiv},
    doi           = {10.1103/PhysRevLett.126.021602},
    journal       = {Phys. Rev. Lett.},
    number        = {2},
    pages         = {021602},
    primaryclass  = {hep-th},
    reportnumber  = {DESY-20-157, DESY 20-157, ZMP-HH/20-25, SAGEX-20-22-E},
    volume        = {126}
}

@article{Buric:2021ywo,
    title         = {{Gaudin models and multipoint conformal blocks: general theory}},
    author        = {Buric, Ilija and Lacroix, Sylvain and Mann, Jeremy A. and Quintavalle, Lorenzo and Schomerus, Volker},
    year          = {2021},
    eprint        = {2105.00021},
    archiveprefix = {arXiv},
    doi           = {10.1007/JHEP10(2021)139},
    journal       = {JHEP},
    pages         = {139},
    primaryclass  = {hep-th},
    reportnumber  = {DESY-21-052, DESY 21-052, SAGEX21-08-E, ZMP-HH/21-8},
    volume        = {10}
}

@article{Buric:2021ttm,
    title         = {{Gaudin models and multipoint conformal blocks. Part II. Comb channel vertices in 3D and 4D}},
    author        = {Buric, Ilija and Lacroix, Sylvain and Mann, Jeremy A. and Quintavalle, Lorenzo and Schomerus, Volker},
    year          = {2021},
    eprint        = {2108.00023},
    archiveprefix = {arXiv},
    doi           = {10.1007/JHEP11(2021)182},
    journal       = {JHEP},
    pages         = {182},
    primaryclass  = {hep-th},
    reportnumber  = {DESY 21-105, SAGEX-21-14-E, ZMP-HH/21-15},
    volume        = {11}
}

@article{Raclariu:2021zjz,
    title         = {{Lectures on Celestial Holography}},
    author        = {Raclariu, Ana-Maria},
    year          = {2021},
    eprint        = {2107.02075},
    archiveprefix = {arXiv},
    month         = {7},
    primaryclass  = {hep-th}
}

@article{Poland:2021xjs,
    title         = {{Recursion relations for 5-point conformal blocks}},
    author        = {Poland, David and Prilepina, Valentina},
    year          = {2021},
    eprint        = {2103.12092},
    archiveprefix = {arXiv},
    doi           = {10.1007/JHEP10(2021)160},
    journal       = {JHEP},
    pages         = {160},
    primaryclass  = {hep-th},
    volume        = {10}
}

@article{Haehl:2021tft,
    title         = {{Six-point functions and collisions in the black hole interior}},
    author        = {Haehl, Felix M. and Streicher, Alexandre and Zhao, Ying},
    year          = {2021},
    eprint        = {2105.12755},
    archiveprefix = {arXiv},
    doi           = {10.1007/JHEP08(2021)134},
    journal       = {JHEP},
    pages         = {134},
    primaryclass  = {hep-th},
    volume        = {08}
}

@article{Hoback:2020pgj,
    title         = {{Towards Feynman rules for conformal blocks}},
    author        = {Hoback, Sarah and Parikh, Sarthak},
    year          = {2021},
    eprint        = {2006.14736},
    archiveprefix = {arXiv},
    doi           = {10.1007/JHEP01(2021)005},
    journal       = {JHEP},
    pages         = {005},
    primaryclass  = {hep-th},
    volume        = {01}
}

@article{Parikh:2019dvm,
    title         = {{A multipoint conformal block chain in $d$ dimensions}},
    author        = {Parikh, Sarthak},
    year          = {2020},
    eprint        = {1911.09190},
    archiveprefix = {arXiv},
    doi           = {10.1007/JHEP05(2020)120},
    journal       = {JHEP},
    pages         = {120},
    primaryclass  = {hep-th},
    volume        = {05}
}

@article{fortin2020all,
    title   = {All global one-and two-dimensional higher-point conformal blocks},
    author  = {Fortin, Jean-Fran{\c{c}}ois and Ma, Wen-Jie and Skiba, Witold},
    year    = {2020},
    eprint  = {arXiv:2009.07674},
    journal = { }
}

@article{Donnay:2020guq,
    title         = {{Asymptotic Symmetries and Celestial CFT}},
    author        = {Donnay, Laura and Pasterski, Sabrina and Puhm, Andrea},
    year          = {2020},
    eprint        = {2005.08990},
    archiveprefix = {arXiv},
    doi           = {10.1007/JHEP09(2020)176},
    journal       = {JHEP},
    pages         = {176},
    primaryclass  = {hep-th},
    reportnumber  = {CPHT-RR022.042020},
    volume        = {09}
}

@article{Fortin:2019zkm,
    title         = {{Higher-Point Conformal Blocks in the Comb Channel}},
    author        = {Fortin, Jean-Fran\c{c}ois and Ma, Wenjie and Skiba, Witold},
    year          = {2020},
    eprint        = {1911.11046},
    archiveprefix = {arXiv},
    doi           = {10.1007/JHEP07(2020)213},
    journal       = {JHEP},
    pages         = {213},
    primaryclass  = {hep-th},
    volume        = {07}
}

@article{Anous:2020vtw,
    title         = {{On the Virasoro six-point identity block and chaos}},
    author        = {Anous, Tarek and Haehl, Felix M.},
    year          = {2020},
    eprint        = {2005.06440},
    archiveprefix = {arXiv},
    doi           = {10.1007/JHEP08(2020)002},
    journal       = {JHEP},
    number        = {08},
    pages         = {002},
    primaryclass  = {hep-th},
    volume        = {08}
}

@article{Fortin:2020bfq,
    title         = {{Seven-point conformal blocks in the extended snowflake channel and beyond}},
    author        = {Fortin, Jean-Fran\c{c}ois and Ma, Wen-Jie and Skiba, Witold},
    year          = {2020},
    eprint        = {2006.13964},
    archiveprefix = {arXiv},
    doi           = {10.1103/PhysRevD.102.125007},
    journal       = {Phys. Rev. D},
    number        = {12},
    pages         = {125007},
    primaryclass  = {hep-th},
    volume        = {102}
}

@article{Fortin:2020yjz,
    title         = {{Six-point conformal blocks in the snowflake channel}},
    author        = {Fortin, Jean-Fran\c{c}ois and Ma, Wen-Jie and Skiba, Witold},
    year          = {2020},
    eprint        = {2004.02824},
    archiveprefix = {arXiv},
    doi           = {10.1007/JHEP11(2020)147},
    journal       = {JHEP},
    pages         = {147},
    primaryclass  = {hep-th},
    volume        = {11}
}

@article{Goncalves:2019znr,
    title         = {{$20'$ Five-Point Function from $AdS_5\times S^5$ Supergravity}},
    author        = {Gon\c{c}alves, Vasco and Pereira, Raul and Zhou, Xinan},
    year          = {2019},
    eprint        = {1906.05305},
    archiveprefix = {arXiv},
    doi           = {10.1007/JHEP10(2019)247},
    journal       = {JHEP},
    pages         = {247},
    primaryclass  = {hep-th},
    reportnumber  = {PUPT-2588},
    volume        = {10}
}

@article{Liu:2018jhs,
    title         = {{$d$-dimensional SYK, AdS Loops, and $6j$ Symbols}},
    author        = {Liu, Junyu and Perlmutter, Eric and Rosenhaus, Vladimir and Simmons-Duffin, David},
    year          = {2019},
    eprint        = {1808.00612},
    archiveprefix = {arXiv},
    doi           = {10.1007/JHEP03(2019)052},
    journal       = {JHEP},
    pages         = {052},
    primaryclass  = {hep-th},
    reportnumber  = {CALT-TH-2018-023},
    volume        = {03}
}

@article{Karateev:2018oml,
    title         = {{Harmonic Analysis and Mean Field Theory}},
    author        = {Karateev, Denis and Kravchuk, Petr and Simmons-Duffin, David},
    year          = {2019},
    eprint        = {1809.05111},
    archiveprefix = {arXiv},
    doi           = {10.1007/JHEP10(2019)217},
    journal       = {JHEP},
    pages         = {217},
    primaryclass  = {hep-th},
    reportnumber  = {CALT-TH 2018-036},
    volume        = {10}
}

@article{Parikh:2019ygo,
    title         = {{Holographic dual of the five-point conformal block}},
    author        = {Parikh, Sarthak},
    year          = {2019},
    eprint        = {1901.01267},
    archiveprefix = {arXiv},
    doi           = {10.1007/JHEP05(2019)051},
    journal       = {JHEP},
    pages         = {051},
    primaryclass  = {hep-th},
    volume        = {05}
}

@article{Rosenhaus:2018zqn,
    title         = {{Multipoint Conformal Blocks in the Comb Channel}},
    author        = {Rosenhaus, Vladimir},
    year          = {2019},
    eprint        = {1810.03244},
    archiveprefix = {arXiv},
    doi           = {10.1007/JHEP02(2019)142},
    journal       = {JHEP},
    pages         = {142},
    primaryclass  = {hep-th},
    volume        = {02}
}

@article{Jepsen:2019svc,
    title         = {{Propagator identities, holographic conformal blocks, and higher-point AdS diagrams}},
    author        = {Jepsen, Christian Baadsgaard and Parikh, Sarthak},
    year          = {2019},
    eprint        = {1906.08405},
    archiveprefix = {arXiv},
    doi           = {10.1007/JHEP10(2019)268},
    journal       = {JHEP},
    pages         = {268},
    primaryclass  = {hep-th},
    reportnumber  = {PUPT-2589},
    volume        = {10}
}

@article{Sleight:2018ryu,
    title         = {{Anomalous Dimensions from Crossing Kernels}},
    author        = {Sleight, Charlotte and Taronna, Massimo},
    year          = {2018},
    eprint        = {1807.05941},
    archiveprefix = {arXiv},
    doi           = {10.1007/JHEP11(2018)089},
    journal       = {JHEP},
    pages         = {089},
    primaryclass  = {hep-th},
    reportnumber  = {PUPT-2567},
    volume        = {11}
}

@article{Lam:2017ofc,
    title         = {{Conformal Basis, Optical Theorem, and the Bulk Point Singularity}},
    author        = {Lam, Ho Tat and Shao, Shu-Heng},
    year          = {2018},
    eprint        = {1711.06138},
    archiveprefix = {arXiv},
    doi           = {10.1103/PhysRevD.98.025020},
    journal       = {Phys. Rev. D},
    number        = {2},
    pages         = {025020},
    primaryclass  = {hep-th},
    volume        = {98}
}

@article{Kravchuk:2018htv,
    title         = {{Light-ray operators in conformal field theory}},
    author        = {Kravchuk, Petr and Simmons-Duffin, David},
    year          = {2018},
    eprint        = {1805.00098},
    archiveprefix = {arXiv},
    doi           = {10.1007/JHEP11(2018)102},
    journal       = {JHEP},
    pages         = {102},
    primaryclass  = {hep-th},
    reportnumber  = {CALT-TH 2018-018},
    volume        = {11}
}

@article{Pasterski:2017kqt,
    title         = {{Conformal basis for flat space amplitudes}},
    author        = {Pasterski, Sabrina and Shao, Shu-Heng},
    year          = {2017},
    eprint        = {1705.01027},
    archiveprefix = {arXiv},
    doi           = {10.1103/PhysRevD.96.065022},
    journal       = {Phys. Rev. D},
    number        = {6},
    pages         = {065022},
    primaryclass  = {hep-th},
    volume        = {96}
}

@article{Hogervorst:2017sfd,
    title         = {{Crossing symmetry in alpha space}},
    author        = {Hogervorst, Matthijs and van Rees, Balt C.},
    year          = {2017},
    eprint        = {1702.08471},
    archiveprefix = {arXiv},
    doi           = {10.1007/JHEP11(2017)193},
    journal       = {JHEP},
    pages         = {193},
    primaryclass  = {hep-th},
    reportnumber  = {YITP-SB-17-7},
    volume        = {11}
}

@article{Pasterski:2016qvg,
    title         = {{Flat Space Amplitudes and Conformal Symmetry of the Celestial Sphere}},
    author        = {Pasterski, Sabrina and Shao, Shu-Heng and Strominger, Andrew},
    year          = {2017},
    eprint        = {1701.00049},
    archiveprefix = {arXiv},
    doi           = {10.1103/PhysRevD.96.065026},
    journal       = {Phys. Rev. D},
    number        = {6},
    pages         = {065026},
    primaryclass  = {hep-th},
    volume        = {96}
}

@article{Alkalaev:2015fbw,
    title         = {{From global to heavy-light: 5-point conformal blocks}},
    author        = {Alkalaev, K. B. and Belavin, V. A.},
    year          = {2016},
    eprint        = {1512.07627},
    archiveprefix = {arXiv},
    doi           = {10.1007/JHEP03(2016)184},
    journal       = {JHEP},
    pages         = {184},
    primaryclass  = {hep-th},
    reportnumber  = {FIAN-TD-2015-15},
    volume        = {03}
}

@article{SimmonsDuffin:2012uy,
    title         = {{Projectors, Shadows, and Conformal Blocks}},
    author        = {Simmons-Duffin, David},
    year          = {2014},
    eprint        = {1204.3894},
    archiveprefix = {arXiv},
    doi           = {10.1007/JHEP04(2014)146},
    journal       = {JHEP},
    pages         = {146},
    primaryclass  = {hep-th},
    volume        = {04}
}

@article{groenevelt2006wilson,
    title         = {Wilson function transforms related to Racah coefficients},
    author        = {Groenevelt, Wolter},
    year          = {2006},
    eprint        = {math/0501511},
    archiveprefix = {arXiv},
    journal       = {Acta Applicandae Mathematica},
    number        = {2},
    pages         = {133--191},
    primaryclass  = {math.CA},
    publisher     = {Springer},
    volume        = {91}
}

@article{groenevelt2003wilson,
    title         = {The Wilson function transform},
    author        = {Groenevelt, Wolter},
    year          = {2003},
    eprint        = {math/0306424},
    archiveprefix = {arXiv},
    journal       = {International Mathematics Research Notices},
    number        = {52},
    pages         = {2779--2817},
    primaryclass  = {math.CA},
    publisher     = {OUP},
    volume        = {2003}
}

@article{Liu:1998th,
    title         = {{Scattering in anti-de Sitter space and operator product expansion}},
    author        = {Liu, Hong},
    year          = {1999},
    eprint        = {hep-th/9811152},
    archiveprefix = {arXiv},
    doi           = {10.1103/PhysRevD.60.106005},
    journal       = {Phys. Rev. D},
    pages         = {106005},
    reportnumber  = {IMPERIAL-TP-98-99-012},
    volume        = {60}
}

@article{wilson1980some,
    title     = {Some hypergeometric orthogonal polynomials},
    author    = {Wilson, James A},
    year      = {1980},
    journal   = {SIAM Journal on Mathematical Analysis},
    number    = {4},
    pages     = {690--701},
    publisher = {SIAM},
    volume    = {11}
}

@article{Dobrev:1976vr,
    title   = {{On the Clebsch-Gordan Expansion for the Lorentz Group in n Dimensions}},
    author  = {Dobrev, V. K. and Mack, G. and Todorov, I. T. and Petkova, V. B. and Petrova, S. G.},
    year    = {1976},
    doi     = {10.1016/0034-4877(76)90057-4},
    journal = {Rept. Math. Phys.},
    pages   = {219--246},
    volume  = {9}
}

@article{Ferrara:1972uq,
    title   = {{The shadow operator formalism for conformal algebra. Vacuum expectation values and operator products}},
    author  = {Ferrara, S. and Grillo, A.F. and Parisi, G. and Gatto, R.},
    year    = {1972},
    doi     = {10.1007/BF02907130},
    journal = {Lett. Nuovo Cim.},
    pages   = {115--120},
    volume  = {4S2}
}

@book{Gelfand1,
    title     = {Generalized Functions, Volume 1},
    author    = {Gel'fand, I.M. and Shilov, G.E.},
    year      = {2014},
    isbn      = {9781483261591},
    publisher = {Elsevier Science},
    url       = {https://books.google.com/books?id=dgnjBQAAQBAJ}
}

@book{Gelfand2,
    title     = {Generalized Functions, Volume 2},
    author    = {Gel'Fand, I.M. and Shilov, G.E.},
    year      = {2013},
    isbn      = {9781483262307},
    publisher = {Elsevier Science},
    url       = {https://books.google.com/books?id=iMo3BQAAQBAJ}
}

@book{andrews1999special,
    title     = {Special Functions},
    author    = {Andrews, G.E. and Askey, R. and Roy, R.},
    year      = {1999},
    isbn      = {9780521789882},
    lccn      = {98025757},
    publisher = {Cambridge University Press},
    series    = {Encyclopedia of Mathematics and its Applications},
    url       = {https://books.google.com/books?id=kGshpCa3eYwC}
}

@book{Dobrev:1977qv,
    title  = {{Harmonic Analysis on the n-Dimensional Lorentz Group and Its Application to Conformal Quantum Field Theory}},
    author = {Dobrev, V. K. and Mack, G. and Petkova, V. B. and Petrova, S. G. and Todorov, I. T.},
    year   = {1977},
    doi    = {10.1007/BFb0009678},
    volume = {63}
}

@inproceedings{Pasterski:2021raf,
    title         = {{Celestial Holography}},
    author        = {Pasterski, Sabrina and Pate, Monica and Raclariu, Ana-Maria},
    year          = {2021},
    eprint        = {2111.11392},
    archiveprefix = {arXiv},
    booktitle     = {{Snowmass 2021}},
    month         = {11},
    primaryclass  = {hep-th}
}
\end{document}